\documentclass[11pt]{article}
\usepackage{amssymb}
\usepackage{graphicx}
\usepackage{amsmath}
\usepackage{makeidx}
\usepackage{indentfirst}
\usepackage[T1]{fontenc}
\usepackage[utf8]{inputenc}
\usepackage{authblk}
\usepackage{rotating}

\newcounter{resultnum}[section]
\newcounter{conclusionnum}[section]
\newcounter{conditionnum}[section]
\newcounter{conjecturenum}[section]
\newcounter{examplenum}[section]
\newcounter{exercisenum}[section]
\newcounter{lemmanum}[section]
\newcounter{notationnum}[section]
\newcounter{theoremnum}[section]
\newcounter{definitionnum}[section]
\newcounter{corollarynum}[section]
\newcounter{remarknum}[section]
\newcounter{propositionnum}[section]
\newcounter{acknowledgementnum}[section]
\newcounter{algorithmnum}[section]
\newcounter{axiomnum}[section]
\newcounter{casenum}[section]
\newcounter{claimnum}[section]
\newcounter{summarynum}[section]
\newcounter{problemnum}[section]
\begin{document}

\title{Heterotic Supergravity with Internal Almost-K\"{a}hler Spaces;\\  Instantons  for  $SO(32)$, or $E_8 \times E_8$, Gauge Groups; and Deformed Black Holes with Soliton,  Quasiperiodic and/or Pattern-forming Structures}
\date{January 25, 2017}
\author{
Lauren\c{t}iu Bubuianu\\
{\small \textit{TVR Ia\c{s}i, \ 33 Lasc\v{a}r Catargi street, 700107 Ia\c{s}i, Romania }} \\
{\small and \textit{University Apollonia, 2 Muzicii street, Ia\c{s}i, Romania }} \\
{\small \textit{email: laurentiu.bubuianu@tvr.ro }}\\ ${}$ \\
Klee Irwin \\
{\small \textit{Quantum Gravity Research; 101 S. Topanga Canyon Blvd \# 1159. Topanga, CA 90290, USA}} \\
{\small \textit{email: klee@quantumgravityresearch.org }}\\ ${}$ \\
Sergiu I. Vacaru\\
{\small \textit{Quantum Gravity Research; 101 S. Topanga Canyon Blvd \#
1159. Topanga, CA 90290, USA}}  \\
{\small and \textit{University "Al. I. Cuza" Ia\c si, Project IDEI }}\footnote{Address for contact:  \textit{Flat 4 Brefney house, Fleet street, Ashton-under-Lyne, OL6 7PG, the UK  }} \\
{\small \textit{email: sergiu.vacaru@gmail.com}}
}
\maketitle


\begin{abstract}
Heterotic supergravity with (1+3)--dimensional domain wall configurations and (warped) internal, six dimensional, almost-K\"{a}hler manifolds $\ ^6\mathbf{X} $ are studied. Considering  ten dimensional spacetimes with nonholonomic distributions and conventional double fibrations, 2+2+...=2+2+3+3, and associated $SU(3)$ structures on internal space, we generalize for real, internal, almost symplectic gravitational structures the constructions with gravitational and gauge instantons of tanh-kink type \cite{lecht1,harl}. They include the first $\alpha ^{\prime }$ corrections to the heterotic supergravity action, parameterized in a form to imply nonholonomic deformations of the Yang-Mills sector and corresponding Bianchi identities. We show how it is possible to construct a variety of solutions depending on the type of nonholonomic distributions and deformations of 'prime' instanton configurations characterized by two real supercharges.
This corresponds to $\mathcal{N}=1/2$ supersymmetric, nonholonomic manifolds from the four dimensional point of view. Our method provides a unified description of embedding nonholonomically deformed tanh-kink-type instantons into half-BPS solutions of heterotic supergravity. This allows us to elaborate new geometric methods of constructing exact solutions of motion equations, with first order $\alpha ^{\prime }$ corrections to the heterotic supergravity. Such a formalism is applied for general and/or warped almost-K\"{a}hler configurations, which allows us to generate nontrivial (1+3)-d domain walls and black hole deformations determined by quasiperiodic internal space structures. This formalism is utilized in our associated publication \cite{partner} in order to construct and study generic off-diagonal nonholonomic deformations of the Kerr metric, encoding contributions from heterotic supergravity.


\vskip3pt MSC2010:\ 8C15, 8D99, 83E99;\  and  PACS2008:\ 04.20.Jb, 04.50.-h, 04.20.Cv
\end{abstract}

\tableofcontents

\section{Introduction}

The majority of different vacua in string gravity theories, including four
dimensional spacetime domains, are elaborated with 6-d internal manifolds
adapted to certain toroidal compactification or warping of extra dimensions.
With the aim of obtaining interesting and realistic models of
lower-dimensional physics, elaborations of 10-d theories with special
Calabi--Yau (and/or more general $SU(3)$ structure) manifolds were used.
Such constructions are related to pseduo-Euclidean 4-d domain configurations
and warped almost-K\"{a}hler internal spaces. Recent results and reviews
related to superstrings, flux compactifications, D-branes, instantons etc.,
are cited respectively \cite{lecht1,harl,gran,wecht, doug,lust, samt}.

Further generalizations with nontrivial solutions in the 4-d domain, such as
reproductions of 4-d black hole solutions and cosmological scenarios related
to modified gravity theories (MGTs) encoding information from extra
dimension internal spaces, are possible if richer geometric structures are
involved. Nonholonomic distributions with splitting on 4-d, 6-d and 10-d
manifolds as well as almost-K\"{a}hler internal manifolds are considered
when bimetric structures, possible nontrivial mass terms for the
graviton, locally anisotropic effects etc. can be reproduced in the
framework of heterotic supergravity theory, see \cite{vjgp,vwitten}. For
MGTs and their applications, we cite \cite%
{ferrara1,kounnas3,kehagias,stavr,saridak,mavromat,basstavr,mavr,odints1,capoz,drg1,hr1,vgrg,vsingl2, vcosmsol2,vcosmsol3,vcosmsol4,vcosmsol5}
and references therein.

In a series of works \cite%
{vpars,sv2001,vapexsol,vex1,vex2,vex3,veym,svvvey,tgovsv,vtamsuper}, the
so-called anholonomic frame deformation method, AFDM, of constructing exact
solutions in commutative and noncommutative (super) gravity and geometric
flow theories has been further developed. By straightforward analytic computations,
it was proven that it is possible to decouple the gravitational field
equations and generate new classes of solutions in various theories of
gravity with metric, nonlinear, N-, and linear connection structures. The
geometric formalism was based on spacetime fibrations determined by
nonholonomic distributions with splitting of dimensions, 2 (or 3) + 2 + 2 +
.... In explicit form, certain classes of N-elongated frames of reference,
considered formal extensions/embeddings of 4-d spacetimes into higher
dimensional spacetimes were introduced and necessary types of adapted linear
connections were defined. Such connections are called distinguished,
d-connections, and defined in some form that preserves the N-connection
splitting. In Einstein gravity, a d-connection is considered as an auxiliary
one which is supplementary to the Levi-Civita, LC, connection. For certain well
defined conditions, the canonical d-connection can be uniquely defined by
the metric structure following the conditions of metric compatibility and
the conditions of zero values for "pure" horizontal and vertical components
but nonzero, nonholonomically-induced, mixed vertical-horizontal torsion
components. Surprisingly, such a canonical d-connection allows us to
decouple the motion equations into general form. As a result, we can
generate various classes of exact solutions in generalized/modified string
and gravity theories. Having constructed a class of generalized solutions in
explicit form (depending on generating and integration functions,
generalized effective sources and integration constants), we then impose
some additional constraints at the end, resulting in zero induced torsion
fields. In this way, we can always "extract" solutions for LC-configurations
and/or Einstein gravity. It should be emphasized that it is important to
impose the zero-torsion conditions at the end, i.e. after we found a class
of generalized solutions. We can not decouple and solve in general forms the
corresponding systems of PDEs if we use the LC-connection from the very
beginning. Here it should be noted that to work with nontrivial torsion
configurations is important in order to find exact solutions in string
gravity and gauge gravity models.

Using the AFDM, a series of exact and/or small parameter depending solutions were constructed, which for small deformations mimic rotoid Kerr - de Sitter like black holes/ellipsoids self--consistently embedded into generic off-diagonal backgrounds of arbitrary finite dimensions. A number of examples for 5,6 and 8 dimensional (non) commutative and/or supersymmetric spacetimes are provided, see examples in \cite{sv2001,vex3,veym,tgovsv,vtamsuper} and references therein. Such backgrounds can be of solitonic/ vertex / instanton type. In this paper, we develop and apply these nonholonomic geometric methods for  constructing  new classes of exact and parametric solutions in  heterotic string theory. The motion equations are re-written in certain nonholonomic variables as generalized (effective) Einstein equations for 4-d spacetimes, encoding nontrivial geometric constructions on extra dimension internal spaces. Here we note that by using nonholonomic distributions and corresponding classes of solutions for heterotic string gravity, it is possible to mimic physically important effects in modified gravity. In a series of works \cite{vgrg,vbranef}), we studied the acceleration of the universe, certain theories with dark energy and dark matter locally anisotropic interactions and effective renormalization of quantum gravity models via nonlinear generic off-diagonal interactions on effective Einstein spaces. In our associated publication \cite{partner}, we explained in more detail the AFDM for constructing general classes of exact solutions with generic off-diagonal metrics in heterotic supergravity and generalized connections depending on all 4-d and 10-d coordinates via corresponding classes of generating and integration functions. In explicit form, we constructed a series of new  solutions in heterotic string gravity describing deformations of the Kerr metric by effective string sources,  ellipsoidal configurations and extra dimensional string modifications by a nontrivial cosmological constant, a NS 3-form and/or 6-d almost -K\"{a}hler  internal spaces.  The main results of this and the associated publications are based on the idea that we can generate physically interesting domain wall configurations (for instance, 4-d deformed black holes) by considering richer geometric structures on the internal space. In this work, we shall construct and analyze physical implications of a series of new classes of solutions with  deformations of Kerr metrics by string solitonic distributions/ waves in internal space; with quasiperiodic (quasicrystal like and/ or with nonlinear three-wave interactions and extra dimensional temporal chaos); and examples of configurations with solitonic and/or quasiperiodic YM and/or almost K\"{a}hler  sections. We shall prove that we can parametrize and generalize all possible off-diagonal solutions in heterotic supergravity in terms of such variables with internal geometric objects that are determined by an almost-K\"{a}hler geometry. This is possible for nonholonomic distributions with conventional splitting $2+2+2=3+3$ (such constructions are based on former results in \cite{vjgp,vwitten,vmedit}).

In this paper, we apply deformation methods used in the geometry of nonholonomic and almost-K\"{a}hler manifolds in order to study heterotic supergravity derived in the low-energy limit of heterotic string theory \cite{grsch,gross1,gross2} . We cite also section 4.4 in \cite{lecht1} for a summary of previous results and certain similar conventions on warped configurations and modified gravitational equations.\footnote{We shall develop a different system of notation with N-connections and auxiliary d--connections which allows us to define geometric objects on higher order shells of nonholonomically decomposed 10-d
spacetimes.} The main goal of this work is to formulate a geometric formalism which allows us to construct new classes of exact and parametric solutions describing nonholonomically deformed black hole solutions encoding various types of  quasiperiodic structures, see details and recent results in Refs. \cite{rucklidge12,rucklidge16,aschheim16,amaral16,asen,sv2013}. Such methods have been  applied in a partner article  \cite{partner} for integrating in generic off-diagonal forms, and for generalized connections, the equations of motion of heterotic supergravity up to and including terms of order $\alpha ^{\prime }.$    The solutions of heterotic supergravity which will be constructed in further sections describe (1+3)--dimensional walls endowed with generic off-diagonal metrics warped to an almost-K\"{a}hler 6-d internal space in the presence of nonholonomically deformed gravitational and gauge instantons induced by nontrivial soliton, quasiperiodic and/or pattern-forming structures. The generalized instanton contributions are adapted to a nontrivial, nonlinear connection structure determined by generic off-diagonal interactions which allows us to solve the Yang--Mills, YM, sector and the corresponding Bianchi identity at order $\alpha ^{\prime }$ (which is related to the gravitational constant in 10-d). Such 10-d solutions preserve two real supercharges, which correspond to the $\mathcal{N}=1/2$ supersymmetry. The almost-K\"{a}hler internal 6-d structure can be defined for various classes of solutions in 10-d gravity, including black hole deformations by quasiperiodic structures in string gravity, if we prescribe an effective Lagrange type generating function. In such an approach, we can work both with real nonholonomic gravitational and YM instanton configurations and/or consider deformed $SU(3)$ structures.

This article is organized as follows:\ In section \ref{s2}, we formulate a geometric approach to heterotic supergravity with almost K\"{a}hler internal 6-d configurations.  The  main geometric conventions are explained in Appendix \ref{as1} and used in that section  for defining noholonomic manifolds with domain--walls, G structures and corresponding BPS equations.   Then in section \ref{s3}, we provide solutions for nonholonomic instanton d-connections, almost-K\"{a}hler manifolds and N-adapted (effective) YM and instanton configurations. The equations of motion of heterotic supergravity are formulated in nonholonomic variables and discussed in detail. Section \ref{s4} provides a series of new and significant contributions in constructing exact and parametric 10-d solutions in string gravity depending, in general, on all 4-d and extra dimensional coordinates. We study examples of solitonic deformations of 4-d Kerr configurations and possible internal space extensions determined by stationary solitonic distributions and nonlinear waves. We construct solutions for black hole deformations by various types of quasiperiodic strictures (quasi-crystal like and three-wave interactions with possible extra dimensional chaos).  Such configurations are then modeled by exact solitonic and/ or quasiperiodic solutions for the YM and amost-K\"{a}hler  internal sectors. Finally, we summarize the results in section \ref{s5}  concluding that the heterotic supergravity theory can be formulated in nonholonomic variables which allows us to integrate in general form (see \cite{partner}) the corresponding modified Einstein equations and study stationary nonholonomic deformations of black hole solutions. In Appendix \ref{as1} we provide necessary geometric preliminaries on 10-d nonholonomic manifolds with 2+2+... splitting and associated nonlinear connection (N-connection) structures related to off-diagonal metric terms and certain classes of N-adapted frames. Such a formalism is necessary for the definition of nontrivial internal space geometric structures and for the decoupling of motions equations.

\section{Almost-K\"{a}hler Internal Configurations in Heterotic Supergravity}

\label{s2}

The goal of this section is to formulate a model of the $\mathcal{N}=1$ and
10-d supergravity coupled to a super Yang--Mills theory for a 10--d
nonholonomic manifold with fibered $4+2s$ structure (where $s=1,2,3$ labels
a geometric  structure with conventional 2-d shells). Readers are referred to
Appendix \ref{as1}, and references therein, containing an introduction into
the geometry of nonholonomic manifolds with nonlinear connection,
N-connection, splitting. Originally, such a formalism was considered for
developing certain methods of classical and quantum deformations in the
Einstein and Finsler modified gravity theories, and for formulating models of
almost-K\"{a}hler geometric flows and Lie algebroid structures, see \cite%
{vjgp,vwitten,vdq1,vmedit}.  In this work, such geometric methods are
developed for (super) spaces enabled with N--connection, structure and a
gauge d--connection $\ _{A}^{s}\widehat{\mathbf{D}}$ distorting in
nonholonomic form a connection $^{A}\check{\nabla}$ $\ $with gauge group $%
SO(32)$ or $E_{8}\times E_{8}.$ We note that in order to generate more general
classes of solutions and provide a more realistic phenomenology in string theory,
with more complex vacua, we shall work with nonholonomic deformations and
internal nonholonomic manifolds with a richer geometric structure (with
nontrivial N--connections, effective extra dimensional gravitational and
matter field interactions etc.). Such geometric constructions and methods
preserve less supersymmetry and lead to more realistic models. This follows
from the holonomy principle which can be formulated for the parallel
d--spinor equations (we shall omit details on spinor considerations for $\
^{s}\widehat{\mathbf{D}},$ which can be found in \cite{vnpfins,vnrflnc,vp,vt}%
). Our model of nonholonomic heterotic supergravity theory will be
determined by $(\mathcal{M},\mathbf{N,}\ ^{s}\mathbf{g,}\widehat{\mathbf{H}},%
\widehat{\phi },\ _{A}^{s}\widehat{\mathbf{D}}),$ where $\mathbf{N}$ and $\
^{s}\mathbf{g}$ are respective, N--connection and d--metric; $\widehat{%
\mathbf{H}}$ is the nonholonomic version of NS 3-form $\check{H};$ $\widehat{%
\phi }$ is a dilaton field on an N--anholonomic manifold which transform
into the standard one $\check{\phi}$ for LC--configurations; $\ _{A}^{s}%
\widehat{\mathbf{D}}$ is an N--adapted version of $^{A}\check{\nabla}$ \ (we
shall describe this construction below).

\subsection{Nonholonomic domain--walls in heterotic supergravity}

We analyze important geometric structures which can be defined for the
decomposition $\ ^{7}\mathbf{X}=\mathbb{R\times }\ ^{6}\mathbf{X.}$ For
holonomic distributions, it is always possible to rewrite the equation (\ref%
{bps2}) in terms of an $SU(3)$ structure defined on $\ ^{6}\mathbf{X}$ and
the domain wall direction. Such constructions are related to the complex
structure and K\"{a}hler geometry \cite{gray,lukas,lecht1}. In order to
develop a heterotic theory with generic off--diagonal metrics $\mathbf{g}=(\ ^{4}\mathbf{\mathbf{g},}\ ^{6}\mathbf{g)}$ (\ref{dm}), we need a richer,
real geometric structure for internal space $\ ^{6}\mathbf{X.}$ Up to
certain classes of frame transforms, any $\ ^{6}\mathbf{g}$ of Euclidean
signature can be uniquely related to an almost-K\"{a}hler geometry.

\subsubsection{The canonical d--connection and BPS equations}

By letting $\ ^{s}\widehat{\mathbf{D}}_{\mid \widehat{\mathcal{T}}%
=0}\rightarrow \ ^{s}\nabla ,$ see (\ref{zerotors}), and formulating, in
N--adapted form, the anomaly cancellation condition of the 10-d super
Yang--Mills, YM, theory coupled to $\mathcal{N}=1,$ 10-d supergravity can be
written as a Bianchi identity on $\widehat{\mathbf{H}},$%
\begin{equation}
\widehat{\mathbf{d}}\widehat{\mathbf{H}}=\frac{\alpha ^{\prime }}{4}Tr(%
\widehat{\mathbf{F}}\wedge \widehat{\mathbf{F}}-\widetilde{\mathbf{R}}\wedge
\widetilde{\mathbf{R}}),  \label{anomalcond}
\end{equation}%
implying two curvature 2-forms: $\widehat{\mathbf{F}}$ is the strength of
the gauge d--connection and $\ _{A}^{s}\widehat{\mathbf{D}};\widetilde{%
\mathbf{R}}$ is the strength of a d--connection $\widetilde{\mathbf{D}}$
which will be defined below; $\widehat{\mathbf{d}}$ is the 10-d exterior
derivative; $Tr$ means the trace on gauge group indices. There are different
connections which can be used in anomaly cancellation conditions of type (%
\ref{anomalcond}). This depends on the type of renormalization scheme and
the preferences of string theory physicists, see discussions in \cite%
{hull,ivanov,berg,becker} and references therein.\footnote{%
It important to note here that the constant $\alpha ^{\prime }$ is related
to the gravitational constant $\ _{10}\kappa $ in 10-d following formulas $\
(\ _{10}\kappa )^{2}=\frac{1}{4\pi }(4\pi ^{2}\alpha ^{\prime })^{4}=\frac{%
\ell _{str}^{8}}{4\pi }.$ The 4-d gravitational constant $\ _{4}\kappa $ is
given by $\ (_{4}\kappa )^{2}=\ell _{str}^{-6}\ (_{10}\kappa )^{2}$ $%
=M_{Pl}^{-2}=8\pi G^{-1},$ where $G$ is the Newton constant.} \ In non
N--adapted form, $\widetilde{\mathbf{D}}$ can be transformed into an almost-K%
\"{a}hler d--connection studied in details in Refs. \cite%
{vwitten,vjgp,vdq1,vmedit}, which allows us to find a number of nontrivial
generic off--diagonal solutions in 10-d following the AFDM. The instanton
equation for (anti) self-dual gravitational fields can be written in nonholonomic form, $\widetilde{\mathbf{R}}\cdot
\epsilon =0$ (for motivations details on holonomic configurations, see details in \cite{harl}). In further sections, we show how $G$ structures \cite{gray,lukas} can be adapted to N--connections. For such configurations, the Killing spinor $\epsilon $ is adapted to
above prime and target d-metric ansatz as $\epsilon (x^{i},t,y^{4},y^{\check{%
a}})=\rho (x^{i},t)\otimes \eta (y^{4},y^{\check{a}})\otimes \check{\theta},$
where the domain wall spinor $\rho $ has two real components corresponding
to the two real supercharges which the holonomic background $\mathbb{R}%
^{2,1} $ preserves; $\eta $ is a covariantly constant Majorana spinor on $\
^{7}\mathbf{X}=(y^{4},\ ^{6}\mathbf{X});$ and $\ \check{\theta}$ is an
eigenvector of the respective Pauli matrix.

In string theory, it is considered that for $\widehat{\mathbf{H}}=0$ the
internal manifold should be Ricci flat and K\"{a}hler and such a condition
does not stablize all K\"{a}hler moduli. A non-zero 3-form flux of the above
breaks scale invariance and provides a corresponding stabilization \cite%
{horow}. In a similar form, the canonical d--connection structure may result
in stable configurations for corresponding nonholonomic constraints. For
instance, the $\alpha ^{\prime }$--solutions with nonzero $\check{H}$
constructed in \cite{lecht1} preserve a $\mathcal{N}=1/2$ supersymmetry
(with two real supercharges instead of four for the usual $\mathcal{N}=1$
supersymmetry) in $1+3$ external dimensions. This is implied by the so
called Bogomol'nyi-Prasad-Sommerfield, BPS, conditions halving the
supersymmetry by the presence of a domain wall, see details in references
\cite{hull,ivanov,berg,becker,lecht1}.

At the zeroth order in $\alpha ^{\prime },$ zero value of the NS 3-form $%
\check{H}$ and a dilaton field $\check{\phi}=const,$ the PBS configurations
are given by $\mathcal{\check{M}}=\mathbb{R}^{2,1}\times \check{c}(\ ^{6}X);\check{H}=0,%
\check{\phi}=const$,
where $\check{c}(\ ^{6}X)$ is the metric cone over a 6-d almost-K\"{a}hler
manifold $\ ^{6}X.$ In this formula, $\mathbb{R}^{2,1}$ is a
pseudo-Euclidean space with signature $(++-)$ and local coordinates $x^{%
\check{i}}=(x^{i},y^{3}=t),$ for $\check{i},\check{j},...=1,2,3$ and $%
i,j,...1,2;$ the local coordinates on $\ ^{6}X$ are those used for the
shells $s=1,2,3,$ when
 $u^{\check{a}}=y^{\check{a}}=%
\{u^{a_{1}}=y^{a_{1}}=(y^{5},y^{6}),u^{a_{2}}=y^{a_{2}}=(y^{7},y^{8}),u^{a_{3}}=y^{a_{3}}=(y^{9},y^{10})\}$,
with indices $\check{a},\check{b},...=5,6,7,8,9,10,$ are labeled in a form
compatible with coordinate conventions (\ref{coordconv}). The space-like
coordinate $u^{4}=y^{4}$ will be used for a 7-d warped extension of $\
^{6}X\rightarrow $ $\ ^{7}X=(y^{4},\ ^{6}X)$ with local coordinates on $\
^{7}X,$ $\ y^{\widetilde{a}%
}=(y^{4},y^{a_{1}},y^{a_{2}},y^{a_{3}})=(y^{4},y^{5},y^{6},y^{7},y^{8},y^{9},y^{10})
$ for indices $\widetilde{a},\widetilde{b},...=4,5,...10.$

\subsubsection{Nonholonomic domain--wall backgrounds}

In Einstein--Cartan and gauge field theories, torsion fields have certain
sources subjected to algebraic equations. If a background with vanishing
fermionic vacuum expectations does not have prescribed nonholonomic
distributions, the supersymmetry transformations of the corresponding
fermionic fields are zero. Such conditions for the holonomic backgrounds are
known as BPS equations. Up to and including terms of order $\alpha ^{\prime
} $ for N--anholonomic backgrouns, the BPS equations are formulated for a
Majorana--Weyl spinor $\epsilon ,$%
\begin{eqnarray}
(\widehat{\mathbf{D}}-\widehat{\mathbf{H}})\cdot \epsilon &=&0,
\label{bpseq} \\
(\widehat{\mathbf{d}}\widehat{\phi }-\frac{1}{2}\widehat{\mathbf{H}})\cdot
\epsilon &=&0,  \notag \\
\widehat{\mathbf{F}}\cdot \epsilon &=&0.  \notag
\end{eqnarray}%
It should be noted that in this work, hatted boldface objects denote 10--d
geometric/physical objects on N--anholonomic manifolds enabled with
d--connection structure $\widehat{\mathbf{D}}.$ We cite here Refs. \cite%
{gray,lukas} for similar details on BPS equations and holonomic heterotic
string theories. For N-adapted constructions and spinors on commutative and
noncommutative (super) manifolds and bundle spaces, the nonolonomic
Mayorana-Weyl spinors and modified Dirac operators are studied in \cite%
{vnpfins,vnrflnc,vp,vt}. In appendix \ref{assclifford}, there are stated
main formulas for N--adapted gamma matrices and Clifford distinguished
algebras  defined on nonholonomic manifolds.

In heterotic string gravity (see details and references in \cite{lecht1}), a geometric background is given by 4-d domain walls with 6 internal directions stating a
compact manifold $\ ^{6}X$ with $SU(3)$ structure for a 10-d metric ansatz and quadratic element:
\begin{eqnarray}
ds^{2}[\mathring{g}] &=&\mathring{g}_{\check{\alpha}\check{\beta}}(y^{4},y^{%
\check{a}})du^{\check{\alpha}}du^{\check{\beta}}=\mathring{g}_{\alpha
_{s}\beta _{s}}(y^{4},y^{\check{a}})du^{\alpha _{s}}du^{\beta _{s}}=
\label{pm1} \\
&=&e^{2A(y^{4},y^{\check{a}})}\left[ \left( dx^{1}\right) ^{2}+\left(
dx^{2}\right) ^{2}-\left( dy^{3}\right) ^{2}+e^{2B(y^{\check{a}})}\left(
dy^{4}\right) ^{2}+\mathring{g}_{\check{b}\check{c}}(y^{4},y^{\check{a}})dy^{%
\check{b}}dy^{\check{c}}\right] ,  \notag
\end{eqnarray}%
where $y^{3}=t$ and $y^{4}$ is chosen to be transverse to the domain wall given with coordinates $(x^{i},t).$ We can consider
orthonormal frames $e^{\check{a}}=e^{\check{a}}(y^{4},y^{\check{c}})e^{\check{a}^{\prime }}$ for a prescribed N--connection structure
$\mathbf{\mathring{N}}=\{\mathring{N}_{i_{s}}^{a_{s}}(y^{4},y^{c_{s}})\rightarrow
\mathring{N}_{i_{s}}^{\check{a}}(y^{4},y^{\check{c}})\}$ defined for a local
system of coordinates in the internal 6-d manifold $\ ^{6}X$ embedded via
warped coordinate $y^{4}$ into higher dimensional ones and transform it into
N-anholonomic manifold $\ ^{6}\mathbf{X\subset }\ ^{7}\mathbf{X\subset }%
\mathcal{M},$ endowed with a d--metric structure of type (\ref{dm}),
\begin{eqnarray}
ds^{2}[\mathbf{\mathring{g}}] &=&e^{2\mathring{A}(y^{4},y^{\check{a}%
})}[\left( dx^{1}\right) ^{2}+\left( dx^{2}\right) ^{2}-\left( \mathbf{%
\mathring{e}}^{3}\right) ^{2}+e^{2\mathring{B}(y^{\check{a}})}\left( \mathbf{%
\mathring{e}}^{4}\right) ^{2}+  \label{pm1d} \\
&&\mathring{g}_{a_{1}}(y^{4},y^{a_{1}})\left( \mathbf{\mathring{e}}%
^{a_{1}}\right) ^{2}+\mathring{g}_{a_{2}}(y^{4},y^{a_{2}})\left( \mathbf{%
\mathring{e}}^{a_{2}}\right) ^{2}+\mathring{g}_{a_{3}}(y^{4},y^{a_{3}})%
\left( \mathbf{\mathring{e}}^{a_{3}}\right) ^{2}],  \notag
\end{eqnarray}%
\begin{eqnarray}
\mbox{where } a_{0} &=&3:\ \mathbf{\mathring{e}}^{3}=dt+\mathring{n}_{i_{0}}dx^{i_{0}},%
\mbox{ for }\mathring{N}_{i_{0}}^{3}=\mathring{n}_{i_{0}}(x^{k},y^{4}),%
\mbox{ for }k,i_{0}=1,2;  \label{pm1dncc} \\
a_{0} &=&4:\ \mathbf{\mathring{e}}^{4}=dy^{4}+\mathring{w}_{i_{0}}dx^{i_{0}},%
\mbox{ for }\mathring{N}_{i_{0}}^{4}=\mathring{w}_{i_{0}}(x^{k},y^{4});
\notag \\
a_{1} &=&5:\ \mathbf{\mathring{e}}^{5}=dy^{5}+\mathring{n}_{i_{1}}dx^{i_{1}},%
\mbox{ for }\mathring{N}_{i_{1}}^{5}=\mathring{n}_{i_{1}}(x^{k},y^{4},y^{6}),%
\mbox{ for }i_{1}=1,2,3,4;  \notag \\
a_{1} &=&6:\ \mathbf{\mathring{e}}^{6}=dy^{6}+\mathring{w}_{i_{1}}dx^{i_{1}},%
\mbox{ for }\mathring{N}_{i_{1}}^{6}=\mathring{w}_{i_{1}}(x^{k},y^{4},y^{6});
\notag \\
a_{2} &=&7:\ \mathbf{\mathring{e}}^{7}=dy^{7}+\mathring{n}_{i_{2}}dx^{i_{2}},%
\mbox{ for }\mathring{N}_{i_{2}}^{7}=\mathring{n}%
_{i_{2}}(x^{k},y^{4},y^{6},y^{8}),\mbox{ for }i_{2}=1,2,3,4,5,6;  \notag \\
a_{2} &=&8:\ \mathbf{\mathring{e}}^{8}=dy^{8}+\mathring{w}_{i_{2}}dx^{i_{2}},%
\mbox{ for }\mathring{N}_{i_{2}}^{8}=\mathring{w}%
_{i_{2}}(x^{k},y^{4},y^{5},y^{6},y^{8});  \notag \\
a_{3} &=&9:\ \mathbf{\mathring{e}}^{9}=dy^{9}+\mathring{n}_{i_{3}}dx^{i_{3}},%
\mbox{ for }\mathring{N}_{i_{3}}^{9}=\mathring{n}%
_{i_{3}}(x^{k},y^{4},y^{5},y^{6},y^{7},y^{8},y^{10}),\mbox{ for }%
i_{3}=1,2,3,4,5,6,7,8;  \notag \\
a_{3} &=&10:\ \mathbf{\mathring{e}}^{10}=dy^{10}+\mathring{w}%
_{i_{3}}dx^{i_{3}},\mbox{ for }\mathring{N}_{i_{3}}^{10}=\mathring{w}%
_{i_{3}}(x^{k},y^{4},y^{5},y^{6},y^{7},y^{8},y^{10}).  \notag
\end{eqnarray}%
In this paper, we shall study stationary configurations with Killing
symmetry on $\partial _{t}$ when the metric/d--metric ansatz do not depend
on coordinate $y^{3}=t.$ We call an ansatz $\mathring{g}$\ (\ref{pm1}) as a
prime off--diagonal metric and an ansatz $\mathbf{\mathring{g}}$ (\ref{pm1d}%
) (prime metrics will be labeled by a circle $\circ $).\footnote{\label{fn7}%
It should be emphasized that indices $a_{1}=(5,6),a_{2}=(7,8),a_{3}=(9,10)$
are shell adapted but indices $\check{a},\check{c},\check{a}^{\prime }$ may
take, in general, values $6,7,...10$ with shell mixing of indices. In order
to apply the AFDM, we shall always consider certain frame/coordinate
transforms with N--adapted shell redefinitions of interior indices and
coordinates. Indices with "inverse" hats are convenient for parametrization
of almost-K\"{a}hler structures but shell indices are important for
constructing exact off-diagonal solutions in 10-d gravity.}

The overall goal of this and associated \cite{partner} articles is to study
nonholonomic deformations of a primary metric $\mathbf{\mathring{g}}$ into a
target metric $\mathbf{\ }^{s}\mathbf{g,}$ {\small
\begin{eqnarray}
\mathbf{\mathring{g}} &\mathbf{=}&\mathbf{[\mathring{g}}_{\alpha _{s}},%
\mathring{N}_{i_{s}}^{a_{s}}\mathbf{]\rightarrow \ }^{s}\mathbf{g}=[\mathbf{g%
}_{\alpha _{s}}=\eta _{\alpha _{s}}\mathbf{\mathring{g}}_{\alpha
_{s}},N_{i_{s}}^{a_{s}}=\eta _{i_{s}}^{a_{s}}\mathring{N}_{i_{s}}^{a_{s}}],%
\rightarrow \mathbf{\ }_{\varepsilon }^{s}\mathbf{g}=[\mathbf{g}_{\alpha
_{s}}=(1+\varepsilon \chi _{\alpha _{s}})\mathbf{\mathring{g}}_{\alpha
_{s}},N_{i_{s}}^{a_{s}}=(1+\varepsilon \chi _{i_{s}}^{a_{s}})\mathring{N}%
_{i_{s}}^{a_{s}}],  \notag \\
\mbox{ for }\eta _{\alpha _{s}} &\simeq &1+\varepsilon \chi _{\alpha
_{s}},\eta _{i_{s}}^{a_{s}},\eta _{i_{s}}^{a_{s}}\simeq 1+\varepsilon \chi
_{i_{s}}^{a_{s}},\mbox{ where }0\leq \varepsilon \ll 1.  \label{nonhdef}
\end{eqnarray}%
} In these formulas, we do not consider summations on repeating indices. Any
target metric $\ ^{s}\mathbf{g}$ will be subjected to these conditions to
define new classes of solutions for certain systems of nonlinear PDEs in
heterotic string gravity (see next section) or in a MGT. It is always
possible to model the internal 6-d space as an almost-K\"{a}hler manifold.
For certain subclasses of solutions with $\lim_{\varepsilon \rightarrow 0}\
^{s}\mathbf{g\rightarrow \mathring{g},}$ the $\eta $--polarization functions
$(\eta _{\alpha _{s}},\eta _{i_{s}}^{a_{s}})$ $\rightarrow 1.$ In general,
such limits with a small parameter $\varepsilon $ may not exist, or can
behave singularly.

It is possible to preserve, in the tangent bundle $T\mathcal{M}$, the $(1+2)$%
--dimensional Lorentz invariance together with N--connection splitting if
restrictions on $\widehat{\phi }$ and $\widehat{\mathbf{H}}_{\alpha
_{s}\beta _{s}\mu _{s}}:$
\begin{equation}
\mathbf{e}_{\check{i}}\widehat{\phi }=0,\widehat{\mathbf{H}}_{\ \check{i}%
\widetilde{a}\widetilde{b}}=0,\widehat{\mathbf{H}}_{\ \check{i}\check{j}%
\widetilde{b}}=0.  \label{cond3}
\end{equation}%
are considered. For a flat 3-d Minkowski spacetime, the only non-zero
components of the NS 3-form flux are $\widehat{\mathbf{H}}_{4\check{a}\check{%
b}},\widehat{\mathbf{H}}_{\check{a}\check{b}\check{c}}$ and $\ \widehat{%
\mathbf{H}}_{\check{i}\check{j}\check{k}}=\ell \sqrt{|\mathbf{g}_{\check{i}%
\check{j}}|}\mathbf{\epsilon }_{\check{i}\check{j}\check{k}}$ for $\ell
=const$ and the totally antisymmetric tensor on $\mathbb{R}^{2,1}$ with
normalization $\mathbf{\epsilon }_{123}=1.$

Let us define $G_{2}$ d--structure adapted to the N--connection splitting in
$\ ^{7}\mathbf{X}=\mathbb{R\times }\ ^{6}\mathbf{X}$ enabled with an
arbitrary d--metric (of the type included in (\ref{pm1}) and (\ref{pm1d})),
respectively,
\begin{eqnarray}
ds^{2}[\ ^{7}\mathring{g}] &=&e^{2B(y^{\check{a}})}\left( dy^{4}\right) ^{2}+%
\mathring{g}_{\check{b}\check{c}}(y^{4},y^{\check{a}})dy^{\check{b}}dy^{%
\check{c}}\mbox{ and }  \label{7dans} \\
ds^{2}[^{7}\mathbf{\mathring{g}}] &=&e^{2\mathring{B}(y^{\check{a}})}\left(
\mathbf{\mathring{e}}^{4}\right) ^{2}+\mathring{g}_{a_{1}}(y^{4},y^{a_{1}})%
\left( \mathbf{\mathring{e}}^{a_{1}}\right) ^{2}+\mathring{g}%
_{a_{2}}(y^{4},y^{a_{2}})\left( \mathbf{\mathring{e}}^{a_{2}}\right) ^{2}+%
\mathring{g}_{a_{3}}(y^{4},y^{a_{2}},y^{a_{3}})\left( \mathbf{\mathring{e}}%
^{a_{3}}\right) ^{2}.  \notag
\end{eqnarray}%
Such holonomic structures were studied in \cite{gray,lukas} for certain
special parameterizations of functions $B,\mathring{B}$ and $\mathring{g}%
_{a_{s}}.$ For nonholonomic deformations (\ref{nonhdef}), we generated
d--metrics on $\ ^{7}\mathbf{X}$ with $\mathring{g}_{4}=$ $e^{2\mathring{B}%
(y^{\check{a}})}\rightarrow g_{4}=e^{2B(y^{\check{a}})}$ (via coordinate
transforms, we can consider parameterizations with $B=0$ but we keep a
nontrivial value of $B$ in order to compare our results with those for
holonomic K\"{a}hler configurations outlined in \cite{lecht1}, where $\Delta
$ is considered for $B$),
\begin{equation}
ds^{2}[^{7}\mathbf{g}]=e^{2B(y^{\check{a}})}\left( \mathbf{e}^{4}\right)
^{2}+g_{a_{1}}(y^{4},y^{a_{1}})\left( \mathbf{e}^{a_{1}}\right)
^{2}+g_{a_{2}}(y^{4},y^{a_{2}})\left( \mathbf{e}^{a_{2}}\right)
^{2}+g_{a_{3}}(y^{4},y^{\check{a}})\left( \mathbf{e}^{a_{3}}\right) ^{2}.
\notag
\end{equation}%
In this formula and (\ref{7dans}), the internal spaces, the indices and
coordinates can be written in any form we need for the definition of almost-K%
\"{a}hler or diadic shell structures as we explained above in footnote \ref%
{fn7}. The d--metric $^{7}\mathbf{g}$ defines the Hodge operator $\ast _{7}.$
The $G_{2}$ d--structure is given by a 3-form $\varpi \in \wedge ^{3}(\ ^{7}%
\mathbf{X})$ and its 7-d Hodge dual $\mathcal{W}:=\ ^{7}\ast \varpi \in
\wedge ^{4}(\ ^{7}\mathbf{X})$ (in a similar form, d-structures can be
introduced for d--spinors and 7-d gamma matrices). In N--adapted form with
absolute differential operator $\ ^{7}\mathbf{d}$ on $\ ^{7}\mathbf{X,}$ the
BPS equations (\ref{bpseq}) imply%
\begin{eqnarray}
\ ^{7}\mathbf{d}\varpi &=&2\ ^{7}\mathbf{d}\widehat{\phi }\wedge \varpi -\
^{7}\ast \widehat{\mathbf{H}}-\ell \mathcal{W},\ ^{7}\mathbf{d}\mathcal{W}%
=2\ ^{7}\mathbf{d}\widehat{\phi }\wedge \mathcal{W},  \notag \\
\ ^{7}\ast \ ^{7}\mathbf{d}\widehat{\phi } &=&-\frac{1}{2}\widehat{\mathbf{H}%
}\wedge \varpi ,\ ^{7}\ast \ell =2\widehat{\mathbf{H}}\wedge \mathcal{W}.
\label{bps2}
\end{eqnarray}%
Such relations can be written in N-adapted form with respect to frames (\ref%
{nadaptb}) and for $\ ^{s}\widehat{\mathbf{D}}.$

\subsection{Almost symplectic structures induced by effective Lagrange
distributions}

In order to enable the internal space with a complex K\"{a}hler structure,
one considers a decomposition of the Majorana spinor $\eta $ into two 6-d
spinors of definite chirality, $\eta =\frac{1}{\sqrt{2}}(\eta _{+}+\eta
_{-}).$ We can specifiy any $SU(3)$ structure on $\ ^{6}\mathbf{X}$ via a
couple of geometric objects $(J,\theta )$ a real 2-form $J$ and a complex
3-form $\theta =\theta _{+}+i\theta _{-},$ where $i^{2}=-1.$ Such values can
be defined for any fixed values $y^{4}$ and by using the chiral spinors $%
\eta _{\pm }.$ The relation between holonomic $G_{2}$ structure $(\varpi ,%
\mathcal{W})$ and $(J,\theta )$ is studied in \cite%
{lecht1,gray,lukas,chiossi}.

The real nonholonomic almost-K\"{a}hler geometry is also determined by a
couple $(\tilde{J},\tilde{\theta})$ which in our work is constructed to be
uniquely determined by a N--connection structure $\tilde{N}$ and a 6-d
metric $\ ^{6}\mathbf{g\rightarrow }\tilde{\theta}$ defined by a Lagrange
type distribution $\tilde{L}(y^{4},y^{\check{a}}).$ In this case, we also
have a nontrivial nonolonomically induced torsion with conventional
splitting of internal coordinates and indices in the form $y^{\check{a}}=(y^{%
\acute{\imath}},y^{\grave{a}}),$ where $\acute{\imath},\acute{j},...=5,6,7$
and (for conventional "vertical", v, indices) $\grave{a},\grave{b}%
,...=8,9,10.$

\subsubsection{The canonical N--connection, d--metric, and almost symplectic
2-forms}

We can re-parametrize a general Riemannian metric $\ ^{6}\mathbf{g\subset g}$
on $\ ^{6}\mathbf{X,}$ when possible dependencies on 4-d spacetime
coordinates $\ ^{0}u=(x^{i},y^{a})$ are considered as parameters \ [for
simplicity, we shall omit writting of 4-d coordinates if it will not result
in ambiguities], \
\begin{equation*}
ds^{2}[\ ^{6}g]=g_{\check{b}\check{c}}(x^{i},y^{a},y^{\check{a}})dy^{\check{b%
}}dy^{\check{c}}\mbox{ and/or }ds^{2}[^{6}\mathbf{\mathring{g}}]=g_{a_{1}}(\
^{0}u,y^{a_{1}})\left( \mathbf{e}^{a_{1}}\right) ^{2}+g_{a_{2}}(\
^{1}u,y^{a_{1}})\left( \mathbf{e}^{a_{2}}\right) ^{2}+g_{a_{3}}(\
^{2}u,y^{a_{2}})\left( \mathbf{e}^{a_{3}}\right) ^{2}.
\end{equation*}%
in the form: \
\begin{eqnarray}
\ ^{6}\mathbf{g} &=&g_{\acute{\imath}\acute{j}}(\ ^{0}u,y^{\acute{\imath}%
},y^{\grave{a}})dy^{\acute{\imath}}\otimes dy^{\acute{j}}+g_{\grave{a}\grave{%
b}}(\ ^{0}u,y^{\acute{\imath}},y^{\grave{a}})e^{\grave{a}}\otimes e^{\grave{b%
}},  \label{gpsmf} \\
\mathbf{e}^{\grave{a}} &=&dy^{\grave{a}}-N_{\acute{\imath}}^{\grave{a}%
}(u)dy^{\acute{\imath}}.  \notag
\end{eqnarray}%
In formluas (\ref{gpsmf}), the vierbein coefficients $e_{\ \underline{\check{%
a}}}^{\check{a}}$ of the dual basis $\ e^{\check{a}}=(e^{\acute{\imath}},e^{%
\grave{a}})=e_{\ \underline{\check{a}}}^{\check{a}}(u)dy^{\underline{\check{a%
}}},$ are parametrized to define a formal $3+3$ splitting with N--connection
structure $\ ^{6}\mathbf{N=\{}N_{\acute{\imath}}^{\grave{a}}\}.$

It is possible to prescribe any generating function $L(u)=L(x^{i},y^{a},y^{%
\check{a}})$ on $\ ^{6}\mathbf{X}$ with nondegenerate Hessian $\det |\
\tilde{g}_{\grave{a}\grave{b}}|\neq 0$ for
\begin{equation}
\tilde{g}_{\grave{a}\grave{b}}:=\frac{1}{2}\frac{\partial ^{2}L}{\partial y^{%
\grave{a}}\partial y^{\grave{b}}}.  \label{elmf}
\end{equation}%
We define a canonical N--connection structure
\begin{equation}
\tilde{N}_{\acute{\imath}}^{\grave{a}}=\frac{\partial G^{\grave{a}}}{%
\partial y^{7+\acute{\imath}}},\mbox{ where }G^{\grave{a}}=\frac{1}{4}\
\tilde{g}^{\grave{a}\ 7+\acute{\imath}}(\frac{\partial ^{2}L}{\partial y^{7+%
\acute{\imath}}\partial y^{\acute{k}}}y^{7+\acute{k}}-\frac{\partial L}{%
\partial y^{\acute{\imath}}}).  \label{clncfa}
\end{equation}%
In these formulas, $\ \tilde{g}^{\grave{a}\grave{b}}$ is inverse to $\
\tilde{g}_{\grave{a}\grave{b}}$ and respective contractions of $h$-- and $v$%
--indices, $\acute{\imath},\acute{j},...=5,6,7$ and $\grave{a},\grave{b}%
,...=8,9,10,$ are performed following this rule: For example, we take an up $%
v$--index $\grave{a}=3+\acute{\imath}$ and contract it with a low index $%
\acute{\imath}=1,2,3.$ Using (\ref{elmf}) and (\ref{clncfa}), we construct
an internal 6-d d--metric
\begin{eqnarray}
^{6}\widetilde{\mathbf{g}} &=&\widetilde{g}_{\acute{\imath}\acute{j}}dy^{%
\acute{\imath}}\otimes dy^{\acute{j}}+\ \widetilde{g}_{\grave{a}\grave{b}}%
\widetilde{\mathbf{e}}^{\grave{a}}\otimes \widetilde{\mathbf{e}}^{\grave{b}},
\label{lfsmf} \\
\ \widetilde{\mathbf{e}}^{\grave{a}} &=&dy^{\grave{a}}+\tilde{N}_{\acute{%
\imath}}^{\grave{a}}dy^{\acute{\imath}},\ \{\ \widetilde{g}_{\grave{a}\grave{%
b}}\}=\ \{\widetilde{g}_{7+\acute{\imath}\ 7+\acute{j}}\}.  \notag
\end{eqnarray}

It should be emphasized that any d-metric $^{6}\mathbf{g}$ (\ref{gpsmf}) can
be parametrized by coefficients $\ ^{6}\mathbf{g}_{\check{a}\check{b}}=[g_{%
\acute{\imath}\acute{j}},g_{\grave{a}\grave{b}},N_{\acute{\imath}}^{\grave{a}%
}]$ computed with respect to a N-adapted basis $\mathbf{e}^{\check{a}}=(e^{%
\acute{\imath}}=dy^{\acute{\imath}},\mathbf{e}^{\grave{a}})$ which is
related to the metric $\ ^{6}\widetilde{\mathbf{g}}_{\check{a}\check{b}}=\ [%
\widetilde{g}_{\acute{\imath}\acute{j}},\widetilde{g}_{\grave{a}\grave{b}},%
\tilde{N}_{\acute{\imath}}^{\grave{a}}]$ (\ref{lfsmf}) with coefficients
defined with respect to a N--adapted dual basis $\ \ \widetilde{\mathbf{e}}^{%
\check{a}}=(e^{\acute{\imath}}=dy^{\acute{\imath}},\widetilde{\mathbf{e}}^{%
\grave{a}})$ if the conditions $\ \ ^{6}\mathbf{g}_{\check{a}^{\prime }%
\check{b}^{\prime }}e_{\ \check{a}}^{\check{a}^{\prime }}e_{\ \check{b}}^{%
\check{b}^{\prime }}=\ ^{6}\widetilde{\mathbf{g}}_{\check{a}\check{b}}$
related to corresponding frame transfoms are satisfied. Fixing any values $\
^{6}\mathbf{g}_{\check{a}^{\prime }\check{b}^{\prime }}$ and $\ ^{6}%
\widetilde{\mathbf{g}}_{\check{a}\check{b}},$ we have to solve a system of
quadratic algebraic equations with unknown variables $e_{\ \check{a}}^{%
\check{a}^{\prime }}.$ A nonholonomic $2+2+2=3+3$ splitting of $\ ^{6}%
\mathbf{X}$ with $\ ^{6}\mathbf{g}_{\check{a}\check{b}}=[g_{\acute{\imath}%
\acute{j}},g_{\grave{a}\grave{b}},N_{\acute{\imath}}^{\grave{a}}]$ $\ $is
convenient for constructing generic off-diagonal solutions but similar
N--connection splitting with equivalent $\ ^{6}\widetilde{\mathbf{g}}_{%
\check{a}\check{b}}=\ [\widetilde{g}_{\acute{\imath}\acute{j}},\widetilde{g}%
_{\grave{a}\grave{b}},\tilde{N}_{\acute{\imath}}^{\grave{a}}]$ will allow us
to define real solutions for effective EYMH systems.

A set of coefficients $\mathbf{\tilde{N}}=\{\tilde{N}_{\acute{\imath}}^{%
\grave{a}}\}$ defines an N--connection splitting as a Whitney sum,%
\begin{equation}
T\ ^{6}\mathbf{X}=h\ ^{6}\mathbf{X}\oplus v\ ^{6}\mathbf{X}  \label{whitneyf}
\end{equation}%
into conventional internal horizontal (h) and vertical (v) subspaces. In
local form, this can be written as
\begin{equation}
\mathbf{\tilde{N}}=\tilde{N}_{\acute{\imath}}^{\grave{a}}(u)dy^{\acute{\imath%
}}\otimes \frac{\partial }{\partial y^{\grave{a}}},  \label{coeffncf}
\end{equation}%
with $\ ^{6}\mathbf{X=}\ ^{3+3}\mathbf{X.}$ As a result, there are
N--adapted frame (vielbein) structures,
\begin{equation}
\widetilde{\mathbf{e}}_{\check{a}}=\left( \widetilde{\mathbf{e}}_{\acute{%
\imath}}=\frac{\partial }{\partial y^{\acute{\imath}}}-\tilde{N}_{\acute{%
\imath}}^{\grave{a}}\frac{\partial }{\partial y^{\grave{a}}},e_{\grave{a}}=%
\frac{\partial }{\partial y^{\grave{a}}}\right) ,  \label{dderf}
\end{equation}%
with dual frame (coframe) structures $\ \widetilde{\mathbf{e}}^{\check{a}},$
see (\ref{lfsmf}). These vielbein structures define the nonholonomy relations%
$\ $
\begin{equation}
\lbrack \widetilde{\mathbf{e}}_{\check{a}},\widetilde{\mathbf{e}}_{\check{b}%
}]=\widetilde{\mathbf{e}}_{\check{a}}\widetilde{\mathbf{e}}_{\check{b}}-%
\widetilde{\mathbf{e}}_{\check{b}}\widetilde{\mathbf{e}}_{\check{a}}=%
\widetilde{w}_{\check{a}\check{b}}^{\check{c}}\widetilde{\mathbf{e}}_{\check{%
c}}  \label{anhrelf}
\end{equation}%
with (antisymmetric) anholonomy coefficients $\widetilde{w}_{\acute{\imath}%
\grave{a}}^{\grave{b}}=\partial _{\grave{a}}\widetilde{N}_{\acute{\imath}}^{%
\grave{b}}$ and $\widetilde{w}_{\acute{j}\acute{\imath}}^{\grave{a}}=%
\widetilde{\Omega }_{\acute{\imath}\acute{j}}^{\grave{a}},$ where $\
\widetilde{\Omega }_{\acute{\imath}\acute{j}}^{\grave{a}}=\widetilde{\mathbf{%
e}}_{\acute{j}}\left( \widetilde{N}_{\acute{\imath}}^{\grave{a}}\right) -%
\widetilde{\mathbf{e}}_{\acute{\imath}}\left( \widetilde{N}_{\acute{j}}^{%
\grave{a}}\right) $ are the coefficients of N--connection curvature (defined
as the Neijenhuis tensor).

Using the canonical N--connection splitting, we introduce a linear operator $%
\widetilde{\mathbf{J}}$ acting on vectors on $\ ^{6}\mathbf{X}$ following
formulas $\widetilde{\mathbf{J}}(\mathbf{e}_{\acute{\imath}})=-e_{7+\acute{%
\imath}}$ and $\widetilde{\mathbf{J}}(e_{7+\acute{\imath}})=\widetilde{%
\mathbf{e}}_{\acute{\imath}},$ where $\widetilde{\mathbf{J}}\mathbf{\circ }%
\widetilde{\mathbf{J}}\mathbf{=-}\mathbb{I},$ for $\mathbb{I}$ being the
unity matrix, and construct a tensor field,
\begin{eqnarray}
\widetilde{\mathbf{J}} &=&\widetilde{\mathbf{J}}_{\ \check{b}}^{\check{a}}\
e_{\check{a}}\otimes e^{\check{b}}=\widetilde{\mathbf{J}}_{\ \underline{%
\check{b}}}^{\underline{\check{a}}}\ \frac{\partial }{\partial u^{\underline{%
\check{a}}}}\otimes du^{\underline{\check{b}}}=\widetilde{\mathbf{J}}_{\
\beta ^{\prime }}^{\check{a}^{\prime }}\ \widetilde{\mathbf{e}}_{\check{a}%
^{\prime }}\otimes \widetilde{\mathbf{e}}^{\beta ^{\prime }}=\mathbf{-}e_{7+%
\acute{\imath}}\otimes e^{\acute{\imath}}+\widetilde{\mathbf{e}}_{\acute{%
\imath}}\otimes \ \widetilde{\mathbf{e}}^{7+\acute{\imath}}  \label{acstrf}
\\
&=&-\frac{\partial }{\partial y^{\acute{\imath}}}\otimes dx^{\acute{\imath}%
}+\left( \frac{\partial }{\partial y^{\acute{\imath}}}-\ \widetilde{N}_{%
\acute{\imath}}^{\grave{a}}\frac{\partial }{\partial y^{\grave{a}}}\right)
\otimes \left( dy^{7+\acute{\imath}}+\widetilde{N}_{\acute{k}}^{7+\acute{%
\imath}}dy^{\acute{k}}\right) .  \notag
\end{eqnarray}%
The corresponding d-tensor field defined globally by an almost complex
structure on\ $\ ^{6}\mathbf{X}$ is completely determined by a prescribed
generating function $\tilde{L}(y^{4},y^{\check{a}})\subset L(u^{\underline{%
\alpha }_{s}}).$ In this subsection, we consider only structures $\mathbf{J=%
\tilde{J}}$ induced by a $\widetilde{N}_{\acute{k}}^{7+\acute{\imath}}.$ In
general, we can define an almost complex structure $\mathbf{J}$ for an
arbitrary N--connection $\mathbf{N,}$ stating a nonholonomic $3+3$ splitting
by using N--adapted bases (\ref{dderf}) which can be included (if necessary)
into respective nonholonomic frames of the 10-d spacetime, see (\ref{nadaptb}%
).

The Neijenhuis tensor field for $\mathbf{\tilde{J}}$ (equivalently, the
curvature of N--connecti\-on $\mathbf{\tilde{N})}$ is
\begin{equation}
\ ^{\mathbf{J}}\widetilde{\mathbf{\Omega }}\mathbf{(X,Y):=-[X,Y]+[\tilde{J}X,%
\tilde{J}Y]-\tilde{J}[\tilde{J}X,Y]-\tilde{J}[X,\tilde{J}Y],}  \label{neijtf}
\end{equation}%
for any d--vectors $\mathbf{X,Y\in }T\ ^{6}\mathbf{X.}$ With respect to
N--adapted bases (\ref{dderf}), a subset of the coefficients of the
Neijenhuis tensor defines the N--connection curvature,
\begin{equation}
\widetilde{\Omega }_{\acute{\imath}\acute{k}}^{\grave{a}}=\frac{\partial
\widetilde{N}_{\acute{\imath}}^{\grave{a}}}{\partial y^{\acute{k}}}-\frac{%
\partial \widetilde{N}_{\acute{k}}^{\grave{a}}}{\partial y^{\acute{\imath}}}+%
\widetilde{N}_{\acute{\imath}}^{\grave{b}}\frac{\partial \widetilde{N}_{%
\acute{k}}^{\grave{a}}}{\partial y^{\grave{b}}}-\widetilde{N}_{\acute{k}}^{%
\grave{b}}\frac{\partial \widetilde{N}_{\acute{\imath}}^{\grave{a}}}{%
\partial y^{\grave{b}}}.  \label{nccurvf}
\end{equation}%
The nonholonomic structure is integrable if $\widetilde{\Omega }_{\acute{%
\imath}\acute{k}}^{\grave{a}}=0.$ We get a complex structure if and only if
both the h-- and v--distributions are integrable, i.e. if and only if $\
\widetilde{\Omega }_{\acute{\imath}\acute{k}}^{\grave{a}}=0$ and $\frac{%
\partial \widetilde{N}_{\acute{\imath}}^{\grave{a}}}{\partial y^{\acute{k}}}-%
\frac{\partial \widetilde{N}_{\acute{k}}^{\grave{a}}}{\partial y^{\acute{%
\imath}}}=0.$

An almost symplectic structure on a manifold is introduced by a
nondegenerate 2--form
\begin{equation*}
\theta =\frac{1}{2}\theta _{\check{a}\check{b}}(u)e^{\check{a}}\wedge e^{%
\check{b}}=\frac{1}{2}\theta _{\acute{\imath}\acute{k}}(u)e^{\acute{\imath}%
}\wedge e^{\acute{k}}+\frac{1}{2}\theta _{\grave{a}\grave{b}}(u)\mathbf{e}^{%
\grave{a}}\wedge \mathbf{e}^{\grave{b}}.
\end{equation*}%
An almost Hermitian model of an internal 6-d Riemannian space equipped with
a N--connection structure $\mathbf{N}$ is defined by a triple $\mathbf{H}%
^{3+3}=(\ ^{6}\mathbf{X},\theta ,\mathbf{J}),$ where $\theta \mathbf{(X,Y)}%
\doteqdot \mathbf{g}\left( \mathbf{JX,Y}\right) $ for any $\mathbf{g}$ (\ref%
{gpsmf}). A space $\mathbf{H}^{3+3}$ is almost-K\"{a}hler, denoted $\mathbf{K%
}^{3+3},$ if and only if $d\theta =0.$

Using $\mathbf{g}=\ \mathbf{\tilde{g}}$ (\ref{lfsmf}) and structures $%
\mathbf{\tilde{N}}$ (\ref{clncfa}) and $\mathbf{\tilde{J}},$ we define $\ \
\widetilde{\theta }\mathbf{(X,Y):}=\widetilde{\mathbf{g}}\left( \widetilde{%
\mathbf{J}}\mathbf{X,Y}\right) ,$ for any d--vectors $\mathbf{X,Y\in }T\ ^{6}%
\mathbf{X.}$ In local N--adapted form, we have
\begin{eqnarray}
\widetilde{\theta } &=&\frac{1}{2}\ \widetilde{\theta }_{\check{a}\check{b}%
}(u)e^{\check{a}}\wedge e^{\check{b}}=\frac{1}{2}\ \widetilde{\theta }_{%
\underline{\check{a}}\underline{\check{b}}}(u)du^{\underline{\check{a}}%
}\wedge du^{\underline{\check{b}}}  \label{asymstrf} \\
&=&\ \tilde{g}_{\acute{\imath}\acute{k}}(x^{i},y^{a},y^{\check{a}})e^{7+%
\acute{\imath}}\wedge dy^{\acute{k}}=\ \tilde{g}_{\acute{\imath}\acute{j}%
}(x^{i},y^{a},y^{\check{a}})(dy^{7+\acute{\imath}}+\ \tilde{N}_{\acute{k}%
}^{7+\acute{\imath}}dy^{\acute{k}})\wedge dy^{\acute{j}}.  \notag
\end{eqnarray}%
Considering the form $\ \widetilde{\ \omega }=\frac{1}{2}\frac{\partial
\widetilde{L}}{\partial y^{7+\acute{\imath}}}dy^{\acute{\imath}},$ we prove
by a straightforward computation that $\ \widetilde{\theta }=d\ \widetilde{%
\omega },$ i.e. $d\ \widetilde{\theta }=dd\ \widetilde{\omega }=0.$ As a
result, any canonical effective Lagrange structure $\left( \mathbf{g}=%
\widetilde{\mathbf{g}},\ \widetilde{\mathbf{N}}\mathbf{,}\widetilde{\mathbf{J%
}}\right) $ induces an almost-K\"{a}hler geometry. The 2--form (\ref%
{asymstrf}) can be written
\begin{eqnarray}
\theta &=&\widetilde{\theta }=\frac{1}{2}\widetilde{\theta }_{\acute{\imath}%
\acute{j}}(u)e^{\acute{\imath}}\wedge e^{\acute{j}}+\frac{1}{2}\ \widetilde{%
\theta }_{\grave{a}\grave{b}}(u)\widetilde{\mathbf{e}}^{\grave{a}}\wedge
\widetilde{\mathbf{e}}^{\grave{b}}  \label{canalmsf} \\
&=&g_{\acute{\imath}\acute{j}}(u)\left[ dy^{\acute{\imath}}+\widetilde{N}_{%
\acute{k}}^{7+\acute{\imath}}(u)dy^{\acute{k}}\right] \wedge dy^{\acute{j}},
\notag
\end{eqnarray}%
where the nontrivial coefficients $\ \widetilde{\theta }_{\grave{a}\grave{b}%
}=\widetilde{\theta }_{7+\acute{\imath}\ 7+\acute{j}}$ are equal to the
N-adapted coefficients $\widetilde{\theta }_{\acute{\imath}\ \acute{j}}$
respectively.

\subsubsection{Almost symplectic connections for N-anholonomic internal
spaces}

Taking a general 2--form $\theta $ constructed for any metric $\mathbf{g}$
and almost complex $\mathbf{J}$\textbf{\ }structures on $\ ^{6}\mathbf{X,}$
one obtains $d\theta \neq 0.$ Nevertheless, we can always define a $3+3$
splitting induced by an effective Lagrange generating function when $d\
\widetilde{\theta }=0.$ Considering frame transforms, $\theta _{\check{a}%
^{\prime }\check{b}^{\prime }}e_{\ \check{a}}^{\check{a}^{\prime }}e_{\
\check{b}}^{\check{b}^{\prime }}=\widetilde{\theta }_{\check{a}\check{b}},$
we can write $d\ \theta =0$ for any set of 2--form coefficients related via
frame transforms to a canonical symplectic structure.

There is a unique normal d--connection
\begin{eqnarray}
\ \widetilde{\mathbf{D}} &=&\left\{ h\widetilde{D}=(\widetilde{D}_{\acute{k}%
},^{v}\widetilde{D}_{\acute{k}}=\widetilde{D}_{\acute{k}});v\widetilde{D}=(%
\widetilde{D}_{\grave{c}},\ ^{v}\widetilde{D}_{\grave{c}}=\widetilde{D}_{%
\grave{c}})\right\}  \label{ndcf} \\
&=&\{\widetilde{\mathbf{\Gamma }}_{\check{b}\check{c}}^{\check{a}}=(%
\widetilde{L}_{\acute{j}\acute{k}}^{\acute{\imath}},\ ^{v}\widetilde{L}_{7+%
\acute{j}\ 7+\acute{k}}^{7+\acute{\imath}}=\widetilde{L}_{\acute{j}\acute{k}%
}^{\acute{\imath}};\ \widetilde{C}_{\acute{j}\grave{c}}^{\acute{\imath}}=\
^{v}\widetilde{C}_{7+\acute{j}\ \grave{c}}^{7+\acute{\imath}},\ ^{v}%
\widetilde{C}_{\grave{b}\grave{c}}^{\grave{a}}=\widetilde{C}_{\grave{b}%
\grave{c}}^{\grave{a}})\},  \notag
\end{eqnarray}%
which is metric compatible, $\widetilde{D}_{\acute{k}}\widetilde{g}_{\acute{%
\imath}\acute{j}}=0$ and $\widetilde{D}_{\grave{b}}\ \widetilde{g}_{\acute{%
\imath}\acute{j}}=0,$ and completely defined by a couple of h-- and
v--components $\ \widetilde{\mathbf{D}}_{\check{a}}=(\widetilde{D}_{\acute{k}%
},\widetilde{D}_{\grave{b}}).$ The corresponding N--adapted coefficients $%
\widetilde{\mathbf{\Gamma }}_{\check{b}\gamma }^{\check{a}}=(\widetilde{L}_{%
\acute{j}\acute{k}}^{\acute{\imath}},\ ^{v}\widetilde{C}_{\grave{b}\grave{c}%
}^{\grave{a}})$ are given by
\begin{equation}
\widetilde{L}_{\acute{j}\acute{k}}^{\acute{\imath}}=\frac{1}{2}\ \widetilde{g%
}^{\acute{\imath}\acute{h}}\left( \widetilde{\mathbf{e}}_{\acute{k}}%
\widetilde{g}_{\acute{j}\acute{h}}+\widetilde{\mathbf{e}}_{\acute{j}}%
\widetilde{g}_{\acute{h}\acute{k}}-\widetilde{\mathbf{e}}_{\acute{h}}%
\widetilde{g}_{\acute{j}\acute{k}}\right) ,\widetilde{C}_{\acute{j}\acute{k}%
}^{\acute{\imath}}=\frac{1}{2}\ \widetilde{g}^{\acute{\imath}\acute{h}%
}\left( \frac{\partial \widetilde{g}_{\acute{j}\acute{h}}}{\partial y^{%
\acute{k}}}+\frac{\partial \widetilde{g}_{\acute{h}\acute{k}}}{\partial y^{%
\acute{j}}}-\frac{\partial \widetilde{g}_{\acute{j}\acute{k}}}{\partial y^{%
\acute{h}}}\right) .  \label{cdccf}
\end{equation}

To develop a differential form calculus on $\ ^{6}\mathbf{X}$ which is
adapted to the canonical N--connection $\widetilde{\mathbf{N}}$, we
introduce the normal d--connection 1--form
\begin{equation}
\widetilde{\mathbf{\Gamma }}_{\acute{j}}^{\acute{\imath}}=\widetilde{L}_{%
\acute{j}\acute{k}}^{\acute{\imath}}e^{\acute{k}}+\widetilde{C}_{\acute{j}%
\acute{k}}^{\acute{\imath}}\mathbf{e}^{\acute{k}}.  \label{dconf}
\end{equation}%
Using this linear connection, we prove that the Cartan structure equations
are satisfied,
\begin{equation}
de^{\acute{k}}-e^{\acute{j}}\wedge \widetilde{\mathbf{\Gamma }}_{\acute{j}}^{%
\acute{\imath}}=-\widetilde{\mathcal{T}}^{\acute{k}},\ d\mathbf{e}^{\acute{k}%
}-\mathbf{e}^{\acute{j}}\wedge \widetilde{\mathbf{\Gamma }}_{\acute{j}}^{%
\acute{\imath}}=-\ ^{v}\widetilde{\mathcal{T}}^{\acute{\imath}},
\label{cart1f}
\end{equation}%
and
\begin{equation}
d\widetilde{\mathbf{\Gamma }}_{\acute{j}}^{\acute{\imath}}-\widetilde{%
\mathbf{\Gamma }}_{\acute{j}}^{\acute{h}}\wedge \widetilde{\mathbf{\Gamma }}%
_{\acute{h}}^{\acute{\imath}}=-\widetilde{\mathcal{R}}_{\ \acute{j}}^{\acute{%
\imath}}.  \label{cart2f}
\end{equation}%
The h-- and v--components of the torsion 2--form $\widetilde{\mathcal{T}}^{%
\check{a}}=\left( \widetilde{\mathcal{T}}^{\acute{k}},\ ^{v}\widetilde{%
\mathcal{T}}^{\acute{\imath}}\right) =\widetilde{\mathbf{T}}_{\ \check{c}%
\check{b}}^{\check{a}}\ \widetilde{\mathbf{e}}^{\check{c}}\wedge \widetilde{%
\mathbf{e}}^{\check{b}}$ from (\ref{cart1f}) are computed
\begin{equation}
\widetilde{\mathcal{T}}^{\acute{\imath}}=\widetilde{C}_{\acute{j}\acute{k}}^{%
\acute{\imath}}e^{\acute{j}}\wedge \mathbf{e}^{\acute{k}},\ ^{v}\widetilde{%
\mathcal{T}}^{\acute{\imath}}=\frac{1}{2}\ \widetilde{\Omega }_{\acute{k}%
\acute{j}}^{\acute{\imath}}e^{\acute{k}}\wedge e^{\acute{j}}+(\frac{\partial
\ \ \widetilde{N}_{\acute{k}}^{\acute{\imath}}}{\partial y^{\acute{j}}}-%
\widetilde{L}_{\acute{j}\acute{k}}^{\acute{\imath}})e^{\acute{k}}\wedge
\mathbf{e}^{\acute{j}}.  \label{tformf}
\end{equation}%
In these formulas, $\ \widetilde{\Omega }_{\acute{k}\acute{j}}^{\acute{\imath%
}}$ are coefficients of the curvature of $\ \widetilde{N}_{\acute{k}}^{%
\acute{\imath}}$ defined by formulas similar to (\ref{nccurvf}). The
formulas (\ref{tformf}) parametrize the h-- and v--components of torsion $%
\widetilde{\mathbf{T}}_{\ \check{c}\check{b}}^{\check{a}}$ in the form
\begin{equation}
\widetilde{T}_{\acute{j}\acute{k}}^{\acute{\imath}}=0,\widetilde{T}_{\acute{j%
}\grave{a}}^{\acute{\imath}}=\widetilde{C}_{\acute{j}\grave{a}}^{\acute{%
\imath}},\widetilde{T}_{\acute{k}\acute{j}}^{\grave{a}}=\ \ \widetilde{%
\Omega }_{\acute{k}\acute{j}}^{\grave{a}},\widetilde{T}_{\acute{\imath}%
\grave{b}}^{\grave{a}}=e_{\grave{b}}\left( \widetilde{N}_{\acute{\imath}}^{%
\grave{a}}\right) -\widetilde{L}_{\ \grave{b}\acute{\imath}}^{\grave{a}},%
\widetilde{T}_{\grave{b}\grave{c}}^{\grave{a}}=0.  \label{cdtorsf}
\end{equation}%
We emphasize that $\widetilde{\mathbf{T}}$ vanishes on h- and v--subspaces,
i.e. $\widetilde{T}_{\acute{j}\acute{k}}^{\acute{\imath}}=0$ and $\widetilde{%
T}_{\grave{b}\grave{c}}^{\grave{a}}=0,$ but other nontrivial
h--v--components are induced by the nonholonomic structure determined
canonically by $\mathbf{g}=\widetilde{\mathbf{g}}$ (\ref{lfsmf}) and $\tilde{%
L}.$

An explicit calculus of the curvature 2--form from (\ref{cart2f}) results in%
\begin{equation}
\widetilde{\mathcal{R}}_{\ \acute{j}}^{\acute{\imath}}=\widetilde{\mathbf{R}}%
_{\ \acute{j}\check{c}\check{b}}^{\acute{\imath}}\ \mathbf{e}^{\check{c}%
}\wedge \ \mathbf{e}^{\check{b}}=\frac{1}{2}\widetilde{R}_{\ \acute{j}\acute{%
k}\acute{h}}^{\acute{\imath}}e^{\acute{k}}\wedge e^{\acute{h}}+\widetilde{P}%
_{\ \acute{j}\acute{k}\grave{a}}^{\acute{\imath}}e^{\acute{k}}\wedge
\widetilde{\mathbf{e}}^{\grave{a}}+\frac{1}{2}\ \widetilde{S}_{\ \acute{j}%
\grave{c}\grave{d}}^{\acute{\imath}}\widetilde{\mathbf{e}}^{\grave{c}}\wedge
\widetilde{\mathbf{e}}^{\grave{d}}.  \label{cformf}
\end{equation}%
The corresponding nontrivial N--adapted coefficients of curvature $%
\widetilde{\mathbf{R}}_{\ \ \check{c}\check{b}\check{e}}^{\check{a}}$ of $%
\mathbf{\tilde{D}}$ are
\begin{eqnarray*}
\widetilde{R}_{\ \acute{j}\acute{k}\acute{h}}^{\acute{\imath}} &=&\widetilde{%
\mathbf{e}}_{\acute{k}}\widetilde{L}_{\ \acute{h}\acute{j}}^{\acute{\imath}}-%
\widetilde{\mathbf{e}}_{\acute{j}}\widetilde{L}_{\ \acute{h}\acute{k}}^{%
\acute{\imath}}+\widetilde{L}_{\ \acute{h}\acute{j}}^{\acute{m}}\widetilde{L}%
_{\ \acute{m}\acute{k}}^{\acute{\imath}}-\widetilde{L}_{\ \acute{h}\acute{k}%
}^{\acute{m}}\widetilde{L}_{\ \acute{m}\acute{j}}^{\acute{\imath}}-%
\widetilde{C}_{\ \acute{h}\grave{a}}^{\acute{\imath}}\ \widetilde{\Omega }%
_{\ \acute{k}\acute{j}}^{\grave{a}} \\
\widetilde{P}_{\ \acute{j}\acute{k}\grave{a}}^{\acute{\imath}} &=&e_{\grave{a%
}}\widetilde{L}_{\ \acute{j}\acute{k}}^{\acute{\imath}}-\widetilde{\mathbf{D}%
}_{\acute{k}}\widetilde{C}_{\ \acute{j}\grave{a}}^{\acute{\imath}},%
\widetilde{S}_{\ \grave{b}\grave{c}\grave{d}}^{\grave{a}}=\ e_{\grave{d}}%
\widetilde{C}_{\ \grave{b}\grave{c}}^{\grave{a}}-e_{\grave{c}}\widetilde{C}%
_{\ \grave{b}\grave{d}}^{\grave{a}}+\widetilde{C}_{\ \grave{b}\grave{c}}^{%
\grave{e}}\widetilde{C}_{\ \grave{e}\grave{d}}^{\grave{a}}-\widetilde{C}_{\
\grave{b}\grave{d}}^{\grave{e}}\widetilde{C}_{\ \grave{e}\grave{c}}^{\grave{a%
}}.
\end{eqnarray*}%
By definition, the Ricci d--tensor $\widetilde{\mathbf{\mathbf{R}}}_{\check{b%
}\check{c}}=\widetilde{\mathbf{\mathbf{R}}}_{\ \check{b}\check{c}\check{a}}^{%
\check{a}}$ is computed
\begin{equation}
\widetilde{\mathbf{\mathbf{R}}}_{\check{b}\check{c}}\mathbf{=}\left(
\widetilde{R}_{\acute{\imath}\acute{j}},\widetilde{R}_{\acute{\imath}\grave{a%
}},\widetilde{R}_{\grave{a}\acute{\imath}},\widetilde{R}_{\grave{a}\grave{b}%
}\right) .  \label{driccif}
\end{equation}%
The scalar curvature $\ \widetilde{R}$ of $\ \widetilde{\mathbf{D}}$ is
given by two h- and v--terms,
\begin{equation}
\ \widetilde{R}=\widetilde{\mathbf{g}}^{\check{b}\check{c}}\widetilde{%
\mathbf{\mathbf{R}}}_{\check{b}\check{c}}=\widetilde{g}^{\acute{\imath}%
\acute{j}}\widetilde{R}_{\acute{\imath}\acute{j}}+\widetilde{g}^{\grave{a}%
\grave{b}}\widetilde{R}_{\grave{a}\grave{b}}\mathbf{.}  \label{scalcf}
\end{equation}

The normal d--connection $\widetilde{\mathbf{D}}$ (\ref{ndcf}) defines a
canonical almost symplectic d--connection, $\widetilde{\mathbf{D}}\equiv \
_{\theta }\widetilde{\mathbf{D}},$ which is N--adapted to the effective
Lagrange and, related to almost symplectic structures, i.e. it preserves the
splitting under parallelism (\ref{whitneyf}),
\begin{equation*}
\widetilde{\mathbf{D}}_{\mathbf{X}}\mathbf{\ }\widetilde{\mathbf{g}}\mathbf{%
=\ }_{\theta }\widetilde{\mathbf{D}}_{\mathbf{X}}\ \widetilde{\theta }%
\mathbf{=}0,
\end{equation*}%
for any $\mathbf{X\in }T\ ^{6}\mathbf{X}$ and its torsion is constrained to
satisfy the conditions $\widetilde{T}_{\acute{j}\acute{k}}^{\acute{\imath}%
}=0 $ and $\widetilde{T}_{\grave{b}\grave{c}}^{\grave{a}}=0.$

We conclude that having chosen a regular generating function $L(x,y)$ on a
Riemannian internal space $\mathbf{V},$ we can always model this spacetime
equivalently as an-almost K\"{a}hler manifold. Using corresponding
nonholonomic frame transforms and deformation of connections, we can work
with equivalent geometric data on the internal space $\ ^{6}\mathbf{X,}$ for
convenience $\ (\ ^{6}\mathbf{g,}\ ^{6}\mathbf{N,}\ ^{6}\widehat{\mathbf{D}}%
)\iff (\ \mathbf{\tilde{g},}\ \mathbf{\tilde{N},}\widetilde{\mathbf{D}},%
\widetilde{L})\iff (\ \ \widetilde{\theta }\mathbf{,}\ \widetilde{\mathbf{J}}%
\mathbf{,}\ _{\theta }\widetilde{\mathbf{D}}).$ The first N-adapted model is
convenient for constructing exact solutions in 6-d and 10-d gravity models
(this will be addressed in the associated paper \cite{partner}, see also
examples in \cite{tgovsv}). The second nonholonomic model with "tilde"
geometric objects (with so--called Lagrange-Finsler variables, in our case,
on a 6--d Riemannian space) is an example of an internal space with
nontrivial nonholonomic $3+3$ splitting by a canonical N--connection
structure determined by an effective Lagrange function $\widetilde{L}.$ The $%
(\widetilde{\theta }\mathbf{,}\ \widetilde{\mathbf{J}},\ _{\theta }%
\widetilde{\mathbf{D}})$ defines an almost-K\"{a}hler geometric model on $\
^{6}\mathbf{X}$ with nontrivial nonholonomically induced d--torsion
structure $\widetilde{\mathcal{T}}^{\check{a}}$. This way, we can mimic a
complex like differential geometry by real values and elaborate on various
applications to quantum gravity, string/brane and geometric flow theories
\cite{vjgp,vwitten,vmedit}. Introducing the complex imaginary unit $%
i^{2}=-1, $ with $\ \widetilde{\mathbf{J}}\thickapprox i...$ and integrable
nonholonomic distributions, we can redefine the geometric constructions for
complex manifolds. Using nonholonomic real $3+3$ distributions, we can
develop gravitational and gauge like models of internal spaces, for
instance, with $SO(3),$ or $SU(3)$ symmetries and their tensor products. Two
different approaches can be unified in a geometric language with double
nonholonomic fibrations $2+2+2=3+3$. Any d--metric with internal $2+2+2$
nonholonomic splitting can be redefined by nonholonomic frame transforms
into an almost symplectic structure with $3+3$ decomposition. Considering
actions of $SO(3),$ or $SU(3)$ on corresponding tangent spaces, we can
reproduce all results with K\"{a}hler internal spaces related 4-d, 6-d and
10-d solutions for holonomic configurations obtained in Refs. \cite%
{lecht1,harl}.

\subsection{N-adapted G$_{2}$ structures on almost-K\"{a}hler internal spaces%
}

\subsubsection{6-d and 7-d almost-K\"{a}hler models}

\label{ss67ak}For any 3--form $\Theta =\Theta _{\check{a}\check{b}\check{c}}%
\widetilde{\mathbf{e}}^{\check{a}}\wedge \widetilde{\mathbf{e}}^{\check{b}%
}\wedge \widetilde{\mathbf{e}}^{\check{c}}$ on $\ ^{6}\mathbf{X}$ endowed
with a canonical almost complex structure $\mathbf{\tilde{J}}$ (\ref{acstrf}%
)
\begin{equation}
\Theta =\ ^{\mathbf{+}}\Theta +\mathbf{\tilde{J}}\ ^{\mathbf{-}}\Theta .
\label{alkdec}
\end{equation}%
We can fix these conditions such that for $\mathbf{\tilde{J}}\rightarrow
i,i^{2}=-1,\Theta $ defines an $SU(3)$ structure defined on $\ ^{6}\mathbf{X}
$ and the tangent space to the domain wall with $\Theta =\ ^{\mathbf{+}%
}\Theta +i\ ^{\mathbf{-}}\Theta $. Defining the gamma matrices $\gamma _{%
\check{a}}$ on $\ ^{6}\mathbf{X}$ from the relation $\widetilde{\gamma }_{%
\check{a}}\widetilde{\gamma }_{\check{b}}+\widetilde{\gamma }_{\check{b}}%
\widetilde{\gamma }_{\check{a}}=2$ $\ ^{6}\widetilde{\mathbf{g}}_{\check{a}%
\check{b}},$ see (\ref{lfsmf}), we can relate the geometric objects in the
almost-K\"{a}hler model of the internal space to the models with $SU(3)$
structure on a typical fiber in the tangent bundle $T\ ^{6}\mathbf{X.}$ For
integrable $SU(3)$ structures and K\"{a}hler \ internal spaces, one works
with the structure forms $(J,\underline{\Theta }),$ when
\begin{equation*}
\underline{\Theta }_{\check{a}\check{b}\check{c}}=(\eta _{+})^{\dagger }%
\underline{\gamma }_{\check{a}\check{b}\check{c}}\eta _{-}\mbox{ and }J_{%
\check{a}\check{b}}=\mp (\eta _{\pm })^{\dagger }\underline{\gamma }_{\check{%
a}\check{b}}\eta _{\pm }
\end{equation*}%
are considered for a K\"{a}hler \ metric $2$ $\ ^{6}\underline{\mathbf{g}}_{%
\check{a}\check{b}}=\underline{\gamma }_{\check{a}}\underline{\gamma }_{%
\check{b}}+\underline{\gamma }_{\check{b}}\underline{\gamma }_{\check{a}}$
with a 6-d Hodge star operator $\underline{\ast }.$ These forms obey the
conditions%
\begin{equation*}
J\wedge \underline{\Theta }=0,\frac{i}{8}\underline{\Theta }\wedge
\underline{\overline{\Theta }}=\frac{1}{3!}J\wedge J\wedge J=\underline{\ast
}1,\underline{\ast }J=\frac{1}{2}J\wedge J,\underline{\ast }\Theta _{\pm
}=\pm \Theta _{\mp },
\end{equation*}%
where $\underline{\overline{\Theta }}$ means complex conjugation of $%
\underline{\Theta }.$

Working with $\mathbf{\tilde{J}}$ instead of $J,$ we can define a similar
3-form $\widetilde{\Theta }$ for an almost-K\"{a}hler model $(\widetilde{%
\theta }\mathbf{,}\ \widetilde{\mathbf{J}}\mathbf{,}\ _{\theta }\widetilde{%
\mathbf{D}})$ and construct the Hodge star operator $\widetilde{\ast }$
corresponding to $\ ^{6}\widetilde{\mathbf{g}}.$ The relation between 6-d $%
\widetilde{\ast }$ and 7-d $\ ^{7}\widetilde{\ast }$ Hodge stars for an
ansatz of type (\ref{7dans}) is
\begin{equation*}
\ ^{7}\widetilde{\ast }(\ _{p}^{6}\omega )=e^{B(y^{\check{a}})}\ \widetilde{%
\ast }(\ _{p}^{6}\omega )\wedge \mathbf{e}^{4}\mbox{ and }\ ^{7}\widetilde{%
\ast }(\mathbf{e}^{4}\wedge \ _{p}^{6}\omega )=e^{-B(y^{\check{a}})}\
\widetilde{\ast }(\ _{p}^{6}\omega ),
\end{equation*}%
where $\ _{p}^{6}\omega $ is a p--form with legs only in the directions on $%
\ ^{6}\mathbf{X.}$ The two exterior derivatives $\ ^{7}d$ and $\widetilde{d}$
are related via $\ \ ^{7}d(\ _{p}^{6}\omega )=\widetilde{d}(\ _{p}^{6}\omega
)+dy^{4}\wedge \frac{\partial }{\partial y^{4}}(\ _{p}^{6}\omega ).$
Applying these formulas, we decompose the 10-d 3-form $\widehat{\mathbf{H}}$
into three N-adapted parts,
\begin{equation*}
\widehat{\mathbf{H}}=\ell vol[\ ^{3}\mathbf{g]+}\ ^{6}\widehat{\mathbf{H}}%
+dy^{4}\wedge \widehat{\mathbf{H}}_{4},
\end{equation*}%
$\ $where \ $\ vol[\ ^{3}\mathbf{g]=}\frac{1}{3}\epsilon _{\check{i}\check{j}%
\check{k}}\sqrt{|^{3}\mathbf{g}_{\check{i}\check{j}}|}\mathbf{e}^{\check{i}%
}\wedge \mathbf{e}^{\check{j}}\wedge \mathbf{e}^{\check{k}},\ ^{6}\widehat{%
\mathbf{H}}=\frac{1}{3!}\widehat{\mathbf{H}}_{\check{a}\check{b}\check{c}}%
\widetilde{\mathbf{e}}^{\check{a}}\wedge \widetilde{\mathbf{e}}^{\check{b}%
}\wedge \widetilde{\mathbf{e}}^{\check{c}},\ \widehat{\mathbf{H}}_{4}=\frac{1%
}{2!}\widehat{\mathbf{H}}_{4\check{b}\check{c}}\widetilde{\mathbf{e}}^{%
\check{b}}\wedge \widetilde{\mathbf{e}}^{\check{c}}.$

The operators $(\mathbf{\tilde{J},}\widetilde{\Theta })$ allow us to
generalize, in almost-K\"{a}hler form, the original constructions for \ K%
\"{a}hler internal spaces provided in \cite{harl,lecht1,gray,lukas,chiossi}
for the $G_{2}$ structure. In our approach, such an N-adapted configuration
is adapted by the data for (\ref{bps2}), $(\varpi ,\mathcal{W})$ which can
be related to the $SU(3)$ almost-K\"{a}hler structure $(\mathbf{\tilde{J},}%
\widetilde{\Theta })$ by expressions%
\begin{equation*}
\varpi =e^{B(y^{\check{a}})}\mathbf{e}^{4}\wedge \mathbf{\tilde{J}+}%
\widetilde{\Theta }_{-}\mbox{ and }\mathcal{W}=e^{B(y^{\check{a}})}\mathbf{e}%
^{4}\wedge \widetilde{\Theta }_{+}+\frac{1}{2}\mathbf{\tilde{J}\wedge \tilde{%
J}.}
\end{equation*}%
For any structure group $SU(3)$ and its Lie algebra $\mathfrak{su}(6)=%
\mathfrak{su}(3)\oplus \mathfrak{su}(3)^{\perp },$ there is a canonical
torsion $^{0}\mathbf{T}_{\ \check{b}^{\prime }\check{c}^{\prime }}^{\check{a}%
^{\prime }}\in \wedge ^{1}\otimes \mathfrak{su}(3)^{\perp },$ where primed
indices refer to an orthonormal basis which can be related to any coordinate
and/or N--adapted basis. For instance we write $^{0}\mathbf{\tilde{T}}_{\
\check{b}^{\prime }\check{c}^{\prime }}^{\check{a}^{\prime }}$ if such a
basis is for an almost-K\"{a}hler structure.
\begin{equation*}
\begin{array}{cccccc}
^{0}\mathbf{T}_{\ \check{b}^{\prime }\check{c}^{\prime }}^{\check{a}^{\prime
}} & = & (3\oplus \overline{3})\otimes & (1\oplus 3\oplus \overline{3}) &  &
\\
& = & (1\oplus 1)\oplus & (8\oplus 8)\oplus & (\overline{6}\oplus 6)\oplus &
2(3\oplus \overline{3}) \\
&  & \mathcal{T}_{1} & \mathcal{T}_{2} & \mathcal{T}_{3} & \mathcal{T}_{4},%
\mathcal{T}_{5}%
\end{array}%
\end{equation*}%
This classification can be N--adapted if we use derivatives of the structure
forms
\begin{eqnarray*}
\widetilde{d}\mathbf{\tilde{J}} &=&\mathbf{-}\frac{3}{2}{Im}(\mathcal{T}_{1}%
\overline{\widetilde{\Theta }})+\mathcal{T}_{4}\wedge \mathbf{\tilde{J}+}%
\mathcal{T}_{3}, \\
\widetilde{d}\widetilde{\Theta } &=&\mathcal{T}_{1}\mathbf{\tilde{J}}\wedge
\mathbf{\tilde{J}}+\mathcal{T}_{2}\wedge \mathbf{\tilde{J}}+\overline{%
\mathcal{T}}_{5}\wedge \widetilde{\Theta },
\end{eqnarray*}%
where ${Im}(\mathcal{T}_{1}\overline{\widetilde{\Theta }})$ should be
treated as the "vertical" part for almost-K\"{a}hler structures (when all
values are real) and as the imaginary part for complex and K\"{a}hler
structures $(\mathbf{\tilde{J}},\widetilde{\Theta })\rightarrow (J,%
\underline{\Theta }).$

\subsubsection{Nonholonomic instanton d--connections nearly almost K\"{a}%
hler manifolds}

The instanton type connections constructed in \cite{harl,lecht1} can be
modelled for almost-K\"{a}hler internal spaces if we work in N--adapted
frames for respective nonholonomically deformed connections. For such
configurations, we set
\begin{equation*}
\mathcal{T}_{2}=\mathcal{T}_{3}=\mathcal{T}_{4}=\mathcal{T}_{5}=0\mbox{ and }%
\mathcal{T}_{1}=\ ^{\mathbf{+}}\mathcal{T}_{1}+\mathbf{\tilde{J}}\ ^{\mathbf{%
-}}\mathcal{T}_{1},
\end{equation*}%
where the last splitting is defined similarly to (\ref{alkdec}). All further
calculations with $(J,\underline{\Theta })$ in \cite%
{harl,lecht1,gray,lukas,chiossi} are similar to those for $(\mathbf{\tilde{J}%
,}\widetilde{\Theta })$ if we work in N--adapted frames on $\ ^{6}\mathbf{X}$
and corresponding model $(\widetilde{\theta }\mathbf{,}\ \widetilde{\mathbf{J%
}}\mathbf{,}\ _{\theta }\widetilde{\mathbf{D}}).$ Hereafter, we shall omit
detailed proofs for almost-K\"{a}hler structures and send readers to
analogous constructions in the aforementioned references.

Considering an ansatz (\ref{7dans}) with possible embedding into a 10-d one
of type (\ref{pm1d}), with $A=B=0$ for simpicity, we can construct solutions
on $\ ^{6}\mathbf{X}$ of the first two nonholonomic BPS equations in (\ref%
{bpseq}) and (\ref{bps2}). Such almost-K\"{a}hler configurations are
determined by the system%
\begin{equation}
\widehat{\mathbf{H}}=\ell vol[\ ^{3}\mathbf{g]-}\frac{1}{2}\partial _{4}\phi
+\left( \frac{3}{2}\ ^{\mathbf{-}}\mathcal{T}_{1}+\frac{7}{8}\ell \right)
+dy^{4}\wedge (2\ ^{\mathbf{-}}\mathcal{T}_{1}+\ell )\widetilde{\mathbf{J}}%
\mbox{ for }\widehat{\phi }=\phi (y^{4})  \label{kheqs}
\end{equation}%
and $(\mathbf{\tilde{J},}\widetilde{\Theta })$ subjected to respective flow
and structure equations:
\begin{eqnarray}
\partial _{4}\widetilde{\mathbf{J}} &=&(\ ^{\mathbf{+}}\mathcal{T}%
_{1}+\partial _{4}\phi )\widetilde{\mathbf{J}},  \label{floweq2} \\
\partial _{4}\ ^{\mathbf{-}}\widetilde{\Theta } &=&-(3\ ^{\mathbf{-}}%
\mathcal{T}_{1}+\frac{15}{8}\ell )\ ^{\mathbf{+}}\widetilde{\Theta }+\frac{3%
}{2}(\ ^{\mathbf{+}}\mathcal{T}_{1}+\partial _{4}\phi )\ ^{\mathbf{-}}%
\widetilde{\Theta },  \notag \\
\partial _{4}\ ^{\mathbf{+}}\widetilde{\Theta } &=&\frac{3}{2}(\ ^{\mathbf{+}%
}\mathcal{T}_{1}+\partial _{4}\phi )\ ^{\mathbf{+}}\widetilde{\Theta }+%
\tilde{\alpha}(y^{4})\ \ ^{\mathbf{-}}\widetilde{\Theta },%
\mbox{ for
arbitrary function }\tilde{\alpha}(y^{4});  \notag
\end{eqnarray}%
\begin{eqnarray}
\mbox{ and }\widetilde{d}\widetilde{\mathbf{J}} &=&-\frac{3}{2}\ ^{\mathbf{-}%
}\mathcal{T}_{1}\ ^{\mathbf{+}}\widetilde{\Theta }+\frac{3}{2}\ ^{\mathbf{+}}%
\mathcal{T}_{1}\ ^{\mathbf{-}}\widetilde{\Theta },  \label{struceq2} \\
\widetilde{d}\widetilde{\Theta } &=&\ \mathcal{T}_{1}\widetilde{\mathbf{J}}%
\wedge \widetilde{\mathbf{J}}.  \notag
\end{eqnarray}

At the zeroth order in $\alpha ^{\prime }$ with the Bianchi identity $%
\widehat{\mathbf{d}}\widehat{\mathbf{H}}=0,$ see (\ref{anomalcond}), we can
choose $\widehat{\mathbf{F}}=0$ in order to solve the third equation in (\ref%
{bpseq}). The time--like component of the equations of motion can be solved
if $\ell =0$ as in the pure K\"{a}hler case \cite{lecht1}. We obtain a
special case of solutions (see \cite{gray} for original K\"{a}hler ones)
when
\begin{equation}
\begin{array}{ccccc}
1. & \phi =const., & \ ^{\mathbf{+}}\mathcal{T}_{1}%
\mbox{ is a free function
}, & \ ^{\mathbf{-}}\mathcal{T}_{1}=0, & \tilde{\alpha}(y^{4})%
\mbox{ is a
free function }; \\
&  &  &  &  \\
2. & \phi =\frac{2}{3}\log (a_{0}y^{4}+b_{0}), & \ ^{\mathbf{+}}\mathcal{T}%
_{1}=0, & \ ^{\mathbf{-}}\mathcal{T}_{1}=0, & \tilde{\alpha}=0,%
\end{array}
\label{table}
\end{equation}%
for integration constants $a_{0}$ and $b_{0}$ corresponding to N--adapted
frames. Respectively, cases 1 and 2 correspond to a nearly almost-K\"{a}hler
geometry, with nonholonomically induced torsion (by off--diagonal N-terms)
and vanishing NS 3-form flux, and a nonholonomic generalized Calabi-Yau with
flux.


\section{The YM Sector and Nonholonomic Heterotic Supergravity}

\label{s3} In this section, we construct a nontrivial gauge d--field $%
\widehat{\mathbf{F}}$ which arises at the first order $\alpha ^{\prime },$
when the YM sector can not be ignored. The approach elaborates on a nonholonomic
and almost-K\"{a}hler version of YM instantons studied in \cite{harl,gray,lecht1}.
The equations of motion of heterotic supergravity are
then re--written in canonical nonholonomic variables, which allows us to
decouple and find general integrals of such systems using methods applied
for nonholonomic EYMH and Einstein--Dirac fields, see \cite%
{sv2001,vex1,svvvey,tgovsv,vtamsuper,vp,vt}.

\subsection{N--adapted YM and instanton configurations}

\label{ssyminst}The nonholonomic instanton equations can be formulated on $\
^{7}\mathbf{X=}\mathbb{R\times }\ ^{6}\mathbf{X}$ with generalized 'h--cone'
d-metric
\begin{eqnarray}
\ _{c}^{7}\mathbf{g} &\mathbf{=}&\mathbf{(e}^{4})^{2}+[\ _{\shortmid
}h(y^{4})]^{2}\ ^{6}\mathbf{g}(y^{\check{c}})=\mathbf{(e}^{4})^{2}+[\
_{\shortmid }h(y^{4})]^{2}\ \ ^{6}\widetilde{\mathbf{g}}_{\check{a}\check{b}%
}(y^{\check{c}}),  \label{warpmetric} \\
\mathbf{e}^{4} &=&dy^{4}+w_{i}(x^{k},y^{a}),  \notag
\end{eqnarray}%
where $\ ^{6}\mathbf{g}$ (\ref{gpsmf}) can be considered for exact solutions
determined in 10-d gravity and $\ ^{6}\widetilde{\mathbf{g}}_{\check{a}%
\check{b}}=\ [\widetilde{g}_{\acute{\imath}\acute{j}},\widetilde{g}_{\grave{a%
}\grave{b}},\tilde{N}_{\acute{\imath}}^{\grave{a}}]$ (\ref{lfsmf}) is for
elaborating respective almost K\"{a}hler models.\footnote{%
The warping factor $h(y^{4})$ can be considered in a more generalized form $%
h(x^{i},y^{3},y^{4})$ because the AFDM also allows us to generate these
classes of solutions. For simplicity, we shall consider factorizations and
frame transforms when the warping factor depends only on the coordinate $%
y^{4}.$ This allows us to reproduce, in explicit form, the results for K\"{a}%
hler internal spaces if the 6-d metric structures do not depend on $y^{4}.$}
We denote by $\ _{\shortmid }\mathbf{e}^{\widetilde{a}\prime }=\{\mathbf{e}%
^{4\prime },\ _{\shortmid }h\cdot \ _{\shortmid }\widetilde{\mathbf{e}}^{%
\check{a}\prime }\}\in T^{\ast }(\ ^{7}\mathbf{X})$ an orthonormal
N--adapted basis on $\widetilde{a}^{\prime },\widetilde{b}^{\prime
},...=4,5,6,7,8,9,10,$ with $\ ^{\shortmid }\widetilde{\mathbf{e}}^{\check{a}%
\prime }=e_{\check{a}}^{\check{a}\prime }\widetilde{\mathbf{e}}^{\check{a}},$
when $\ ^{6}\widetilde{\mathbf{g}}_{\check{a}\check{b}}=\delta _{\check{a}%
\prime \check{b}^{\prime }}e_{\check{a}}^{\check{a}\prime }e_{\check{b}}^{%
\check{b}^{\prime }}.$

We can relate the N--adapted configuration to the orthonormal frame $\mathbf{%
e}^{\check{a}\prime }$ by introducing certain K\"{a}hler operators in
standard form instead of $(\mathbf{\tilde{J},}\widetilde{\Theta })$,%
\begin{equation*}
\ \ _{\shortmid }\mathbf{\tilde{J}}:=\ \ _{\shortmid }\widetilde{\mathbf{e}}%
^{5}\wedge \ _{\shortmid }\widetilde{\mathbf{e}}^{6}+\ _{\shortmid }%
\widetilde{\mathbf{e}}^{7}\wedge \ _{\shortmid }\widetilde{\mathbf{e}}^{8}+\
_{\shortmid }\widetilde{\mathbf{e}}^{9}\wedge \ _{\shortmid }\widetilde{%
\mathbf{e}}^{10},\ _{\shortmid }\widetilde{\Theta }:=(\ _{\shortmid }%
\widetilde{\mathbf{e}}^{5}+i\ _{\shortmid }\widetilde{\mathbf{e}}^{6})\wedge
(\ _{\shortmid }\widetilde{\mathbf{e}}^{7}+i\ _{\shortmid }\widetilde{%
\mathbf{e}}^{8})\wedge (\ _{\shortmid }\widetilde{\mathbf{e}}^{9}+\
_{\shortmid }\widetilde{\mathbf{e}}^{10}),
\end{equation*}%
where $i^{2}=-1$ is used for $SU(3).$ Using properties of such
orthonormalized N-adapted bases, we can verify that standard conditions for
the nearly K\"{a}hler internal spaces are satisfied \cite{harl,gray,lecht1},
but also mimic almost-K\"{a}hler manifolds with
\begin{equation*}
d(\ _{\shortmid }^{+}\widetilde{\Theta })=2\ \ _{\shortmid }\mathbf{\tilde{J}%
}\wedge \ _{\shortmid }\mathbf{\tilde{J}}\mbox{ and }d\ _{\shortmid }\mathbf{%
\tilde{J}}=3(\ _{\shortmid }^{-}\widetilde{\Theta }).
\end{equation*}

As a result, we can consider the same reduction of the instanton equations
(third formula in (\ref{bpseq})) as for holonomic K\"{a}hler structures with
nontrivial torsion structure encoded in N--adapted bases for differential
forms on nonholonomic internal manifolds $\ ^{7}\mathbf{X}$. Using two types
of warping variables,
\begin{equation}
dy^{4}=e^{f(\tau )}d\tau ,\mbox{ for }e^{f(\tau )}=\ _{\shortmid
}h(y^{4}(\tau ))  \label{tauparam}
\end{equation}%
and two equivalent d--metrics, $\ _{c}^{7}\mathbf{g}=e^{2f}\ _{z}^{7}\mathbf{%
g}$ and $\ _{z}^{7}\mathbf{g}=d\tau ^{2}+\ ^{6}\widetilde{\mathbf{g}}_{%
\check{a}\check{b}},$ we obtain nonholonomic instanton equations
\begin{equation}
\ast _{z}\widetilde{\mathbf{F}}=-(\ast _{z}\mathbf{Q}_{z})\wedge \widetilde{%
\mathbf{F}},  \label{nhinsteqcyl}
\end{equation}%
where $\mathbf{Q}_{z}=d\tau \wedge \ _{\shortmid }^{+}\widetilde{\Theta }+%
\frac{1}{2}\ _{\shortmid }\mathbf{\tilde{J}}\wedge \ _{\shortmid }\mathbf{%
\tilde{J}}$ and $\ast _{z}$is the Hodge star with respect to the cylinder
metric $\ _{z}^{7}\mathbf{g.}$ The almost-K\"{a}hler structure of $\ ^{7}%
\mathbf{X}$ is encoded into boldface operators $\mathbf{Q}_{z},\ _{\shortmid
}\mathbf{\tilde{J}}$ and canonical tilde like for $\ _{\shortmid }^{+}%
\widetilde{\Theta }.$ This does not allow us to solve such equations with an
ansatz for the canonical connection on $\ ^{6}\mathbf{X}$ determined by the
LC--connection as in \cite{harl,lecht1} but imposes the necessity to involve
the normal (almost symplectic) d--connection $\ \widetilde{\mathbf{D}}_{%
\check{a}}=(\widetilde{D}_{\acute{k}},\widetilde{D}_{\grave{b}})=\{%
\widetilde{\omega }_{\ \check{a}\check{c}^{\prime }}^{\check{b}^{\prime
}}\}, $ (\ref{cdccf}). Let us consider $\ \ _{A}\widetilde{\mathbf{D}}=\
^{can}\widetilde{\mathbf{D}}+\psi (\tau )\ _{\shortmid }\mathbf{e}^{%
\widetilde{a}\prime }I_{\widetilde{a}\prime },$ where the canonical
d--connection on $\ ^{6}\mathbf{X}$ enabled with almost-K\"{a}hler structure
is  $\ ^{can}\widetilde{\mathbf{D}}=\{\ ^{can}\widetilde{\omega }_{\ \check{a}%
\check{c}^{\prime }}^{\check{b}^{\prime }}:=\widetilde{\omega }_{\ \check{a}%
\check{c}^{\prime }}^{\check{b}^{\prime }}+\frac{1}{2}(\ _{\shortmid }^{+}%
\widetilde{\Theta })_{\ \check{c}^{\prime }\check{a}^{\prime }}^{\check{b}%
^{\prime }}e_{\check{a}}^{\check{a}^{\prime }}\}$.
In these formulas, the matrices $I_{\widetilde{a}\prime }=(I_{\widetilde{i}%
^{\prime }},I_{\check{a}^{\prime }})$ split into a basis/generators $I_{%
\widetilde{i}^{\prime }}$ $\subset \mathfrak{s0}(3)$ and generators $I_{%
\check{a}^{\prime }}$ for the orthogonal components of $\mathfrak{su}(3)$ in
$\mathfrak{g}\subset \mathfrak{s0}(7)$ satisfy the Lie algebra commutator $%
[I_{\check{a}^{\prime }},I_{\check{b}^{\prime }}]=f_{\check{a}^{\prime }%
\check{b}^{\prime }}^{\widetilde{i}^{\prime }}I_{\widetilde{i}^{\prime }}+f_{%
\check{a}^{\prime }\check{b}^{\prime }}^{\check{c}^{\prime }}I_{\check{c}%
^{\prime }},$ with respective structure constants $f_{\check{a}^{\prime }%
\check{b}^{\prime }}^{\widetilde{i}^{\prime }}$ and $f_{\check{a}^{\prime }%
\check{b}^{\prime }}^{\check{c}^{\prime }}$ (see formulas (3.9) in \cite%
{lecht1} for explicit parameterizations).

We can define and compute the curvature d--form
\begin{eqnarray*}
\ _{A}\widetilde{\mathbf{F}} &=&\frac{1}{2}[\ _{A}\widetilde{\mathbf{D}},\
_{A}\widetilde{\mathbf{D}}]:=\mathcal{F}(\psi ) \\
&=&\ ^{can}\widetilde{R}+\frac{1}{2}\psi ^{2}f_{\check{a}^{\prime }\check{b}%
^{\prime }}^{\widetilde{i}^{\prime }}\ I_{\widetilde{i}^{\prime }}\
_{\shortmid }\mathbf{e}^{\check{a}^{\prime }}\wedge \ _{\shortmid }\mathbf{e}%
^{\check{b}^{\prime }}+\frac{\partial \psi }{\partial \tau }d\tau \wedge I_{%
\check{c}^{\prime }}\ _{\shortmid }\mathbf{e}^{\check{c}^{\prime }}+\frac{1}{%
2}(\psi -\psi ^{2})I_{\check{b}^{\prime }}(\ _{\shortmid }^{+}\widetilde{%
\Theta })_{\ \check{c}^{\prime }\check{a}^{\prime }}^{\check{b}^{\prime }}\
_{\shortmid }\mathbf{e}^{\check{c}^{\prime }}\wedge \ _{\shortmid }\mathbf{e}%
^{\check{a}^{\prime }},
\end{eqnarray*}%
with parametric dependence on $\tau $ (\ref{tauparam}) via $\psi (\tau ).$
Such an $\ _{A}\widetilde{\mathbf{F}}$ is a solution of the nonholonomic
instanton equations  (\ref{nhinsteqcyl}) for any solution of the 'kink
equation' $\frac{\partial \psi }{\partial \tau }=2\psi (\psi -1).$ For the
aforementioned types of d--connections, and for the LC-connection, we can
consider two fixed points, $\psi =0$ and $\psi =1$ and a non-constant
solution $\ \psi (\tau )=\frac{1}{2}\left( 1-\tanh |\tau -\tau _{0}|\right)
, $ where the integration constant $\tau _{0}$ fixes the position of the
instanton in the $\tau $ direction but such an instanton also encodes an
almost-K\"{a}hler structure.

In heterotic supergravity, we can consider two classes of nonholonomic
instanton configurations. The first one is for the gauge-like curvature, $\
_{A}\widetilde{\mathbf{F}}$ $=\mathcal{F}(\ ^{1}\psi )$ and $\ \widetilde{%
\mathbf{R}}$ $=\mathcal{R}(\ ^{2}\psi )$, [in the second case, we also solve
the condition $\ \widetilde{\mathbf{R}}$ $\cdot \epsilon =0$] where the
values $\ ^{1}\psi $ and $\ ^{2}\psi $ will be defined below.

\subsection{Static and/or dynamic $SU(3)$ nonholonomic structures on almost K%
\"{a}hler configurations}

The transforms $\ ^{6}\widetilde{\mathbf{g}}_{\check{a}\check{b}}=$ $[\
_{\shortmid }h]^{2}\ \ ^{6}\widetilde{\mathbf{g}}_{\check{a}\check{b}}$ in
d--metric (\ref{warpmetric}) impose certain relations on the two pairs $(\
_{\shortmid }\mathbf{\tilde{J},}\ _{\shortmid }\widetilde{\Theta })$ and $(%
\mathbf{\tilde{J},}\widetilde{\Theta })$, where the last couple is subjected
to the respective flow and structure equations, (\ref{floweq2}) and (\ref%
{struceq2}), and define the 3--form $\widehat{\mathbf{H}}$ (\ref{kheqs}). We
note that such relations are a mixing between real and imaginary parts and
nonholonomically constrained in order to adapt the Lie algebra symmetries to
the almost-K\"{a}hler structure. Like in \cite{lecht1}, it is considered a $%
y^{4}$ depending mixing angle $\ _{\shortmid }\beta (y^{4})\in \lbrack
0,2\pi )$ when%
\begin{equation*}
\mathbf{\tilde{J}}\mathbf{=}[\ _{\shortmid }h]^{2}\ _{\shortmid }\mathbf{%
\tilde{J},}\ ^{+}\widetilde{\Theta }=[\ _{\shortmid }h]^{3}(\ _{\shortmid
}^{+}\widetilde{\Theta }\cos \ _{\shortmid }\beta +\ _{\shortmid }^{-}%
\widetilde{\Theta }\sin \ _{\shortmid }\beta ),\ ^{-}\widetilde{\Theta }=[\
_{\shortmid }h]^{3}(-\ _{\shortmid }^{+}\widetilde{\Theta }\sin \
_{\shortmid }\beta +\ _{\shortmid }^{-}\widetilde{\Theta }\cos \ _{\shortmid
}\beta ).
\end{equation*}%
Introducing such values into the relations for $(\mathbf{\tilde{J},}\ ^{+}%
\widetilde{\Theta },\ ^{-}\widetilde{\Theta }),$ we obtain (compare to (\ref%
{table}))
{\small
\begin{eqnarray*}
\widehat{\mathbf{H}} &=&\ell vol[\ ^{3}\mathbf{g]}+\ _{\shortmid
}hdy^{4}\wedge (\ell \ _{\shortmid }h-4\sin \ _{\shortmid }\beta )\
_{\shortmid }\mathbf{\tilde{J}}+[\ _{\shortmid }h]^{2}\left[ -\frac{1}{2}\ _{\shortmid }h(\partial
_{4}\phi )\cos \ _{\shortmid }\beta +3\sin ^{2}\ _{\shortmid }\beta -\frac{7%
}{8}\ell \ _{\shortmid }h\sin \ _{\shortmid }\beta \right] \ _{\shortmid
}^{+}\widetilde{\Theta }  \\
&& +[\ _{\shortmid }h]^{2}\left[ -\frac{1}{2}\ _{\shortmid }h(\partial
_{4}\phi )\sin \ _{\shortmid }\beta -3\sin \ _{\shortmid }\beta \cos \
_{\shortmid }\beta +\frac{7}{8}\ell \ _{\shortmid }h\cos \ _{\shortmid
}\beta \right] \ _{\shortmid }^{-}\widetilde{\Theta }.
\end{eqnarray*}%
}
The above formula involves the conditions $\ \ ^{\mathbf{+}}\mathcal{T}%
_{1}=2(\ _{\shortmid }h)^{-1}\cos \ _{\shortmid }\beta $ and$\ ^{\mathbf{-}}%
\mathcal{T}_{1}=-2(\ _{\shortmid }h)^{-1}\sin \ _{\shortmid }\beta $ which
allows us to fix $\ \tilde{\alpha}(y^{4})=3\ ^{\mathbf{-}}\mathcal{T}_{1}+%
\frac{15}{8}\ell .$

There are additional conditons on the scalar functions $\ _{\shortmid }h,\
_{\shortmid }\beta ,\ell $ and $\phi $ which must be imposed on coupled
nonholonomic instanton solutions $\ ^{1}\psi $ and $\ ^{2}\psi $ satisfying
the Bianchi conditions, the nonholonomic BPS equations and the time--like
components of the equations of motion. We omit such an analysis because it
is similar to that of the pure K\"{a}hler configurations, see sections 4 and
5 in \cite{lecht1}. The purpose of nonholonomic almost-K\"{a}hler variables
is so that we can work with respect to N-adapted frames in a form which is
very similar to that for complex and symplectic structures.

\subsection{Equations of motion of heterotic supergravity in nonholonomic
variables}

In order to apply the AFDM, we have to rewrite the equations of motion
(generalized Einstein equations) in nonholonomic variables. The formal
procedure is to take such equations written for the LC--connection with
respect to coordinate frames and re--write them for the same metric
structure, but for corresponding geometric objects with "hats " and "waves"
and with respect to N--adapted frames on corresponding shells. In this way,
\ including terms of order $\alpha ^{\prime },$ the N--adapted equations of
motion of heterotic nonholonomic supergravity considered in \cite{lecht1}
can be written in such a form:%
\begin{eqnarray}
\widehat{\mathbf{R}}_{\mu _{s}\nu _{s}}+2(\ ^{s}\widehat{\mathbf{D}}\widehat{%
\mathbf{d}}\widehat{\phi })_{\mu _{s}\nu _{s}}-\frac{1}{4}\widehat{\mathbf{H}%
}_{\alpha _{s}\beta _{s}\mu _{s}}\widehat{\mathbf{H}}_{\nu _{s}}^{\quad
\alpha _{s}\beta _{s}}+\frac{\alpha ^{\prime }}{4}\left[ \widetilde{\mathbf{R%
}}_{\mu _{s}\alpha _{s}\beta _{s}\gamma _{s}}\widetilde{\mathbf{R}}_{\nu
_{s}}^{\quad \alpha _{s}\beta _{s}\gamma _{s}}-tr\left( \widehat{\mathbf{F}}%
_{\mu _{s}\alpha _{s}}\widehat{\mathbf{F}}_{\nu _{s}}^{\quad \alpha
_{s}}\right) \right] &=&0,  \label{hs1} \\
\ ^{s}\widehat{R}+4\widehat{\square }\widehat{\phi }-4|\widehat{\mathbf{d}}%
\widehat{\phi }|^{2}-\frac{1}{2}|\widehat{\mathbf{H}}|^{2}+\frac{\alpha
^{\prime }}{4}tr\left[ |\widetilde{\mathbf{R}}|^{2}-|\widehat{\mathbf{F}}|%
\right] &=&0,  \label{hs2} \\
e^{2\widehat{\phi }}\widehat{\mathbf{d}}\widehat{\ast }(e^{-2\widehat{\phi }}%
\widehat{\mathbf{F}})+\widehat{\mathbf{A}}\wedge \widehat{\ast }\widehat{%
\mathbf{F}}-\widehat{\ast }\widehat{\mathbf{F}}\wedge \widehat{\mathbf{A}}+%
\widehat{\ast }\widehat{\mathbf{H}}\wedge \widehat{\mathbf{F}} &=&0,
\label{hs3} \\
\widehat{\mathbf{d}}\widehat{\ast }(e^{-2\widehat{\phi }}\widehat{\mathbf{H}}%
) &=&0,  \label{hs4}
\end{eqnarray}%
where the Hodge operator $\widehat{\ast },$ $\ ^{s}\widehat{\mathbf{D}}=\{%
\widehat{\mathbf{D}}_{\mu _{s}}\}$ (\ref{candcon}), the canonical
nonholonomic d'Alambert wave operator $\widehat{\square }:=\widehat{\mathbf{g%
}}^{\mu _{s}\nu _{s}}\widehat{\mathbf{D}}_{\mu _{s}}\widehat{\mathbf{D}}%
_{\nu _{s}},$ $\widehat{\mathbf{R}}_{\mu _{s}\nu _{s}}$ (\ref{dricci}),$\
^{s}\widehat{R}$ (\ref{rdsc}) are all determined by a d--metric $\widehat{%
\mathbf{g}}$ (\ref{dm}). \ The curvature d--tensor $\widetilde{\mathbf{R}}%
_{\mu _{s}\alpha _{s}\beta _{s}\gamma _{s}}$ is taken for an almost-K\"{a}%
hler structure $\widetilde{\theta }$ (\ref{canalmsf}) defined by
corresponding nonholonomic distributions which are stated up to frame
transforms by the N--connection structure and components of d--metric on
shells $s=1,2,3$ as we described above. The gauge field $\widehat{\mathbf{A}}
$ corresponds to the N--adapted operator
\begin{equation*}
\ _{A}^{s}\widehat{\mathbf{D}}=\ ^{s}\widehat{\mathbf{D}}+\ ^{1}\psi (y^{4})[%
\mathbf{e}^{a_{1}}I_{a_{1}}+\mathbf{e}^{a_{2}}I_{a_{2}}+\mathbf{e}%
^{a_{3}}I_{a_{3}}]=\ ^{s}\widehat{\mathbf{D}}+\ ^{1}\psi (y^{4})I_{\check{c}%
^{\prime }}\ _{\shortmid }\mathbf{e}^{\check{c}^{\prime }}=\widehat{\mathbf{d%
}}+\widehat{\mathbf{A}}
\end{equation*}%
and curvature $\widehat{\mathbf{F}}=\mathcal{F}(\ ^{1}\psi )$ via a map
constructed above (for $^{s}\widehat{\mathbf{D}}_{\mid \widehat{\mathcal{T}}%
=0}\rightarrow \ ^{s}\nabla ,$ see details in \cite{lecht1}). \ For
instance, the LC--configurations of (\ref{hs1})\ are determined by equations
\begin{equation}
R_{\mu \nu }+2(\nabla d\phi )_{\mu \nu }-\frac{1}{4}H_{\alpha \beta \mu
}H_{\nu }^{\quad \alpha \beta }+\frac{\alpha ^{\prime }}{4}\left[ \widetilde{%
R}_{\mu \alpha \beta \gamma }\widetilde{R}_{\nu }^{\quad \alpha \beta \gamma
}-tr\left( \widehat{F}_{\mu \alpha }\widehat{F}_{\nu }^{\quad \alpha
}\right) \right] =0  \label{hs1a}
\end{equation}%
with standard 10-d indices $\alpha ,\beta ,...=0,1,2,...9$ and geometric
values determined by $\nabla .$ Unfortunately, the system of nonlinear PDEs (%
\ref{hs1a}) can not be decoupled and integrated in any general form if we do
not consider shell N--adapted frames/coordinates and generalized connections
which can be nonholonomically constrained to LC-configurations.

The equations (\ref{hs3}) and (\ref{hs4}) can be solved for arbitrary
almost-K\"{a}hler internal spaces by introducing corresponding classes of
N-adapted variables as we proved in previous sections. Such solutions can be
classified as for pure K\"{a}hler spaces, for simplicity, considering a
special case with $\ ^{\mathbf{-}}\mathcal{T}_{1}=0$ and $\ell =0$ and in
terms of 'kink' solutons with $\ e^{2f}=e^{2(\tau -\tau _{0})}+\frac{\alpha
^{\prime }}{4}[(\ ^{1}\psi )^{2}-(\ ^{2}\psi )^{2}],\tau _{0}=const.$ The NS
3--form flux is given by a simple formula%
\begin{equation}
\widehat{\mathbf{H}}(\tau ,y^{\check{c}})=-\frac{1}{2}=[\ _{\shortmid
}h]^{3}(\partial _{4}\phi )(\ _{\shortmid }^{+}\widetilde{\Theta })=\frac{%
\alpha ^{\prime }}{4}[(\ ^{1}\psi )^{2}(2\ ^{1}\psi -3)-(\ ^{2}\psi )^{2}(2\
^{2}\psi -3)](\tau )[\ _{\shortmid }^{+}\widetilde{\Theta }(y^{\check{c}})].
\label{kinkhconf}
\end{equation}%
Here we reproduce the classification of 8 cases with fixed and/or kink
configurations for almost-K\"{a}hler configurations,%
\begin{equation*}
\begin{array}{ccc}
\mbox{ Case } & \ ^{1}\psi ,\ ^{2}\psi ; & e^{2f}=e^{2(\tau -\tau _{0})}+%
\frac{\alpha ^{\prime }}{4}[(\ ^{1}\psi )^{2}-(\ ^{2}\psi )^{2}],\phi (\tau
)=\phi _{0}+2(f-\tau ) \\
1. & \ ^{1}\psi =\ ^{2}\psi ; & f=\tau -\tau _{0},\phi =\phi _{0}-2\tau \\
2. & \ ^{1}\psi =\ ^{2}\psi =0; & f=f_{0}:=\frac{1}{2}\log (\frac{\alpha
^{\prime }}{4}),\phi =\phi _{0}+2(f_{0}-\tau ) \\
3. & \ ^{1}\psi =1,\ ^{2}\psi =0; & e^{2f}=e^{2(\tau -\tau _{0})}-\frac{%
\alpha ^{\prime }}{4},e^{\phi -\phi _{0}}=e^{-2\tau _{0}}-\frac{\alpha
^{\prime }}{4}e^{-2\tau } \\
4. & \ ^{1}\psi =0,\ ^{2}\psi =\mbox{ kink }; & e^{2f}=e^{2(\tau -\tau
_{0})}+\frac{\alpha ^{\prime }}{16}[1-\tanh (\tau -\tau _{1})]^{2},\phi
=\phi _{0}+2(f-\tau ) \\
5. & ^{1}\psi =\mbox{ kink },\ ^{2}\psi =0; & e^{2f}=e^{2(\tau -\tau _{0})}-%
\frac{\alpha ^{\prime }}{16}[1-\tanh (\tau -\tau _{1})]^{2},\phi =\phi
_{0}+2(f-\tau ) \\
6. & ^{1}\psi =1,\ ^{2}\psi =\mbox{ kink }; & e^{2f}=e^{2(\tau -\tau _{0})}+%
\frac{\alpha ^{\prime }}{16}[\tanh (\tau -\tau _{1})+1][\tanh (\tau -\tau
_{1})-3] \\
7. & ^{1}\psi =\mbox{ kink },\ ^{2}\psi =1; & e^{2f}=e^{2(\tau -\tau _{0})}-%
\frac{\alpha ^{\prime }}{16}[\tanh (\tau -\tau _{1})+1][\tanh (\tau -\tau
_{1})-3] \\
8. &
\begin{array}{c}
^{1}\psi =\mbox{ kink },\ ^{2}\psi =\mbox{ kink } \\
\mbox{ with }\tau _{1}\neq \tau _{2};%
\end{array}
&
\begin{array}{c}
e^{2f}=e^{2(\tau -\tau _{0})}+\frac{\alpha ^{\prime }}{16}[\tanh ^{2}(\tau
-\tau _{1})-2\tanh (\tau -\tau _{1})- \\
\tanh ^{2}(\tau -\tau _{2})+2\tanh (\tau -\tau _{2})]%
\end{array}%
\end{array}%
\end{equation*}%
Such a classification can be used for parametrizing certain effective
sources of Einstein--Yang-Mills-Higgs type and preserved for constructing
generic off-diagonal solutions following the AFDM \cite%
{sv2001,vapexsol,vex3,vpars,svvvey,tgovsv,vgrg,vsingl2}. We emphasize that
classes 1-8 distinguish different almost-K\"{a}hler structures encoding
corresponding assumptions, that for identification of almost complex
structure $I$ with the complex unity $i$ in the typical fiber of tangent
space use the same classification as for symplectic configurations
introduced in \cite{lecht1,harl}. A unified classification for internal
(almost) K\"{a}hler spaces is possible with respect to corresponding
N--adapted frames generated by a conventional Lagrange function as we
considered in formulas (\ref{clncfa}) and (\ref{canalmsf}).

Finally, we note that we shall construct and analyze various classes of
generic off-diagonal exact solutions of the equations (\ref{hs1}) and (\ref%
{hs2}) in our associated work \cite{partner}. The main goal of that work is
to prove that it is possible to reproduce certain types of off--diagonal
deformations of the Kerr metric in heterotic supergravity for the 4-d sector
if the internal 6-d space is endowed with richer (than in K\"{a}hler
geometry) structures. In section \ref{s3}, we proved that it is always
possible to introduce such nonholonomic variables when the internal space
geometric data is parameterized via geometric objects of an effective
almost-K\"{a}hler geometry. This allows us to solve the
equations (\ref{hs3}) and (\ref{hs4}) following the same procedure and
classification as in \cite{lecht1,harl} when the domain walls were endowed
with trivial pseudo-Euclidean structure warped nearly K\"{a}hler internal
spaces. The off--diagonal deformation techniques defined by the AFDM allow
us to generalize the constructions for nontrivial exact and parametric
solutions in 4-d, 6-d and 10-d MGTs and string gravity.

\section{String Deformations of Kerr Metrics with Quasiperiodic or Pattern--Form\-ing Structure}
\label{s4}
We provide and analyze explicit examples of solutions in heterotic string gravity defining deformations of the Kerr metric with pattern-forming and spatiotemporal chaos \cite{rucklidge12} in the 4-d spacelike part and/or for the internal almost-K\"{a}hler structure. In brief, we shall refer to such configurations with possible additional solitonic distributions and nonlinear waves as quasiperiodic structures. It should be noted that the generic off-diagonal solutions for the effective 4-d and string 10-d gravity considered in this section are  characterized by effective free energy functionals as in \cite{rucklidge12,rucklidge16}, see also recent results on quasiperiodic (quasicrystal like)  cosmological models \cite{aschheim16,amaral16}.\ The geometric method for constructing such solutions in string gravity is described in details in the partner and recent works on locally anisotropic cosmology \cite{partner,sv2001,veym,tgovsv,vex1,vex2,vex3} and references therein.

\subsection{Solutions with small parametric deformations in 10-d string gravity}

The motion equations for heterotic gravity (\ref{hs1}) - (\ref{hs4}) with
nontrivial sources determined by fields $\widehat{\mathbf{H}}_{\alpha
_{s}\beta _{s}\mu _{s}},$ $\widehat{\mathbf{F}}_{\mu _{s}\alpha _{s}},$ and $%
\widehat{\phi },$ and internal almost-K\"{a}hler curvature $\widetilde{%
\mathbf{R}}_{\mu _{s}\alpha _{s}\beta _{s}\gamma _{s}}$ can be integrated
in general form for certain parameterizations of such fields with respect to
N--adapted frames. By explicit computations, we can check this for
\begin{equation*}
\widehat{\mathbf{H}}_{\alpha _{s}\beta _{s}\mu _{s}}=\ ^{H}s\sqrt{|\mathbf{g}%
_{\beta _{s}\mu _{s}}|}\epsilon _{\alpha _{s}\beta _{s}\mu _{s}},\ \widehat{%
\mathbf{F}}_{\mu _{s}\alpha _{s}}=\ ^{F}s\sqrt{|\mathbf{g}_{\beta _{s}\mu
_{s}}|}\epsilon _{\mu _{s}\alpha _{s}},\ \widetilde{\mathbf{R}}_{\mu
_{s}\alpha _{s}\beta _{s}\gamma _{s}}=\ ^{R}s\sqrt{|\mathbf{g}_{\beta
_{s}\mu _{s}}|}\epsilon _{\mu _{s}\alpha _{s}\beta _{s}\gamma _{s}},
\end{equation*}%
with absolute anti-symmetric $\epsilon $-tensors, which are similar to
ansatz considered for (effective) Einstein Yang--Mills Higgs systems and
heterotic string gravity \cite{svvvey,partner}. The corresponding effective
energy--momentum tensors are computed
\begin{eqnarray}
\mathbf{\Upsilon }_{ij} &=&\ _{K}\mathbf{\Upsilon }=\ _{K}\Lambda \mathbf{g}%
_{ij}  \label{effsources} \\
\ ^{H}\mathbf{\Upsilon }_{\mu _{s}\nu _{s}} &=&-\frac{10}{2}(\ ^{H}s)^{2}%
\mathbf{g}_{\beta _{s}\mu _{s}},\mbox{ for }\ ^{H}\Lambda =-5(\ ^{H}s)^{2},
\notag \\
\ ^{F}\mathbf{\Upsilon }_{\mu _{s}\nu _{s}} &=&-\frac{10\alpha ^{\prime }}{2}%
n_{F}(\ ^{F}s)^{2}\mathbf{g}_{\beta _{s}\mu _{s}},\mbox{ for
}\ ^{F}\Lambda =-5n_{F}(\ ^{F}s)^{2},  \notag \\
\ ^{R}\mathbf{\Upsilon }_{\mu _{s}\nu _{s}} &=&\frac{10\alpha ^{\prime }}{2}%
n_{R}(\ ^{R}s)^{2}\mathbf{g}_{\beta _{s}\mu _{s}},\mbox{ for }\ ^{R}\Lambda
=-5trn_{R}(\ ^{R}s)^{2},  \notag
\end{eqnarray}%
where respectively associated cosmological constants are determined, for
instance, by the numbers $n_{F}=tr[internal$ $F]$ and $n_{R}=tr[internal$ $%
\widetilde{R}]$ depending on the representation of the Lie algebra for $F$
and on the representation of Lie groups on the internal space, and on real
constants $^{H}s,\ ^{F}s$ and $\ ^{R}s.$ We can introduce more general
sources with coefficients in N--adapted form determined by redefinition of
generating functions and sources but all resulting in an effective
cosmological constant $\Lambda =\ ^{H}\Lambda +\ ^{F}\Lambda +\ ^{R}\Lambda
$ (see formula (104) and relevant details for sections 2.3.4 and 3.4.2 in
the partner work \cite{partner}).

The quadratic elements for such classes of 10-d off-diagonal solutions with
parametric dependence on a small parameter $\varepsilon ,$ ($0\leq
\varepsilon \ll 1$) in heterotic supergravity split in conventional 4-d
spacetime and 6-d internal space components,%
\begin{equation}
ds_{10d}^{2}=ds^{2}[^{4}\mathbf{g}]+ds^{2}[^{6}\mathbf{g}].  \label{10dqe}
\end{equation}%
In this formula, the 4-d spacetime part of the d-metric is written \
\begin{eqnarray}
ds^{2}[^{4}\mathbf{g}] &=&(1+\varepsilon e^{\ ^{0}q}\frac{\ ^{1}q}{\Lambda }%
)[(dx^{1\prime })^{2}+(dx^{2\prime })^{2}]+\left( \overline{A}-\varepsilon
\frac{\mathring{\Phi}^{2}\chi }{4\Lambda }\right) [dy^{3^{\prime }}+\left(
\varepsilon \partial _{k^{\prime }}\ ^{\chi }n(x^{i^{\prime }})-\partial
_{k^{\prime }}(\widehat{y}^{3^{\prime }}+\varphi \frac{\overline{B}}{%
\overline{A}})\right) dx^{k^{\prime }}]^{2}  \notag \\
&&+\left[ 1+\varepsilon \ \left( 2(\chi +\frac{\mathring{\Phi}}{\partial
_{\varphi }\mathring{\Phi}}\partial _{\varphi }\chi )+\frac{\mathring{\Phi}%
^{2}\chi }{4\overline{A}\Lambda }\right) \right] (\overline{C}-\frac{%
\overline{B}^{2}}{\overline{A}})[d\varphi +\varepsilon \partial _{i^{\prime
}}\check{A})dx^{i^{\prime }}]^{2},  \label{4dqe}
\end{eqnarray}%
where $\partial _{4}=\partial /\partial y^{4}=\partial _{\varphi }=\partial
/\partial \varphi $. The functions $\chi (x^{k},y^{4})$ and $\mathring{\Phi}%
(x^{k},y^{4})$ define a generating function $\ ^{\varepsilon }\Phi =%
\mathring{\Phi}(x^{k},\varphi )[1+\varepsilon \chi (x^{k},\varphi )]$
subjected to the condition
\begin{equation}
\varepsilon \partial _{i^{\prime }}\check{A}=\partial _{i^{\prime }}(\
^{\varepsilon }\Phi )/\partial _{\varphi }(\ ^{\varepsilon }\Phi ),
\label{acoef}
\end{equation}%
which for any functions $\check{A}(x^{k},\varphi )$ and $^{\chi
}n(x^{i^{\prime }}),$ (for $i^{\prime }=1,2$) result in zero
nonholonomically induced torsion. The values $\widetilde{\zeta },\omega
_{0},\varphi _{0}$ are some integration constants and the functions $\
^{0}q(x^{i^{\prime }})$ and $\ ^{1}q(x^{i^{\prime }})$ are determined by a
solution of the 2-d Poisson equation $\partial _{1}^{2}q+\partial
_{2}^{2}q=2\Lambda ,$ when
\begin{equation*}
(1+\varepsilon e^{\ ^{0}q}\frac{\ ^{1}q}{\Lambda })[(dx^{1\prime
})^{2}+(dx^{2\prime })^{2}]=e^{q(x^{k})}[(dx^{1})^{2}+(dx^{2})^{2}].
\end{equation*}%
For $\varepsilon =0,$ the 4-d metric (\ref{4dqe}) transforms into the Kerr
metric in the so--called Boyer--Linquist coordinates $(r,\vartheta ,\varphi
,t),$ for $r=m_{0}(1+p\widehat{x}_{1}),\widehat{x}_{2}=\cos \vartheta $.
Such a parametrization is convenient for applications of the AFDM in order
to construct nonholonomic and extra dimensional deformations of solutions
and generalizations. The Boyer--Linquist coordinates can be
expressed via certain parameters $p,q$ (related to the total black hole
mass, $m_{0}$ and the total angular momentum, $am_{0},$ for the
asymptotically flat, stationary and anti-symmetric Kerr spacetime), when $%
m_{0}=Mp^{-1}$ and $a=Mqp^{-1}$ when $p^{2}+q^{2}=1$ implies $%
m_{0}^{2}-a^{2}=M^{2}$. In these variables, the vacuum 4-d Kerr black hole
solution can be written in the form%
\begin{eqnarray*}
ds_{[0]}^{2} &=&(dx^{1^{\prime }})^{2}+(dx^{2^{\prime }})^{2}+\overline{A}(%
\mathbf{e}^{3^{\prime }})^{2}+(\overline{C}-\overline{B}^{2}/\overline{A})(%
\mathbf{e}^{4^{\prime }})^{2}, \\
\mathbf{e}^{3^{\prime }} &=&dt+d\varphi \overline{B}/\overline{A}%
=dy^{3^{\prime }}-\partial _{i^{\prime }}(\widehat{y}^{3^{\prime }}+\varphi
\overline{B}/\overline{A})dx^{i^{\prime }},\mathbf{e}^{4^{\prime
}}=dy^{4^{\prime }}=d\varphi ,
\end{eqnarray*}%
where $\ x^{1^{\prime }}(r,\vartheta ),\ x^{2^{\prime }}(r,\vartheta ),\
y^{3^{\prime }}=t+\widehat{y}^{3^{\prime }}(r,\vartheta ,\varphi )+\varphi
\overline{B}/\overline{A},y^{4^{\prime }}=\varphi ,\ \partial _{\varphi }%
\widehat{y}^{3^{\prime }}=-\overline{B}/\overline{A}$ are any coordinate
functions for which $(dx^{1^{\prime }})^{2}+(dx^{2^{\prime }})^{2}=\Xi
\left( \Delta ^{-1}dr^{2}+d\vartheta ^{2}\right) $, and the coefficients are%
\begin{eqnarray}
\overline{A} &=&-\Xi ^{-1}(\Delta -a^{2}\sin ^{2}\vartheta ),\overline{B}%
=\Xi ^{-1}a\sin ^{2}\vartheta \left[ \Delta -(r^{2}+a^{2})\right] ,\overline{%
C}=\Xi ^{-1}\sin ^{2}\vartheta \left[ (r^{2}+a^{2})^{2}-\Delta a^{2}\sin
^{2}\vartheta \right] ,\mbox{ and }  \notag \\
\Delta &=&r^{2}-2m_{0}+a^{2},\ \Xi =r^{2}+a^{2}\cos ^{2}\vartheta .
\label{kerrcoef}
\end{eqnarray}

The effective d-metric for the 6-d internal space in (\ref{10dqe}) is{\small
\begin{eqnarray}
&&ds^{2}[^{6}\mathbf{g}]=[1-\varepsilon \frac{\ ^{1}\mathring{\Phi}^{2}\
_{1}^{\Lambda }\chi }{4\Lambda \underline{\mathring{g}}_{5}}]\underline{%
\mathring{g}}_{5}[dy^{6}+[1+\varepsilon \ \ ^{1}\widetilde{n}_{i_{1}}\int
dy^{6}(\ _{1}^{\Lambda }\chi +\frac{\ ^{1}\mathring{\Phi}}{\partial _{6}(\
^{1}\mathring{\Phi})}\partial _{6}(\ _{1}^{\Lambda }\chi )-\frac{5}{8\Lambda
\underline{\mathring{g}}_{6}}\partial _{6}(\ ^{1}\mathring{\Phi}^{2}\
_{1}^{\Lambda }\chi ))]\underline{\mathring{n}}_{i_{1}}d\widetilde{x}%
^{i_{1}}]^{2}  \notag \\
&&+[1+\varepsilon \ \left( 2(\ _{1}^{\Lambda }\chi +\frac{\ ^{1}\mathring{%
\Phi}}{\partial _{6}(\ ^{1}\mathring{\Phi})}\partial _{6}(\ _{1}^{\Lambda
}\chi )-\frac{\ ^{1}\mathring{\Phi}}{4\Lambda \underline{\mathring{g}}_{6}}\
_{1}^{\Lambda }\chi \right) ]\underline{\mathring{g}}_{6}\left[
dy^{6}+[1+\varepsilon (\frac{\partial _{i_{1}}(\ _{1}^{\Lambda }\chi \ \ ^{1}%
\mathring{\Phi})}{\partial _{i_{1}}\ \ ^{1}\mathring{\Phi}}-\frac{\partial
_{6}(\ _{1}^{\Lambda }\chi \ \ ^{1}\mathring{\Phi})}{\partial _{6}(\ ^{1}%
\mathring{\Phi})})]\widetilde{\mathring{w}}_{i_{1}}d\widetilde{x}^{i_{1}}%
\right] ^{2}  \label{6dqe}
\end{eqnarray}%
\begin{eqnarray*}
&&+[1-\varepsilon \frac{^{2}\mathring{\Phi}^{2}\ _{2}^{\Lambda }\chi }{%
4\Lambda \underline{\mathring{g}}_{7}}\ ]\underline{\mathring{g}}%
_{7}[dy^{7}+[1+\varepsilon \ \ ^{2}\widetilde{n}_{i_{2}}\int dy^{8}(\
_{2}^{\Lambda }\chi +\frac{\ ^{2}\mathring{\Phi}}{\partial _{7}(\ ^{2}%
\mathring{\Phi})}\partial _{8}(\ _{2}^{\Lambda }\chi )-\frac{5}{8\Lambda
\underline{\mathring{g}}_{8}}\partial _{8}(\ ^{2}\mathring{\Phi}^{2}\
_{2}^{\Lambda }\chi ))]\underline{\mathring{n}}_{i_{2}}d\widetilde{x}%
^{i_{2}}]^{2} \\
&&+[1+\varepsilon \ \left( 2(\ _{2}^{\Lambda }\chi +\frac{\ ^{2}\mathring{%
\Phi}}{\partial _{8}(\ ^{2}\mathring{\Phi})}\partial _{8}(\ _{2}^{\Lambda
}\chi )-\frac{\ ^{2}\mathring{\Phi}}{4\Lambda \underline{\mathring{g}}_{8}}\
_{2}^{\Lambda }\chi \right) ]\underline{\mathring{g}}_{8}\left[
dy^{8}+[1+\varepsilon (\frac{\partial _{i_{2}}(\ _{2}^{\Lambda }\chi \ \ ^{2}%
\mathring{\Phi})}{\partial _{i_{2}}\ \ ^{2}\mathring{\Phi}}-\frac{\partial
_{8}(\ _{2}^{\Lambda }\chi \ \ ^{2}\mathring{\Phi})}{\partial _{8}(\ ^{2}%
\mathring{\Phi})})]\widetilde{\mathring{w}}_{i_{2}}d\widetilde{x}^{i_{2}}%
\right] ^{2}
\end{eqnarray*}%
\begin{eqnarray*}
&&+[1-\varepsilon \frac{\ ^{3}\mathring{\Phi}^{2}\ _{3}^{\Lambda }\chi }{%
4\Lambda \underline{\mathring{g}}_{9}}]\underline{\mathring{g}}_{9}(x^{k}(%
\widetilde{x}^{k^{\prime }}))[dy^{10}+[1+\varepsilon \ \ ^{3}\widetilde{n}%
_{i_{3}}\int dy^{10}(\ _{3}^{\Lambda }\chi +\frac{\ ^{3}\mathring{\Phi}}{%
\partial _{10}(\ ^{3}\mathring{\Phi})}\partial _{10}(\ _{3}^{\Lambda }\chi )-%
\frac{5}{8\Lambda \underline{\mathring{g}}_{10}}\partial _{10}(\ ^{3}%
\mathring{\Phi}^{2}\ _{3}^{\Lambda }\chi ))]\underline{\mathring{n}}_{i_{3}}d%
\widetilde{x}^{i_{3}}]^{2} \\
&&+[1+\varepsilon \ \left( 2(\ _{3}^{\Lambda }\chi +\frac{\ 3\mathring{\Phi}%
}{\partial _{10}(\ ^{3}\mathring{\Phi})}\partial _{10}(\ _{3}^{\Lambda }\chi
))-\frac{\ ^{3}\mathring{\Phi}}{4\Lambda \underline{\mathring{g}}_{10}}\
_{3}^{\Lambda }\chi \right) ]\underline{\mathring{g}}_{10}\left[
dy^{10}+[1+\varepsilon (\frac{\partial _{i_{3}}(\ _{3}^{\Lambda }\chi \ \
^{3}\mathring{\Phi})}{\partial _{i_{3}}\ \ ^{3}\mathring{\Phi}}-\frac{%
\partial _{10}(\ _{3}^{\Lambda }\chi \ \ ^{3}\mathring{\Phi})}{\partial
_{10}(\ ^{3}\mathring{\Phi})})]\widetilde{\mathring{w}}_{i_{3}}d\widetilde{x}%
^{i_{3}}\right] ^{2}.
\end{eqnarray*}%
}

In these formulas, we consider generating functions of the form
\begin{eqnarray}
&&\ ^{1}\mathring{\Phi}(r,\vartheta ,\varphi ,y^{6}),\ ^{2}\mathring{\Phi}%
(r,\vartheta ,\varphi ,y^{6},y^{8}),\ ^{3}\mathring{\Phi}(r,\vartheta
,\varphi ,y^{6},y^{8},y^{10})\mbox{ and }  \notag \\
&&\ _{1}^{\Lambda }\chi \left( r,\vartheta ,\varphi ,t,y^{6}\right) ,\
_{2}^{\Lambda }\chi \left( r,\vartheta ,\varphi
,t,y^{5},y^{6},y^{7},y^{8}\right) ,\ _{2}^{\Lambda }\chi \left( r,\vartheta
,\varphi ,t,y^{5},y^{6},y^{7},y^{8},y^{10}\right) ,  \label{genfkerrd}
\end{eqnarray}%
where the coordinates are parameterized
\begin{eqnarray*}
u^{\mu } &=&(x^{1}=x^{1^{\prime }}=r,x^{2}=x^{2^{\prime }}=\vartheta
,x^{3}=t,x^{4}=x^{4^{\prime }}=\varphi , \\
y^{a_{1}}
&=&\{u^{5}=y^{5},u^{6}=y^{6}\},y^{a_{2}}=\{u^{7}=y^{7},u^{8}=y^{8}%
\},y^{a_{3}}=\{u^{9}=y^{9},u^{10}=y^{10}\}).
\end{eqnarray*}%
In above solutions and generating functions, the left symbols label the
"shells" $s=1,2,3,$ and emphasize that such metrics are defined as solutions
in 10-d heterotic supergravity with effective sources (\ref{effsources})
wtih $\Lambda =\ ^{H}\Lambda +\ ^{F}\Lambda +\ ^{R}\Lambda .$

The target metric $\mathbf{g}$ (\ref{10dqe}) describes a nonholonomic
deformation $\mathbf{\mathring{g}}=[\mathring{g}_{i},\mathring{h}_{a_{s}},%
\mathring{N}_{b_{s}}^{j_{s}}]$ $\rightarrow \mathbf{g=}\ ^{s}\mathbf{g}%
=[g_{i},h_{a_{s}},N_{b_{s}}^{j_{s}}],$ when the prime metric is
parameterized
\begin{eqnarray*}
ds^{2} &=&\mathring{g}_{i}(x^{k})(dx^{i})^{2}+\mathring{h}%
_{a}(x^{k},y^{3})(dy^{a})^{2}(\mathbf{\mathring{e}}^{a})^{2}+\mathring{g}%
_{a_{1}}(x^{k},y^{3})\left( \mathbf{\mathring{e}}^{a_{1}}\right) ^{2} \\
&&+\mathring{g}_{a_{2}}(x^{k},y^{3},y^{a_{1}},y^{7})\left( \mathbf{\mathring{%
e}}^{a_{2}}\right) ^{2}+\mathring{g}%
_{a_{3}}(x^{k},y^{3},y^{a_{1}},y^{a_{2}},y^{9})\left( \mathbf{\mathring{e}}%
^{a_{3}}\right) ^{2},
\end{eqnarray*}%
where $\mathbf{\mathring{e}}^{a_{s}}$ are N--elongated differentials. Such a
metric contains a 4-d spacetime part describing a Kerr black hole solution
and can be diagonalizable if there is a coordinate transform $u^{\alpha
_{s}^{\prime }}=u^{\alpha _{s}^{\prime }}(u^{\alpha _{s}})$ for which $%
ds^{2}=\mathring{g}_{i^{\prime }}(x^{k\prime })(dx^{i^{\prime }})^{2}+%
\mathring{h}_{a_{s}^{\prime }}(x^{k\prime })(dy^{a_{s}^{\prime }})^{2}$,
with $\ ^{s}\mathring{w}_{i_{s}}=\ ^{s}\mathring{n}_{i_{s}}=0,$ i.e. the
anholonomy coefficients vanish, $\mathring{W}_{\beta _{s}\gamma
_{s}}^{\alpha _{s}}(u^{\mu _{s}})=0,$ see (\ref{anhrelf}). The extra
dimensional coefficients can be arbitrary ones describing a nontrivial
internal space sturcture, or considered for trivial diagonalizable data, for
instance, $\mathring{g}_{a_{s}}=1,$ for $s=1,2,3.$ The target metrics will
be constructed for small  generic off--diagonal parametric
deformations of the Kerr metric embedded in a 10-d prime spacetime into as a
solution of nonholonomic motion equations in heterotic string gravity (\ref%
{hs1}). Conventionally, the N-adapted coefficients are parameterized
\begin{eqnarray}
ds^{2} &=&\eta _{i}\mathring{g}_{i}(dx^{i})^{2}+\eta _{a_{s}}\mathring{g}%
_{a_{s}}(\mathbf{e}^{a_{s}})^{2},  \label{targm} \\
\mathbf{e}^{3} &=&dt+\ ^{n}\eta _{i}\mathring{n}_{i}dx^{i},\mathbf{e}%
^{4}=dy^{4}+\ ^{w}\eta _{i}\mathring{w}_{i}dx^{i},\ \mathbf{e}^{5}=dy^{5}+\
^{n}\eta _{i_{1}}\mathring{n}_{i_{1}}dx^{i_{1}},\mathbf{e}^{6}=dy^{6}+\
^{w}\eta _{i_{1}}\mathring{w}_{i_{1}}dx^{i_{1}},  \notag \\
\mathbf{e}^{7} &=&dy^{7}+\ ^{n}\eta _{i_{2}}\mathring{n}_{i_{2}}dx^{i_{2}},%
\mathbf{e}^{8}=dy^{8}+\ ^{w}\eta _{i_{2}}\mathring{w}_{i_{s}}dx^{i_{s}},\
\mathbf{e}^{9}=dy^{9}+\ ^{n}\eta _{i_{3}}\mathring{n}_{i_{3}}dx^{i_{3}},%
\mathbf{e}^{10}=dy^{10}+\ ^{w}\eta _{i_{3}}\mathring{w}_{i_{3}}dx^{i_{3}},
\notag
\end{eqnarray}%
where the coefficients $(g_{\alpha _{s}}=\eta _{\alpha _{s}}\mathring{g}%
_{\alpha _{s}},^{w}\eta _{i_{s}}\mathring{w}_{i_{s}},\ ^{n}\eta
_{i_{s}}n_{i_{s}})$ are determined by polarization $\eta $--coefficients \
with \thinspace $\varepsilon $--parameter linearization,
\begin{eqnarray}
\eta _{i} &=&1+\varepsilon \chi _{i}(x^{k}),\eta _{a}=1+\varepsilon \chi
_{a}(x^{k},y^{4}),\eta _{a_{1}}=1+\varepsilon \chi
_{a_{1}}(x^{k},y^{a},y^{6}),  \notag \\
\eta _{a_{2}} &=&1+\varepsilon \chi
_{a_{2}}(x^{k},y^{a},y^{a_{1}},y^{8}),\eta _{a_{3}}=1+\varepsilon \chi
_{a_{3}}(x^{k},y^{a},y^{a_{1}},y^{a_{2}},y^{10});  \notag \\
\ ^{n}\eta _{i} &=&1+\varepsilon \ \ ^{n}\chi _{i}(x^{k},y^{4}),\ ^{w}\eta
_{i}=1+\varepsilon \ ^{w}\chi _{i}(x^{k},y^{4}),  \label{polarfunct} \\
\ ^{n}\eta _{i_{1}} &=&1+\varepsilon \ \ ^{n}\chi
_{i_{1}}(x^{k},y^{a},y^{6}),\ ^{w}\eta _{i_{1}}=1+\varepsilon \ ^{w}\chi
_{i_{1}}(x^{k},y^{a},y^{6}),  \notag \\
\ ^{n}\eta _{i_{2}} &=&1+\varepsilon \ \ ^{n}\chi
_{i_{2}}(x^{k},y^{a},y^{a_{1}},y^{8}),\ ^{w}\eta _{i_{2}}=1+\varepsilon \
^{w}\chi _{i_{2}}(x^{k},y^{a},y^{a_{1}},y^{8}),  \notag \\
\ ^{n}\eta _{i_{3}} &=&1+\varepsilon \ \ ^{n}\chi
_{i_{3}}(x^{k},y^{a},y^{a_{1}},y^{a_{2}},y^{10}),\ ^{w}\eta
_{i_{3}}=1+\varepsilon \ ^{w}\chi
_{i_{3}}(x^{k},y^{a},y^{a_{1}},y^{a_{2}},y^{10}).  \notag
\end{eqnarray}%
We note that such nonholonomic \thinspace $\varepsilon $--parameters
deformations embed and nonholonomically transform a 4-d stationary Kerr
metric into a 10-d solutions of the nonholonomic motion equations in
heterotic gravity (\ref{hs1})-(\ref{hs4}) with an effective scalar field
encoded into the N--connection structure and non-stationary internal
spacetimes. Both 4-d and 6-d components, respectively, (\ref{4dqe}) and (\ref%
{6dqe}), of generic off-diagonal metrics (\ref{10dqe}) written in terms of
polarization functions (\ref{targm}), can ve re-defined in almost K\"{a}hler
variables.

\subsection{Solitonic deformations of 4-d Kerr configurations and internal
space extensions}
Off-diagonal terms and nontrivial effective sources in heterotic string
gravity result in various effects with solitonic like distributions in the
4-d spacetime and 6-d internal spaces and various possible nonlinear
solitonic waves on mixed spacetime and internal space coordinates.

\subsubsection{String solitonic polarizations of the mass of 4-d Kerr black
holes}

We can construct off-diagonal analogs of the Kerr-Sen rotoid configurations
\cite{asen} if the generating function $\ ^{sld}\chi =\chi (r,\vartheta
,\varphi )$ \ in (\ref{4dqe}) is a solution of the solitonic equation
\begin{equation}
\partial _{rr}^{2}\chi +\epsilon \partial _{\varphi }(\partial _{\vartheta
}\chi +6\chi \partial _{\varphi }\chi +\partial _{\varphi \varphi \varphi
}^{3}\chi )=0,  \label{solitdistr}
\end{equation}%
for $\epsilon =\pm 1$. In this formula, the left label $sld$ emphasizes that
such a function is defined as a "solitonic distribution". In gravity theories, such Kadomtzev--Petviashvili, KP, type equations were studied in
details \cite{sv2013,vp,vt}, see references therein (with examples of 2-d
and 3-d sign-Gordon equations are studied). The corresponding $tt$%
-coefficient of the d-metric, $h_{3}=\overline{A}-\varepsilon \frac{%
\mathring{\Phi}^{2}\chi }{4\Lambda }=\widetilde{A}(r,\vartheta ,\varphi )$,
can be re-written in a form with an effective solitonic polarization of
mass,
\begin{equation}
m_{0}\rightarrow \widetilde{m}(r,\vartheta ,\varphi )=m_{0}[1-\varepsilon
\chi (r,\vartheta ,\varphi )\frac{\mathring{\Phi}(r,\vartheta ,\varphi )}{%
8\Lambda m_{0}}(r^{2}+a^{2}\cos ^{2}\vartheta )].  \label{mpolar}
\end{equation}%
This formula describes possible modifications of Kerr black holes, BH, for a
nontrivial cosmological constant $\Lambda =\ ^{H}\Lambda +\ ^{F}\Lambda +\
^{R}\Lambda $ induced effectively by "matter field" sources (\ref{effsources}%
). For small values of parameter $\varepsilon ,$ we can consider that
contributions from heterotic string gravity result also into generic
off--diagonal nonholonomic deformations of the vacuum BH solutions to
certain Kerr-type rotoid configurations with nontrivial vacuum modifications
of the gravitational media, which is contained explicitly in the diagonal
coefficients, $h_{a}[\ ^{slt}\chi ],$ and the N-coefficients, computed using
the formula (\ref{acoef}) for $w_{i}[\ ^{slt}\chi ],$ in (\ref{4dqe}). In
explicit form, the function $\widetilde{m}(r,\vartheta ,\varphi )$ can be
computed for certain prescribed values, i.e. boundary conditions, for the
solitons determined as solutions (\ref{solitdistr}) and choosing explicit
frame and coordinate systems with respective $\mathring{\Phi}(r,\vartheta
,\varphi )$ and $\Lambda .$ Such values should be determined by explicit
experimental / observational data in modern astrophysics and cosmology.

\subsubsection{Stationary solitonic distributions of in the internal space}

On the first shell, we can generate such solutions if the generating
function $\ _{1}^{\Lambda }\chi $ is taken in the form $\ _{1}\chi =\
_{1}^{sld}\chi (r,\vartheta ,y^{6})$ and constrained to be a solution of the
3-d solitonic equation
\begin{equation*}
\partial _{rr}^{2}(_{1}\chi )+\epsilon \partial _{6}[\partial _{\vartheta
}(_{1}\chi )+6(_{1}\chi )\partial _{6}(_{1}\chi )+\partial
_{666}^{3}(_{1}\chi )]=0.
\end{equation*}%
In result, we obtain soltionic gravitational configurations with N-adapted coefficients (\ref{6dqe}) being functionals of
 solitonic functions, when $h_{a_{1}}[\ _{1}^{slt}\chi ],$ (where $a_{1}=5,6),$ and nontrivial $N_{i}^{a_{1}}[\ _{1}^{sld}\chi ].$\footnote{%
An alternative class of solitonic distributions can be generated if we
consider $\ _{1}^{\Lambda }\chi $ in the form $\ _{1}\chi =\ _{1}^{slt}\chi
(r,\varphi ,y^{6})$ and constrained to be a solution of the 3-d solitonic
equation $\ \partial _{rr}^{2}(_{1}\chi )+\epsilon \partial _{6}[\partial
_{\varphi }(_{1}\chi )+6(_{1}\chi )\partial _{6}(_{1}\chi )+\partial
_{666}^{3}(_{1}\chi )]=0.$}

In similar forms, we can generate 3-d solitonic configurations, with mixed
dependence on 4-d space coordinates and an extra-dimension coordinate, on
all 3 shells of the internals space of the heterotic supergravity. Such
solutions are determined by $\ _{s}\chi =\ _{s}^{sld}\chi (r,\vartheta
,y^{2s+2}),$ (when dependencies on coordinates $y^{6},y^{8},y^{10}$ are
obtained correspondingly for $s=1,2,3$), as respective solutions of
\begin{equation}
\partial _{rr}^{2}(_{s}\chi )+\epsilon \frac{\partial }{\partial y^{2s+2}}%
[\partial _{\vartheta }(_{s}\chi )+6(_{s}\chi )\frac{\partial }{\partial
y^{2s+2}}(_{s}\chi )+\frac{\partial ^{3}}{(\partial y^{2s+2})^{3}}(_{s}\chi
)]=0.  \label{solitdistrs}
\end{equation}%
Prescribing generating functions (\ref{genfkerrd}), we generate internal
space d-metrics, and noholonomic solitonic off-diagonal deformations of the
Kerr 4-d BH into stationary solutions with 10-d metrics determined as
functionals (\ref{10dqe}), i.e. functionals of type $\mathbf{g=g}$ $%
[Kerr4d,\ _{s}^{sld}\chi (r,\vartheta ,y^{2s+2})].$ In respective N-adapted
frames and coordinate systems, such 10-d solitonic gravitational
configurations do not depend on the time like coordinate $t.$ They can be,
in general, with nontrivial nonholonomically induced torsion coefficients.
Imposing additional nonhlonomic contstraints, we can extract torsionless
Levi-Civita configurations as in \cite{tgovsv,vex3}.

\subsubsection{Solitonic waves in extra-dimensional shells on the internal space}

We can construct solutions of the equations of motion in heterotic supergravity
generated in the internal space by solitonic waves with explicit dependence
on time like coordinate $t.$ For such internal metrics, the generating
function $\ _{s}^{\Lambda }\chi $ can be taken, for instance, in such forms%
{\small
\begin{equation}
\ _{s}\chi =\left\{
\begin{array}{ccc}
\ _{s}^{sw}\chi (t,\vartheta ,y^{2s+2}) & \mbox{ as a solution of } &
\partial _{tt}^{2}(_{s}\chi )+\epsilon \frac{\partial }{\partial y^{2s+2}}%
[\partial _{\vartheta }(_{s}\chi )+6(_{s}\chi )\frac{\partial }{\partial
y^{2s+2}}(_{s}\chi )+\frac{\partial ^{3}}{(\partial y^{2s+2})^{3}}(_{s}\chi
)]=0; \\
\ _{s}^{sw}\chi (\vartheta ,t,y^{2s+2}) & \mbox{ as a solution of } &
\partial _{\vartheta \vartheta }^{2}(_{s}\chi )+\epsilon \frac{\partial }{%
\partial y^{2s+2}}[\partial _{t}(_{s}\chi )+6(_{s}\chi )\frac{\partial }{%
\partial y^{2s+2}}(_{s}\chi )+\frac{\partial ^{3}}{(\partial y^{2s+2})^{3}}%
(_{s}\chi )]=0; \\
\ _{s}^{sw}\chi (t,r,y^{2s+2}) & \mbox{ as a solution of } & \partial
_{tt}^{2}(_{s}\chi )+\epsilon \frac{\partial }{\partial y^{2s+2}}[\partial
_{r}(_{s}\chi )+6(_{s}\chi )\frac{\partial }{\partial y^{2s+2}}(_{s}\chi )+%
\frac{\partial ^{3}}{(\partial y^{2s+2})^{3}}(_{s}\chi )]=0; \\
\ _{s}^{sw}\chi (r,t,y^{2s+2}) & \mbox{ as a solution of } & \partial
_{rr}^{2}(_{s}\chi )+\epsilon \frac{\partial }{\partial y^{2s+2}}[\partial
_{t}(_{s}\chi )+6(_{s}\chi )\frac{\partial }{\partial y^{2s+2}}(_{s}\chi )+%
\frac{\partial ^{3}}{(\partial y^{2s+2})^{3}}(_{s}\chi )]=0; \\
\ _{s}^{sw}\chi (t,\varphi ,y^{2s+2}) & \mbox{ as a solution of } & \partial
_{tt}^{2}(_{s}\chi )+\epsilon \frac{\partial }{\partial y^{2s+2}}[\partial
_{\varphi }(_{s}\chi )+6(_{s}\chi )\frac{\partial }{\partial y^{2s+2}}%
(_{s}\chi )+\frac{\partial ^{3}}{(\partial y^{2s+2})^{3}}(_{s}\chi )]=0; \\
\ _{s}^{sw}\chi (\varphi ,t,y^{2s+2}) & \mbox{ as a solution of } & \partial
_{\varphi \varphi }^{2}(_{s}\chi )+\epsilon \frac{\partial }{\partial
y^{2s+2}}[\partial _{t}(_{s}\chi )+6(_{s}\chi )\frac{\partial }{\partial
y^{2s+2}}(_{s}\chi )+\frac{\partial ^{3}}{(\partial y^{2s+2})^{3}}(_{s}\chi
)]=0.%
\end{array}%
\right.  \label{swaves}
\end{equation}%
} Introducing certain values $\ _{s}\chi $ (\ref{swaves}) instead of
respective $\ _{s}^{\Lambda }\chi $ in (\ref{6dqe}), we generate solutions
of type (\ref{10dqe}) as functionals $\mathbf{g=g}$ $[Kerr4d,\ _{s}^{sw}\chi
(.,.,y^{2s+2})].$ Such metrics in 10-d gravity define nonstationary generic
off-diagonal deformations of Kerr BHs with nonlinear soliton interactions in
the internal space.

\subsection{Quasiperiodic configurations in 10-d heterotic supergravity}

Recently, quasi-crystal like cosmological structures defined as exact
solutions in modified and Einstein gravity, and/or with loop quantum
corrections have been studied in Refs. \cite{aschheim16,amaral16}. One of
the main goals of this paper is to show that such quasiperiodic models can
be elaborated for stationary and nonstationary off--diagonal deformations of
BH solutions in string and modified gravity theories \cite%
{partner,tgovsv,vex3,vex1}. In this section, we provide examples of BH
deformations by quasi-crystal structures with possible three-wave
interactions and extra dimensional temporal chaos in internal space of
heterotic gravity. The analogs of solutions of PDEs considered for
quasiperiodic structures in condensed matter physics \cite%
{rucklidge16,rucklidge12} are derived for the system of motion equations (%
\ref{hs1}) - (\ref{hs4}) in heterotic string gravity. We shall express the
N-adapted coefficients in terms of $\eta $-functions as in (\ref{targm})
which for $\varepsilon $--decompostions can be defined in terms of $\chi $%
-functions using (\ref{polarfunct}).

\subsubsection{Quasi-crystal structures in the 4-d spacetime and higher
dimension internal spaces}

For the internal spaces, we consider analogous of 3-d phase field crystal models,  PFC,
which  were elaborated for soft matter systems generated by modulations with two length scales  \cite{rucklidge16}.   The polarization
generating functions are parameterized in a form $\eta
_{a_{s}}(r,t,y^{a_{s}}),$ or $\eta _{a_{s}}(\vartheta ,t,y^{a_{s}}),$ or $%
\eta _{a_{s}}(t,\varphi ,y^{a_{s}}),$ emphasizing different examples of
possible anisotropies on radial, angular, time like and extra dimension
coordinates. In explicit form, the constructions in this subsection will be
performed by prescribing generating functions \ $_{s}\eta =\{\eta
_{a_{s}}(t,\varphi ,y^{2s+2})\},$ for $s=1,2,3$. Such values specify
respective shells  certain nonholonomic deformation of the metric and
N-connection coefficients in a point $(\varphi ,y^{2s+2})$ at time $t.$ This
model possesses conserved dynamics described by respective nonlinear PDEs
describing the evolution of nonholonomic deformations over diffusive time
scales. In particular, we shall use a toy construction when our PFC includes
all the resonant interactions for a correlated 4-d and internal space
deformations of certain BH prime metrics, for simplicity, approximated by
models with icosahedral symmetry. Such gravitational off-diagonal models
with nontrivial vacuum extends previous works for soft matter systems \cite%
{rucklidge16} to BH configurations with quasipersiodic internal space
structure in certain sence completing the results on quasiperiodic
cosmological models \cite{aschheim16,amaral16} with local anisotropy. This
allows an independent control over the growth rates of gravitational string
waves with two wavelengths. As a result, one proves that just as for 2-d
quasi-crystals, the resonant interactions between the two wavelengths do
stabilize 3-d quasiperiodic heterotic string structures.

The analogous PFC models are determined by shell labeled effective free
energies,%
\begin{equation}
\mathcal{F}[\ _{s}\eta ]=\int \sqrt{|\ _{s}\mathbf{g}|}dt\delta \varphi
\delta y^{2s+2}\left[ -\frac{1}{2}(\ _{s}\eta )\mathcal{L}(\ _{s}\eta )-%
\frac{Q}{3}(\ _{s}\eta )^{3}+\frac{1}{4}(\ _{s}\eta )^{4}\right] ,
\label{qcryst}
\end{equation}%
where the operator \ $\mathcal{L}$ and parameter $Q$ can be chosen in
certain forms which stabilize the $t$-evolution to be compatible with
possible experimental data for soft matter, or cosmology, see \cite%
{rucklidge16,aschheim16,amaral16}. Such values in heterotic string gravity
can be chosen for certain limits of off-diagonal deformations of 4-d Kerr
holes in order to be compatible with astrophysical/ cosmological data. Above
functionals result into a conserved dynamics with respective evolution
equations on shells of internal space,%
\begin{equation}
\frac{\partial (\ _{s}\eta )}{\partial t}=(\ _{\mid }^{s}\widehat{\mathbf{D}}%
)^{2}\left[ \mathcal{L}(\ _{s}\eta )+Q(\ _{s}\eta )^{2}-(\ _{s}\eta )^{3}%
\right]  \label{qcevol}
\end{equation}%
where $\ _{\mid }^{s}\widehat{\mathbf{D}}$ $\ $\ is the canonical
d-connection restricted on respective $s$-shell with spacelike signature $%
(++..+)$ and the system is stabilized nonlinearly by the cubic term. The
resonant term is set by the value $Q.$

The PDEs (\ref{qcevol}) \ describe a quasicrystal like structure with stable
evolution of BHs into extra-dimensional 10-d spacetime, which is  determined by exact
solutions of the motion equations in heterotic supergravity. Such
quasiperiodic structures are characterized by an effective free energy
functional (\ref{qcryst}) \ which can be related to geometric flows and
Perelman's functionals as in Refs. \cite{vnrflnc,vmedit}.

Introducing the values $\ _{s}\eta $ into (\ref{polarfunct}) and (\ref{targm})
we construct 10-d exact solutions of type (\ref{10dqe}) describing 4-d
Kerr BHs with extra dimensional evolving locally anisotropic hear and/or
quasiperiodic like structure. Such solutions can be characterized both by
Hawking-Bekenstein entropy values for BHs and by Perelman's functionals for
entropy.

\subsubsection{Three-wave interactions and extra dimensional temporal chaos}

We can consider another class of evolution PDEs generalizing (\ref{qcevol})
to
\begin{equation}
\frac{\partial (\ _{s}\eta )}{\partial t}=\mathcal{L}(\ _{s}\eta )+Q_{1}(\
_{s}\eta )^{2}+Q_{2}(\ _{s}\eta )(\quad _{\mid }^{s}\widehat{\mathbf{D}}%
)^{2}(\ _{s}\eta )+Q_{3}|\quad _{\mid }^{s}\widehat{\mathbf{D}}(\ _{s}\eta
)|^{2}-(\ _{s}\eta )^{3}.  \label{3waveint}
\end{equation}%
Such models are studied in \cite{rucklidge12} for respective values of the
operator $\mathcal{L}$ and constants $Q_{1},Q_{2},Q_{3}.$ In heterotic
string gravity, this evolution equation describes nonholonomic deformations
of the BHs by three-wave interactions involving internal space coordinates.
Such many pattern-forming gravitational systems encapsulate the basic
properties of nonlinear interactions of 4-d Kerr parameters and internal
space nonlinear interactions. We can consider configurations with two
comparable length scales when two gravitational waves of the shorter
wavelength nonlinearly interact with one wave of the longer. For a different
set of parameters, two waves of the longer wavelength to interact with one
wave of the shorter extra-dimensional gravitational waves. Such nonlinear
three-wave interactions with respective parameters may result in the
presence of complex patterns and/or spatiotemporal chaos, see similar
constructions in \cite{vapexsol}.

\subsection{Solitonic and/or quasiperiodic YM and almost-K\"{a}hler sectors}

All solutions in heterotic supergravity theories defining BH, solitonic and
quasi-periodic configurations, constructed in this section and partner works
\cite{partner,tgovsv,vex3,vex1}, can be reformulated in nonholonomic
variables distinguishing in explicit form almost-K\"{a}hler internal
configurations and the YM sector. There are two approaches: In the first
case, we can take any solution (\ref{10dqe}) defining nontrivial
quasiperiodic deformations of BH solutions and perform respective frame
transoforms which allows us to distinguish respective almost-K\"{a}hler and/or
YM variables. In the second case, we have to introduce some effective
Lagrange functions in the internal spaces which generate respective
quasiperiodic almost symplectic geometric structures and YM fields.

\subsubsection{Qusiperiodic canonical connections, metrics, and
almost symplectic configurations}

Let us consider any d-metric (\ref{10dqe}) for a Kerr deformed BH of type $%
\mathbf{g=g}[Kerr4d,\ _{s}^{sld}\chi (r,\vartheta ,y^{2s+2})],$ with $\ \
_{s}\chi =\ _{s}^{sld}\chi (r,\vartheta ,y^{2s+2})$ and being a solution of (%
\ref{solitdistrs}); or $\mathbf{g=g}[Kerr4d,\ _{s}^{sw}\chi ]$ , with any $\
_{s}^{sw}\chi $ (\ref{swaves}) subjected to respective evolution equations; $%
\mathbf{g=g}[Kerr4d,\ \eta _{a_{s}}]$ with $\eta _{a_{s}}(r,t,y^{a_{s}}),$
or $\eta _{a_{s}}(\vartheta ,t,y^{a_{s}}),$ or $\eta _{a_{s}}(t,\varphi
,y^{a_{s}}),$ emphasizing different examples of possible anisotropies on
radial, angular, time like and extra dimension coordinates, and when $%
_{s}\eta =\{\eta _{a_{s}}(t,\varphi ,y^{2s+2})\},$ for $s=1,2,3,$ are
solutions of evolution equations of type (\ref{qcevol}), or (\ref{3waveint}%
). The corresponding component of the quadratic element for the internal
space is determined by $\ ^{6}\overrightarrow{\mathbf{g}}\mathbf{=}\ ^{6}%
\mathbf{g}[\overrightarrow{\eta }]$ (\ref{6dqe}) on $\ ^{6}\mathbf{X}$ for
any set of quasiperiodic generating functions
\begin{equation}
\quad \overrightarrow{\eta }=\left\{
\begin{array}{cccc}
\ _{s}^{sld}\eta & = & 1+\varepsilon \ _{s}^{sld}\chi ; &  \\
\ _{s}^{sw}\eta & = & 1+\varepsilon \ _{s}^{sw}\chi ; &  \\
_{s}\eta & = & \{\eta _{a_{s}}(t,\varphi ,y^{2s+2})\}, & \mbox{ for }s=1,2,3.%
\end{array}%
\right.  \label{qpgenf}
\end{equation}%
In the following section, the symbols with "up arrow" will be used in order to
emphasize that polarization functions of type (\ref{polarfunct}) are
determined  for a set of data with quasiperiodic structure,  when $\
^{6}\overrightarrow{\mathbf{g}}\mathbf{\subset }\overrightarrow{\mathbf{g}}$
(computed as in (\ref{10dqe}) for a respective $\overrightarrow{\eta })$ is
a solution of the motion equations in heterotic supergravity. One parameterizes
\begin{eqnarray}
ds^{2}[\ ^{6}\overrightarrow{\mathbf{g}}] &=&\overrightarrow{g}_{\check{b}%
\check{c}}(x^{i},y^{a},y^{\check{a}})dy^{\check{b}}dy^{\check{c}}%
\mbox{ in
off-diagonal coordiante form and/or (in N-adapted form)  }  \label{qpqel} \\
ds^{2}[\ ^{6}\overrightarrow{\mathbf{g}}] &=&\overrightarrow{g}_{a_{1}}(\
^{0}u,y^{a_{1}})\left( \mathbf{e}^{a_{1}}\right) ^{2}+\quad \overrightarrow{g%
}_{a_{2}}(\ ^{1}u,y^{a_{1}})\left( \mathbf{e}^{a_{2}}\right) ^{2}+\quad
\overrightarrow{g}_{a_{3}}(\ ^{2}u,y^{a_{2}})\left( \mathbf{e}%
^{a_{3}}\right) ^{2}.  \notag
\end{eqnarray}

On $\ ^{6}\mathbf{X,}$ we can define a form $3+3$ splitting determined by
N--adapted frame transforms with vierbein coefficients $e_{\ \underline{%
\check{a}}}^{\check{a}}$ of the dual basis $\ e^{\check{a}}=(e^{\acute{\imath%
}},e^{\grave{a}})=e_{\ \underline{\check{a}}}^{\check{a}}(u)dy^{\underline{%
\check{a}}},$ when for any data determining a quasiperiodic structure, the
d-metric (\ref{qpqel}) is expressed in the form (\ref{gpsmf}) with
coefficients being functionals of a set of quasiperiodic generating
functions of type (\ref{qpgenf})
\begin{eqnarray}
\ ^{6}\mathbf{g} &=&g_{\acute{\imath}\acute{j}}[\overrightarrow{\eta }(\
^{0}u,y^{\acute{\imath}},y^{\grave{a}})]dy^{\acute{\imath}}\otimes dy^{%
\acute{j}}+g_{\grave{a}\grave{b}}[\overrightarrow{\eta }(\ ^{0}u,y^{\acute{%
\imath}},y^{\grave{a}})]e^{\grave{a}}\otimes e^{\grave{b}},  \label{qpqela}
\\
\mathbf{e}^{\grave{a}} &=&dy^{\grave{a}}-N_{\acute{\imath}}^{\grave{a}}[%
\overrightarrow{\eta }(\ ^{0}u,y^{\acute{\imath}},y^{\grave{a}})]dy^{\acute{%
\imath}},  \notag
\end{eqnarray}%
with quasiperiodic N--connection structure, $\ ^{6}\mathbf{N}[%
\overrightarrow{\eta }]=\{N_{\acute{\imath}}^{\grave{a}}[%
\overrightarrow{\eta }]\}$ determining a conventional splitting of
coordinates $y^{\acute{\imath}}$ (for $\acute{\imath}=5,6,7)$ and $y^{\grave{a%
}}$ (for $\grave{a}=8,9,10),$ when $\ ^{0}u=(x^{i},y^{a}).$ Prescribing any
generating function $L[\overrightarrow{\eta }]=\overrightarrow{L}%
(x^{i},y^{a},y^{\acute{\imath}},y^{\grave{a}}),$ we can define canonical
effective Lagrange variables on $\ ^{6}\mathbf{X.}$ We can relate $%
\overrightarrow{L},$ for instance, to any effective Lagrange of soliton
type, or free energy (\ref{qcryst}) for quasiperiodic structures. $\ $Using
formulas (\ref{elmf}), (\ref{clncfa}) and (\ref{lfsmf}) with $L\rightarrow
\overrightarrow{L},$ we enable the internal space with a \ 6-d canonical
d-metric ("tilde" on symbols are changed into "$\rightarrow $" in order to
state that such physical and geometric objects are induced by quasiperiodic
structures),
\begin{eqnarray}
\quad ^{6}\overrightarrow{\mathbf{g}} &=&\overrightarrow{g}_{\acute{\imath}%
\acute{j}}dy^{\acute{\imath}}\otimes dy^{\acute{j}}+\ \overrightarrow{g}_{%
\grave{a}\grave{b}}\ \overrightarrow{\mathbf{e}}^{\grave{a}}\otimes \
\overrightarrow{\mathbf{e}}^{\grave{b}},  \label{qplfsmf} \\
\ \overrightarrow{\mathbf{e}}^{\grave{a}} &=&dy^{\grave{a}}+\overrightarrow{N%
}_{\acute{\imath}}^{\grave{a}}dy^{\acute{\imath}},\ \{\ \overrightarrow{g}_{%
\grave{a}\grave{b}}\}=\ \{\overrightarrow{g}_{7+\acute{\imath}\ 7+\acute{j}%
}\}.  \notag
\end{eqnarray}%
Any $\ ^{6}\mathbf{X}$ can be enabled with canonical d-metric and
N-connection quasiperiodic structures $[\quad ^{6}\overrightarrow{\mathbf{g}}%
\mathbf{,}\overrightarrow{\mathbf{N}}]$ and denoted $\ ^{6}\overrightarrow{%
\mathbf{X}}.$

We note that any d-metric $^{6}\mathbf{g}$ (\ref{6dqe}) generated by
quasiperiodic generating functions $\overrightarrow{\eta }$ in any
equivalent form (\ref{qpqel}), or (\ref{qpqela}), for instance, by data $\
^{6}\mathbf{g}_{\check{a}\check{b}}[\overrightarrow{\eta }]=[g_{\acute{\imath%
}\acute{j}}[\overrightarrow{\eta }],g_{\grave{a}\grave{b}}[\overrightarrow{%
\eta }],N_{\acute{\imath}}^{\grave{a}}[\overrightarrow{\eta }]]$ computed
with respect to a N-adapted basis $\mathbf{e}^{\check{a}}[\overrightarrow{%
\eta }]=(e^{\acute{\imath}}=dy^{\acute{\imath}},\mathbf{e}^{\grave{a}}[%
\overrightarrow{\eta }])$ can be transformed into a d-metric $\ ^{6}%
\overrightarrow{\mathbf{g}}_{\check{a}\check{b}}=\ [\overrightarrow{g}_{%
\acute{\imath}\acute{j}},\overrightarrow{g}_{\grave{a}\grave{b}},%
\overrightarrow{N}_{\acute{\imath}}^{\grave{a}}]$ (\ref{qplfsmf}) with
coefficients defined with respect to a canonical N--adapted dual basis $\ \
\overrightarrow{\mathbf{e}}^{\check{a}}=(e^{\acute{\imath}}=dy^{\acute{\imath%
}},\overrightarrow{\mathbf{e}}^{\grave{a}}).$ Such nonholonomic frame
transforms are subjected to the conditions $\ \ ^{6}\mathbf{g}_{\check{a}%
^{\prime }\check{b}^{\prime }}[\overrightarrow{\eta }]e_{\ \check{a}}^{%
\check{a}^{\prime }}e_{\ \check{b}}^{\check{b}^{\prime }}=\ ^{6}%
\overrightarrow{\mathbf{g}}_{\check{a}\check{b}}.$ Fixing any quasiperiodic
data $\ ^{6}\mathbf{g}_{\check{a}^{\prime }\check{b}^{\prime }}[%
\overrightarrow{\eta }]$ and $\ ^{6}\overrightarrow{\mathbf{g}}_{\check{a}%
\check{b}},$ we can find $e_{\ \check{a}}^{\check{a}^{\prime }}$ as
solutions of corresponding system of quadratic algebraic equations.

A quasi-periodic canonical N--connection structure $\overrightarrow{\mathbf{N%
}}$ defines an almost complex structure as a linear operator $\overrightarrow{%
\mathbf{J}}$ acting on vectors on $\ ^{6}\overrightarrow{\mathbf{X}}$
following formulas $\overrightarrow{\mathbf{J}}(\ \overrightarrow{\mathbf{e}}%
_{\acute{\imath}})=-e_{7+\acute{\imath}}$ and $\overrightarrow{\mathbf{J}}%
(e_{7+\acute{\imath}})=\ \overrightarrow{\mathbf{e}}_{\acute{\imath}},$
where $\overrightarrow{\mathbf{J}}\mathbf{\circ \overrightarrow{\mathbf{J}}=-%
}\mathbb{I},$ for $\mathbb{I}$ being the unity matrix, and construct a
tensor field following formulas (\ref{acstrf}) with $\widetilde{N}_{\acute{%
\imath}}^{\grave{a}}\rightarrow \overrightarrow{N}_{\acute{\imath}}^{\grave{a%
}}.$ The operator $\overrightarrow{\mathbf{J}}$ and the canonical d-metric $%
\mathbf{g}=\ \overrightarrow{\mathbf{g}}$ (\ref{qplfsmf}) allow to define $\
\ \overrightarrow{\theta }\mathbf{(X,Y):}=\left( \overrightarrow{\mathbf{J}}%
\mathbf{X,Y}\right) ,$ for any d--vectors $\mathbf{X,Y\in }T\ \ ^{6}%
\overrightarrow{\mathbf{X}}\mathbf{.}$ Considering a quasiperiodic form $\
\ \overrightarrow{\omega }=\frac{1}{2}\frac{\partial \overrightarrow{L}}{%
\partial y^{7+\acute{\imath}}}dy^{\acute{\imath}}$ (see formulas (\ref%
{asymstrf}) and (\ref{canalmsf}), redefined for quasiperiodic
configurations), we conclude that any data $\left( \mathbf{g}=\overrightarrow{\mathbf{g}},\ \overrightarrow{%
\mathbf{N}}\mathbf{,}\overrightarrow{\mathbf{J}}\right) $ induces an almost-K%
\"{a}hler geometry characterized by a 2--form of type (\ref{asymstrf}),
\begin{equation}
\theta =\overrightarrow{\theta }=\frac{1}{2}\overrightarrow{\theta }_{\acute{%
\imath}\acute{j}}(u)e^{\acute{\imath}}\wedge e^{\acute{j}}+\frac{1}{2}\
\overrightarrow{\theta }_{\grave{a}\grave{b}}(u)\ \overrightarrow{\mathbf{e}}%
^{\grave{a}}\wedge \ \overrightarrow{\mathbf{e}}^{\grave{b}}=\overrightarrow{%
g}_{\acute{\imath}\acute{j}}(u)\left[ dy^{\acute{\imath}}+\overrightarrow{N}%
_{\acute{k}}^{7+\acute{\imath}}(u)dy^{\acute{k}}\right] \wedge dy^{\acute{j}%
}.  \label{qptwof}
\end{equation}

Using $\quad ^{6}\overrightarrow{\mathbf{g}}$ (\ref{qplfsmf}) \ and $%
\overrightarrow{\theta }$ (\ref{qptwof}), we can construct a quasiperiodic
normal d--connection $\overrightarrow{\mathbf{D}}$ of type (\ref{ndcf}).
This induces a respective canonical almost symplectic d--connection, $%
\overrightarrow{\mathbf{D}}\equiv \ _{\theta }\overrightarrow{\mathbf{D}},$
satisfying the compatibility conditions
\begin{equation*}
\overrightarrow{\mathbf{D}}_{\mathbf{X}}\mathbf{\ (}\quad ^{6}\mathbf{%
\overrightarrow{\mathbf{g}})=\ }_{\theta }\overrightarrow{\mathbf{D}}_{%
\mathbf{X}}\ \overrightarrow{\theta }\mathbf{=}0,
\end{equation*}%
for any $\mathbf{X\in }T\ ^{6}\overrightarrow{\mathbf{X}}.$

Any solution with quasiperiodic structure $\overrightarrow{\eta }$ (\ref%
{qpgenf}) for the internal space in heterotic supergravity can be
parameterized in nonhlonomic varialbes determined by a regular generating
effective function $\overrightarrow{L}.$ Such a geometric model can be
reformulated equivalently as an almost--K\"{a}hler manifold $\ ^{6}%
\overrightarrow{\mathbf{X}}$ which allows us to work equivalently with such
quasiperiodic geometric data on any $\ ^{6}\overrightarrow{\mathbf{X}},$
\begin{equation*}
(\ ^{6}\mathbf{g}[\overrightarrow{\eta }]\mathbf{,}\ ^{6}\mathbf{N}[%
\overrightarrow{\eta }]\mathbf{,}\ ^{6}\widehat{\mathbf{D}}[\overrightarrow{%
\eta }])\iff (\overrightarrow{\mathbf{g}},\ \overrightarrow{\mathbf{%
N}}\mathbf{,}\overrightarrow{\mathbf{D}},\overrightarrow{L})\iff (\ \
\widetilde{\theta }\mathbf{,}\ \widetilde{\mathbf{J}}\mathbf{,}\ _{\theta }%
\widetilde{\mathbf{D}}).
\end{equation*}
Such constructions generalize for quasiperiodic almost K\"{a}hler
structures the results of \cite{lecht1,harl}. We can prove this by  introducing the complex imaginary unit $i^{2}=-1,$ with $\ \overrightarrow{%
\mathbf{J}}\thickapprox i...$ and considering integrable nonholonomic distributions
which state respective complex geometries and models of K\"{a}hler and
complex manifolds.

\subsubsection{Quasiperiodic N-adapted $G_{2}$ and almost-K\"{a}hler
structures in internal spaces}

Let us consider a 3-form $\Theta $ defining an $SU(3)$ structure on the
internal space $\ ^{6}\mathbf{X}$ and related on the tangent space to the
domain wall with a decomposition  $\Theta =\ ^{\mathbf{+}}\Theta +i\ ^{%
\mathbf{-}}\Theta $ as in formula (\ref{alkdec}). Substituting the complex
unity by  $\overrightarrow{\mathbf{J}}$ on $\ ^{6}\overrightarrow{\mathbf{X}}
$ endowed with quasiperiodic structure, we can express
\begin{equation*}
\overrightarrow{\Theta }=\Theta _{\check{a}\check{b}\check{c}}%
\overrightarrow{\mathbf{e}}^{\check{a}}\wedge \overrightarrow{\mathbf{e}}^{%
\check{b}}\wedge \overrightarrow{\mathbf{e}}^{\check{c}}=\ ^{\mathbf{+}%
}\Theta +\overrightarrow{\mathbf{J}}\ ^{\mathbf{-}}\Theta .
\end{equation*}%
The contributions of quasiperiodic structures can be included also via
Clifford structures with  gamma matrices $\overrightarrow{\gamma }_{\check{a}%
}$ on $\ ^{6}\overrightarrow{\mathbf{X}}.$ Such geometric objects are
determined by the relation $\overrightarrow{\gamma }_{\check{a}}%
\overrightarrow{\gamma }_{\check{b}}+\overrightarrow{\gamma }_{\check{b}}%
\overrightarrow{\gamma }_{\check{a}}=2$ $\ ^{6}\overrightarrow{\mathbf{g}}_{%
\check{a}\check{b}},$ see (\ref{qplfsmf})  and  appendix \ref{assclifford}.

Considering  $\overrightarrow{\mathbf{J}}$ instead of $J,$ taken for a
integrable $SU(3)$ structures and K\"{a}hler \ internal spaces,  we can
construct  a similar 3-form $\overrightarrow{\Theta }$ for any
quasiperiodic almost-K\"{a}hler model $(\overrightarrow{\theta }\mathbf{,}\
\overrightarrow{\mathbf{J}}\mathbf{,}\ _{\theta }\overrightarrow{\mathbf{D}})
$ and construct the Hodge star operator $\overrightarrow{\ast }$
corresponding to $\ ^{6}\overrightarrow{\mathbf{g}}.$ Here we note the
relation between 6-d  $\overrightarrow{\ast }$ and 7-d $\ ^{7}$ $%
\overrightarrow{\ast }$ Hodge stars which can be constructed similarly to
ansatz (\ref{7dans}), is
\begin{equation*}
\ ^{7}\overrightarrow{\ast }(\ _{p}^{6}\omega )=e^{B(y^{\check{a}})}\
\overrightarrow{\ast }(\ _{p}^{6}\omega )\wedge \mathbf{e}^{4}\mbox{ and }\
^{7}\overrightarrow{\ast }(\mathbf{e}^{4}\wedge \ _{p}^{6}\omega )=e^{-B(y^{%
\check{a}})}\ \overrightarrow{\ast }(\ _{p}^{6}\omega ).
\end{equation*}
The two exterior derivatives $\ ^{7}d$ and $\overrightarrow{d}$ are related
via $\ \ ^{7}d(\ _{p}^{6}\omega )=\overrightarrow{d}(\ _{p}^{6}\omega
)+dy^{4}\wedge \frac{\partial }{\partial y^{4}}(\ _{p}^{6}\omega ),$ where $%
\ _{p}^{6}\omega $ is a p--form with legs only in the directions on $\ ^{6}%
\overrightarrow{\mathbf{X}}\mathbf{.}$ On quasiperiodic internals spaces,
these formulas allow us to decompose the 10-d 3-form $\widehat{\mathbf{H}}$
into three N-adapted parts,
\begin{equation*}
\widehat{\mathbf{H}}[\overrightarrow{\eta }]=\ell vol[\ ^{3}\mathbf{g]+}\
^{6}\widehat{\mathbf{H}}[\overrightarrow{\eta }]+dy^{4}\wedge \widehat{%
\mathbf{H}}_{4}[\overrightarrow{\eta }],
\end{equation*}%
$\ $where \ $\ vol[\ ^{3}\mathbf{g]=}\frac{1}{3}\epsilon _{\check{i}\check{j}%
\check{k}}\sqrt{|^{3}\mathbf{g}_{\check{i}\check{j}}|}\mathbf{e}^{\check{i}%
}\wedge \mathbf{e}^{\check{j}}\wedge \mathbf{e}^{\check{k}}$ and the set of
polarization functions determining exact solutions in heterotic supergravity
are encoded into
\begin{equation*}
\ ^{6}\widehat{\mathbf{H}}[\overrightarrow{\eta }]=\frac{1}{3!}\widehat{%
\mathbf{H}}_{\check{a}\check{b}\check{c}}\overrightarrow{\mathbf{e}}^{\check{%
a}}[\overrightarrow{\eta }]\wedge \overrightarrow{\mathbf{e}}^{\check{b}}[%
\overrightarrow{\eta }]\wedge \overrightarrow{\mathbf{e}}^{\check{c}}[%
\overrightarrow{\eta }]\mbox{ and }\ \widehat{\mathbf{H}}_{4}=\frac{1}{2!}%
\widehat{\mathbf{H}}_{4\check{b}\check{c}}\overrightarrow{\mathbf{e}}^{%
\check{b}}[\overrightarrow{\eta }]\wedge \overrightarrow{\mathbf{e}}^{\check{%
c}}[\overrightarrow{\eta }].
\end{equation*}

The operators $(\overrightarrow{\mathbf{J}}\mathbf{,}\overrightarrow{\Theta }%
)$ provide examples of  quasiperiodic structures derived as almost-K\"{a}hler
geometries in section \ref{ss67ak}.  For trivial holonomic smooth
configurations with $\overrightarrow{\mathbf{J}}\thickapprox i$ and $G_{2}$
structure, we can reproduce the results for \ K\"{a}hler internal spaces
provided in \cite{harl,lecht1,gray,lukas,chiossi}. In this section,
quasiperiodic nonholonomic configurations are adapted to the data for (\ref%
{bps2}), $(\varpi ,\mathcal{W})$ which can be related to the $SU(3)$ almost-K%
\"{a}hler structure by expressions%
\begin{equation*}
\varpi =e^{B(y^{\check{a}})}\mathbf{e}^{4}\wedge \overrightarrow{\mathbf{J}}%
\mathbf{+}\overrightarrow{\Theta }_{-}\mbox{ and }\mathcal{W}=e^{B(y^{\check{%
a}})}\mathbf{e}^{4}\wedge \overrightarrow{\Theta }_{+}+\frac{1}{2}%
\overrightarrow{\mathbf{J}}\mathbf{\wedge \overrightarrow{\mathbf{J}}.}
\end{equation*}%
Modifications by quasiperiodic structures are computed following the
procedure of changing geometric objects with "tilde" into geometric objects
with "arrow" but with the same Lie algebra $\mathcal{T}$--classification.
This classification can be N--adapted if we use derivatives of the structure
forms
\begin{equation*}
\overrightarrow{d}\overrightarrow{\mathbf{J}}=\mathbf{-}\frac{3}{2}{Im}(%
\mathcal{T}_{1}\overline{\overrightarrow{\Theta }})+\mathcal{T}_{4}\wedge
\overrightarrow{\mathbf{J}}\mathbf{+}\mathcal{T}_{3}\mbox{ and }%
\overrightarrow{d}\overrightarrow{\Theta }=\mathcal{T}_{1}\overrightarrow{%
\mathbf{J}}\wedge \overrightarrow{\mathbf{J}}+\mathcal{T}_{2}\wedge
\overrightarrow{\mathbf{J}}+\overline{\mathcal{T}}_{5}\wedge \overrightarrow{%
\Theta }.
\end{equation*}%
The vertical part for real quasiperiodic almost-K\"{a}hler structures, i.e.
${Im}(\mathcal{T}_{1}\overline{\overrightarrow{\Theta }})$, is computed as
the impaginary part of respective forms if  $(\overrightarrow{\mathbf{J}}%
\mathbf{,}\overrightarrow{\Theta })\rightarrow (J,\underline{\Theta }).$

\subsubsection{Quasiperiodic  instantons nearly almost K\"{a}hler manifolds}

We extend on  almost-K\"{a}hler quasiperiodic internal spaces the instanton
type connections \cite{harl,lecht1,gray,lukas,chiossi} for canonical
N--adapted frames, when  $\ \mathcal{T}_{2}=\mathcal{T}_{3}=\mathcal{T}_{4}=%
\mathcal{T}_{5}=0$ and $\mathcal{T}_{1}=\ ^{\mathbf{+}}\mathcal{T}_{1}+%
\overrightarrow{\mathbf{J}}\ ^{\mathbf{-}}\mathcal{T}_{1}.$ Let us consider
geometric data $(\overrightarrow{\mathbf{J}}\mathbf{,}\overrightarrow{\Theta
})$ and $(\overrightarrow{\theta }\mathbf{,}\ \overrightarrow{\mathbf{J}}%
\mathbf{,}\ _{\theta }\overrightarrow{\mathbf{D}})$ for  an ansatz (\ref%
{7dans}) generated by quasi-periodic structures with  embedding into a 10-d
ones of type (\ref{10dqe}), with $A=B=0$ (for simplicity). \ This way, we
construct solutions on $\ ^{6}\overrightarrow{\mathbf{X}}$ of the first two
nonholonomic BPS equations in (\ref{bpseq}) and (\ref{bps2}) defining
quasiperiodic almost-K\"{a}hler configurations for
\begin{equation*}
\widehat{\mathbf{H}}=\ell vol[\ ^{3}\mathbf{g]-}\frac{1}{2}\partial _{4}\phi
+\left( \frac{3}{2}\ ^{\mathbf{-}}\mathcal{T}_{1}+\frac{7}{8}\ell \right)
+dy^{4}\wedge (2\ ^{\mathbf{-}}\mathcal{T}_{1}+\ell )\overrightarrow{\mathbf{%
J}}\mbox{ for }\widehat{\phi }=\phi (y^{4}).
\end{equation*}%
The data  $(\overrightarrow{\mathbf{J}}\mathbf{,}\overrightarrow{\Theta })$
are subjected to  respective quasiperiodic flow and structure equations
(encoding (\ref{qcryst})  and possible nonholonomic deformations with
generalized evolution equations):
\begin{eqnarray}
\partial _{4}\overrightarrow{\mathbf{J}} &=&(\ ^{\mathbf{+}}\mathcal{T}%
_{1}+\partial _{4}\phi )\overrightarrow{\mathbf{J}},  \label{qpflows} \\
\partial _{4}\ ^{\mathbf{-}}\overrightarrow{\Theta } &=&-(3\ ^{\mathbf{-}}%
\mathcal{T}_{1}+\frac{15}{8}\ell )\ ^{\mathbf{+}}\overrightarrow{\Theta }+%
\frac{3}{2}(\ ^{\mathbf{+}}\mathcal{T}_{1}+\partial _{4}\phi )\ ^{\mathbf{-}}%
\overrightarrow{\Theta },  \notag \\
\partial _{4}\ ^{\mathbf{+}}\overrightarrow{\Theta } &=&\frac{3}{2}(\ ^{%
\mathbf{+}}\mathcal{T}_{1}+\partial _{4}\phi )\ ^{\mathbf{+}}\overrightarrow{%
\Theta }+\tilde{\alpha}(y^{4})\ \ ^{\mathbf{-}}\overrightarrow{\Theta },%
\mbox{ for
arbitrary function }\tilde{\alpha}(y^{4});  \notag
\end{eqnarray}%
\begin{equation}
\mbox{ for  }\overrightarrow{d}\overrightarrow{\mathbf{J}}=-\frac{3}{2}\ ^{%
\mathbf{-}}\mathcal{T}_{1}\ ^{\mathbf{+}}\overrightarrow{\Theta }+\frac{3}{2}%
\ ^{\mathbf{+}}\mathcal{T}_{1}\ ^{\mathbf{-}}\overrightarrow{\Theta }\mbox{
and  }\overrightarrow{d}\overrightarrow{\Theta }=\ \mathcal{T}_{1}\widetilde{%
\mathbf{J}}\wedge \widetilde{\mathbf{J}}.  \notag
\end{equation}

Similarly to \cite{lecht1}, there are two cases corresponding  to a nearly
almost-K\"{a}hler  quasiperiodic geometric (i.e. with nonholonomically
induced torsion by off--diagonal N-terms and vanishing NS 3-form flux) and a
nonholonomic generalized Calabi-Yau with flux of quasiperiodic structures.

\subsubsection{ YM and instanton configurations induced by quasiperiodic
structures}

Nonholonomic instanton equations with quasiperiodic structure can be
formulated on $\ ^{7}\overrightarrow{\mathbf{X}}:=\mathbb{R\times }%
\ ^{6}\overrightarrow{\mathbf{X}}$ by  generalized 'h--cone' d-metrics of type
\begin{eqnarray*}
\ _{c}^{7}\mathbf{g} &\mathbf{=}&\mathbf{(e}^{4})^{2}+[\ _{\shortmid
}h(y^{4})]^{2}\ ^{6}\mathbf{g}[\overrightarrow{\eta }(y^{\check{c}})]=%
\mathbf{(e}^{4})^{2}+[\ _{\shortmid }h(y^{4})]^{2}\ \ ^{6}\overrightarrow{%
\mathbf{g}}_{\check{a}\check{b}}[\overrightarrow{\eta }(y^{\check{c}})], \\
\mathbf{e}^{4} &=&dy^{4}+w_{i}(x^{k},y^{a}).
\end{eqnarray*}%
In this formula, \ $^{6}\overrightarrow{\mathbf{g}}$ (\ref{qplfsmf}) is
taken for an exact solution  in 10-d gravity, as we considered  in section %
\ref{ssyminst}. In this subsection, we use the local frames adapted to the
quasiperiodic structure as   $\ _{\shortmid }\overrightarrow{\mathbf{e}}^{%
\widetilde{a}\prime }=\{\mathbf{e}^{4\prime },\ _{\shortmid }h\cdot \
_{\shortmid }\overrightarrow{\mathbf{e}}^{\check{a}\prime }\}\in T^{\ast }(\
^{7}\mathbf{X}),$ i.e. for an orthonormal N--adapted basis on $\widetilde{a}%
^{\prime },\widetilde{b}^{\prime },...=4,5,6,7,8,9,10,$ with $\ ^{\shortmid }%
\overrightarrow{\mathbf{e}}^{\check{a}\prime }=e_{\check{a}}^{\check{a}%
\prime }\overrightarrow{\mathbf{e}}^{\check{a}},$ when $\ ^{6}%
\overrightarrow{\mathbf{g}}_{\check{a}\check{b}}=\delta _{\check{a}\prime
\check{b}^{\prime }}e_{\check{a}}^{\check{a}\prime }e_{\check{b}}^{\check{b}%
^{\prime }}.$

Considering  almost-K\"{a}hler operators in standard form (instead
of $(\overrightarrow{\mathbf{J}}\mathbf{,}\overrightarrow{\Theta })$ but
with respect to orthonormal frames), we construct
\begin{equation*}
\ \ _{\shortmid }\overrightarrow{\mathbf{J}}:=\ \ _{\shortmid }%
\overrightarrow{\mathbf{e}}^{5}\wedge \ _{\shortmid }\overrightarrow{\mathbf{%
e}}^{6}+\ _{\shortmid }\overrightarrow{\mathbf{e}}^{7}\wedge \ _{\shortmid }%
\overrightarrow{\mathbf{e}}^{8}+\ _{\shortmid }\overrightarrow{\mathbf{e}}%
^{9}\wedge \ _{\shortmid }\overrightarrow{\mathbf{e}}^{10}\mbox{ and  }\
_{\shortmid }\overrightarrow{\Theta }:=(\ _{\shortmid }\overrightarrow{%
\mathbf{e}}^{5}+i\ _{\shortmid }\overrightarrow{\mathbf{e}}^{6})\wedge (\
_{\shortmid }\overrightarrow{\mathbf{e}}^{7}+i\ _{\shortmid }\overrightarrow{%
\mathbf{e}}^{8})\wedge (\ _{\shortmid }\overrightarrow{\mathbf{e}}^{9}+\
_{\shortmid }\overrightarrow{\mathbf{e}}^{10}),
\end{equation*}%
where $i^{2}=-1$ is used for $SU(3).$ For such orthonormalized quasi-period
N-adapted bases, we can verify that there are satisfied conditions which are
similar to those  for the nearly K\"{a}hler internal spaces, see Refs.  \cite%
{harl,gray,lecht1}. There are reproduced nonholonomic varaiants of  almost-K%
\"{a}hler manifold structure equations,
\begin{equation*}
d(\ _{\shortmid }^{+}\overrightarrow{\Theta })=2\ \ _{\shortmid }%
\overrightarrow{\mathbf{J}}\wedge \ _{\shortmid }\overrightarrow{\mathbf{J}}%
\mbox{ and }d\ _{\shortmid }\overrightarrow{\mathbf{J}}=3(\ _{\shortmid }^{-}%
\overrightarrow{\Theta }).
\end{equation*}

Using formulas (\ref{tauparam}) and (\ref{nhinsteqcyl}) in terms of "arrow"
operators, we define \ on $^{7}\overrightarrow{\mathbf{X}}$ two equivalent
d--metrics, $\ _{c}^{7}\mathbf{g}[\overrightarrow{\eta }]=e^{2f}\ _{z}^{7}%
\mathbf{g}[\overrightarrow{\eta }]$ and $\ _{z}^{7}\mathbf{g}[%
\overrightarrow{\eta }]=d\tau ^{2}+\ ^{6}\overrightarrow{\mathbf{g}}_{\check{%
a}\check{b}}.$ The respective  nonholonomic instanton equations with
quasiperiodic structure are
\begin{equation}
\overrightarrow{\ast }_{z}\overrightarrow{\mathbf{F}} = -(\overrightarrow{%
\ast }_{z}\overrightarrow{\mathbf{Q}}_{z})\wedge \overrightarrow{\mathbf{F}},
 \mbox{ for }
\overrightarrow{\mathbf{Q}}_{z} = d\tau \wedge \ _{\shortmid }^{+}%
\overrightarrow{\Theta }+\frac{1}{2}\ _{\shortmid }\overrightarrow{\mathbf{J}%
}\wedge \ _{\shortmid }\overrightarrow{\mathbf{J}},  \label{qpinsteq}
\end{equation}%
where  $\overrightarrow{\ast }_{z}$ is the Hodge star with respect to the
cylinder metric $\ _{z}^{7}\mathbf{g[\overrightarrow{\eta }].}$ We encode
the quasiperiodic almost-K\"{a}hler structure of $\ ^{7}\overrightarrow{%
\mathbf{X}}$ into boldface arrow operators $\overrightarrow{\mathbf{Q}}%
_{z},\ _{\shortmid }\overrightarrow{\mathbf{J}}$ and  $\ _{\shortmid }^{+}%
\overrightarrow{\Theta }$ and consider the  normal (quasiperiodic almost
symplectic) d--connection $\ \overrightarrow{\mathbf{D}}_{\check{a}}=(%
\overrightarrow{D}_{\acute{k}},\overrightarrow{D}_{\grave{b}})=\{%
\overrightarrow{\omega }_{\ \check{a}\check{c}^{\prime }}^{\check{b}^{\prime
}}\},$ computed for formulas similar to (\ref{cdccf}) but with arrow values.

In terms of arrow-operators, we construct the canonical values $\ \
_{A}\overrightarrow{\mathbf{D}}=\ ^{can}\overrightarrow{\mathbf{D}}+\psi
(\tau )\ _{\shortmid }\mathbf{e}^{\widetilde{a}\prime }I_{\widetilde{a}%
\prime },$ where the canonical d--connection on $\ ^{6}\overrightarrow{%
\mathbf{X}}$ is  $\ ^{can}\overrightarrow{\mathbf{D}}=\{\ ^{can}\overrightarrow{\omega }_{\
\check{a}\check{c}^{\prime }}^{\check{b}^{\prime }}:=\overrightarrow{\omega }%
_{\ \check{a}\check{c}^{\prime }}^{\check{b}^{\prime }}+\frac{1}{2}(\
_{\shortmid }^{+}\overrightarrow{\Theta })_{\ \check{c}^{\prime }\check{a}%
^{\prime }}^{\check{b}^{\prime }}e_{\check{a}}^{\check{a}^{\prime }}\}$.
  The the curvature d--form \ induced by quasiperiodic structures is computed%
\begin{eqnarray*}
\ _{A}\overrightarrow{\mathbf{F}} &=&\frac{1}{2}[\ _{A}\widetilde{\mathbf{D}}%
,\ _{A}\widetilde{\mathbf{D}}]:=\mathcal{F}(\psi ) \\
&=&\ ^{can}\overrightarrow{R}+\frac{1}{2}\psi ^{2}f_{\check{a}^{\prime }%
\check{b}^{\prime }}^{\widetilde{i}^{\prime }}\ I_{\widetilde{i}^{\prime }}\
_{\shortmid }\overrightarrow{\mathbf{e}}^{\check{a}^{\prime }}\wedge \
_{\shortmid }\overrightarrow{\mathbf{e}}^{\check{b}^{\prime }}+\frac{%
\partial \psi }{\partial \tau }d\tau \wedge I_{\check{c}^{\prime }}\
_{\shortmid }\overrightarrow{\mathbf{e}}^{\check{c}^{\prime }}+\frac{1}{2}%
(\psi -\psi ^{2})I_{\check{b}^{\prime }}(\ _{\shortmid }^{+}\overrightarrow{%
\Theta })_{\ \check{c}^{\prime }\check{a}^{\prime }}^{\check{b}^{\prime }}\
_{\shortmid }\overrightarrow{\mathbf{e}}^{\check{c}^{\prime }}\wedge \
_{\shortmid }\overrightarrow{\mathbf{e}}^{\check{a}^{\prime }},
\end{eqnarray*}%
with a similar parametric dependence on $\tau $ (\ref{tauparam}) via $\psi
(\tau ).$ \ For instance, \ a $\ _{A}\overrightarrow{\mathbf{F}}$ is a
quasiperiodic solution of the nonholonomic instanton equations \ (\ref%
{qpinsteq}) if $\psi $ is a  solution of the 'kink equation' $\frac{\partial
\psi }{\partial \tau }=2\psi (\psi -1).$

In heterotic supergravity, it is possible to derive  two classes of
quasiperiodic instanton configurations. In the first case, one considers
gauge-like curvatures of type $\ _{A}\overrightarrow{\mathbf{F}}$ $=\mathcal{%
F}(\ ^{1}\psi )$ and $\ \widetilde{\mathbf{R}}$ $[\overrightarrow{\eta }]=%
\mathcal{R}(\ ^{2}\psi ).$ In the second case, we have to  solve also the
equations  $\ \widetilde{\mathbf{R}}[\overrightarrow{\eta }]$ $\cdot
\epsilon =0.$

\section{Summary of Results and Conclusions}

\label{s5}In this work, we have studied how to generalize certain classes of
\ 'prime' solutions in heterotic string gravity constructed in \cite%
{lecht1,harl} for arbitrary 'target' ten dimensional, 10-d, spaces with metrics depending on all 4-d spacetime and 6-d internal space coordinates.  The
structure and classification of prime configurations [with pseudo-Euclidean
(1+3)--dimensional domain walls and 6-d warped nearly K\"{a}hler manifolds
in the presence of gravitational and gauge instantons] can be preserved for
nontrivial curved 4-d spacetime configurations. For instance, we can
generate black hole/ellipsoid configurations if certain types of
nonholonomic variables with conventional 2+2+....=10 splitting are defined
and respective almost-K\"{a}hler internal configurations are associated for
 2+2+2=3+3 double fibrations. The diadic "shell by shell"
nonholonomic decomposition of 4-d, 6-d and 10-d pseudo-Riemannian manifolds
allows us to integrate the motion equations (\ref{hs1})--(\ref{hs4}) in very
general forms using the AFDM, see further developments of this paper in \cite%
{partner}. The 3+3 decomposition is used for constructing nonholonomic
deformed instanton configurations which are necessary for solving the
Yang-Mills sector and the generalized Bianchi identity at order $\alpha
^{\prime },$ when certain generalized classes of solutions may contain
internal configurations depending, in principle, on all 9 space-like
coordinates for a 10-d effective gravity theory. The triadic formalism is
also important for associating $SU(3)$ structures in certain holonomic
limits to well known solutions (with more "simple" domain wall and an
internal structures) in heterotic string gravity.

The geometric techniques explained in this paper (which is a development
for the heterotic string gravity results of a series of studies \cite%
{vpars,sv2001,vapexsol,vex1,vex2,vex3,veym,svvvey,tgovsv,vtamsuper,vjgp,vwitten,vmedit}%
) allows us to work with arbitrary stationary generic off-diagonal metrics
on 10-d spacetimes. Effective MGTs spacetimes can be enabled with
generalized connections, depending on all possible 4-d and extra dimension
space coordinates. One of the main results of the presented work is that the
system (\ref{hs3})--(\ref{hs4}) admits subclasses of solutions with warping
on coordinate $y^{4}$ nearly almost-K\"{a}hler 6-d internal manifolds in the
presence of nonholonomically deformed gravitational and gauge instantons.
The almost-K\"{a}hler structure is necessary if we want to generate in the
4-d spacetime part, for instance, the Kerr metric with possible (off-)
diagonal and nonholonomic deformations to black ellipsoid type
configurations characterized by locally anisotropic polarized physical
constants, small deformations of horizons, embedding into nontrivial extra
dimension vacuum gravitational fields and/or gauge configurations [which are
considered in detail in \cite{vpars,sv2001,vapexsol,vex1,vex2,vex3,veym,svvvey,tgovsv,vtamsuper,partner}].  In this work, we concentrate on 10-d configurations preserving two real
supercharges corresponding to $N=1/2$ supersymmetry from the viewpoint of
four non-compact dimensions and various nonholonomic deformations.

Following methods of the geometry of nonholonomic manifolds and almost-K\"{a}%
hler spaces, applied for deformation and/or A-brane quantization and
geometric flows of gravity theories \cite{vjgp,vwitten,vmedit}, we defined
two pairs of related $(\ _{\shortmid }\mathbf{\tilde{J},}\ _{\shortmid }%
\widetilde{\Theta })$ and $(\mathbf{\tilde{J},}\widetilde{\Theta })$
structures depending both on warping and other space coordinates. Such
constructions encode possible almost symplectic configurations determined by
the 6-d internal sector of heterotic gravity and BPS equations re-written in
nonholonomic variables. This involves and mixes certain gauge like, i.e. a
NS 3-form for flux, gravitational solitons, and effective scalar fields. For
additional constraints, such stationary configurations transform into static
$SU(3)$ structures as in \cite{lecht1,harl}. This assumption is crucial for
classifying new types of nonolonomically deformed solutions following the
same principles, as it was done originally for K\"{a}hler internal spaces
and standard instanton constructions.

Let us outline the most important results obtained in this paper which provide significant contributions in  heterotic supergravity and string theories and for development of new geometric methods for constructing exact solutions in (super) string and gravity modified theories:
\begin{enumerate}
\item We developed a nonholonomic geometric  approach to heterotic supergravity using the formalism of nonlinear connections  on (super)  manifolds with fibred structure and  bundle superspaces. This allows us to define almost symplectic structures canonically induced by effective Lagrange distributions and  construct real solutions with generalized connections for BPS equations. We have defined nonholonomic domain-wall backgrounds and N--adapted $G_2$ structures on almost K\"{a}hler internal spaces and constructed various classes of nonholonomic instanton solutions.

\item We found new classes of exact solutions for the N-adapted YM and instanton configurations, and for static/ dynamic $SU(3)$ nonholonomic structures on  almost K\"{a}hler configurations.

\item A fundamental result is that the equations of motion in heterotic supergravity (formulated in terms of a well-defined class of nonholonomic variables and auxiliary connections) can be decoupled in general forms. This allows us to construct exact solutions in 4-d and 10-d gravity parameterized by generic off-diagonal metrics and generalized connections depending, in general, on all spacetime and internal space coordinates via respective classes of generating and integration functions, effective matter sources, integration constants, and prescribed symmetries. Detailed proofs and applications of such geometric methods have been presented in a series of recent papers \cite{tgovsv,vtamsuper,sv2013,svvvey,vpars} and in the partner work \cite{partner}. Those results are completed with examples of  nine new classes of solutions studied in section \ref{s4} of this paper.

\item Section \ref{s4}  is devoted to string deformations of Kerr metrics determined by solitonic, quasiperiodic and/or pattern-forming structures and generic off-diagonal terms of 4-d spacetime metrics and 6-d internal spaces. Such new classes, exact and parametric solutions provide very important contributions to modern cosmology, astrophysics and modified gravity (see also recent results obtained by applying similar geometric methods in Refs. \cite{vcosmsol2,vcosmsol3,vcosmsol4,vcosmsol5,aschheim16,amaral16}) and concern such fundamental issues:
\begin{enumerate}
\item String solitonic interactions result in effective anisotropic polarizations of the mass of 4-d Kerr black holes. Such physical effects can be determined by generic off-diagonal terms of metrics in an effective 4-d gravity (encoding string contributions)  and nonholonomic deformations determined by stationary solitonic distributions and/or solitonic waves in extra-dimensional shells on the internal space.
\item Prescribing certain classes of generating functions and effective sources, we can generate off-diagonal deformations of black hole solutions resulting in  quaisperiodic configurations in 10-d supergravity. Such quasicrystal like and/or pattern-forming structures are characterized by respective free-energy and evolution type equations of astrophysical (and cosmological \cite{vrgeomfl,vrajgf,vmedit,vnrflnc}) objects. Such free energy type values  are similar to Lyapunov type functionals and entropy  functionals  considered by G. Perelman in his, and R. Hamilton, approach to Ricci flow theory. The possibility to construct exact solutions in hetereotic supergravity with nonlinear evolution on mixed 4-d spacetime and internal spaces variables presents a new fundamental result for the theory of (super) geometric flows and applications in modern physics and geometric mechanics.
\item A series of new geometric and physically important solutions consists of 10-d, 7-d, and 6-d metrics encoding almost symplectic models with quasiperiodic canonical connections; N-adapted $G_2$  and/or almost complex structures; YM and instanton configurations induced by quasiperiodic structures.
\end{enumerate}

\end{enumerate}

Finally, we note that a series of recent works  \cite{tgovsv,vtamsuper,vrgeomfl,vrajgf,vmedit,vnrflnc} prove that  the AFDM allows us to construct various  classes of physically important solutions of the motion equations (\ref{hs1})--(\ref{hs2}) (like black holes, cosmological metrics, wormholes etc.) in the 4-d sector, using  nonholonomic solutions of (\ref{hs3})--(\ref{hs4}).  In the associated paper \cite{partner}, the motion equations in heterotic string gravity resulting in stationary metrics, in particular, in generic off-diagonal deformations of the Kerr solution to certain ellipsoid like configurations are solved by integrating modified Einstein equations in very general off-diagonal forms. Further developments in string MGTs with cosmological solutions of type \cite{ferrara1,kounnas3,saridak,mavromat,odints1,vcosmsol2,vcosmsol3,vcosmsol4,vcosmsol5}, with quasiperiodic and other type nonlinear structures are left for future work.

\vskip3pt \textbf{Acknowledgments:} The SV research is for the QGR--Topanga with a former partial support by IDEI, PN-II-ID-PCE-2011-3-0256 and DAAD. This work contains also a summary of results presented in a talk for GR21 at NY.

\appendix

\setcounter{equation}{0} \renewcommand{\theequation}
{A.\arabic{equation}} \setcounter{subsection}{0}
\renewcommand{\thesubsection}
{A.\arabic{subsection}}

\section{Nonholonomic Manifolds with 2+2+... Splitting}

\label{as1}We summarize necessary results from the geometry of nonholonomic
manifolds which will then be applied in the heterotic supergravity
(modelled in the low-energy limit of heterotic string theory as a $\mathcal{N%
}=1$ and 10-d supergravity coupled to super Yang-Mills theory). A geometric
formalism with nonholonomic variables and conventional 2+2+... splitting
defined in such forms, that allows a general decoupling of the motion equations will
be elaborated upon, see our associated work \cite{partner}. Such a higher
dimensional pseudo--Riemannian spacetime is modelled as a 10-d manifold $%
\mathcal{M}$, equipped with a Lorentzian metric $\check{g}$ of signature $%
(++-+++++++)$ with a time like third coordinate\footnote{%
Working with such a signature is convenient for deriving recurrent formulas
for exact generic off-diagonal solutions in 4d to 10d spacetimes. By
re-defining at the end of the frame/coordinate systems, we can consider
"standard " coordinates and signatures of type $(-++++...+).$}. In our
approach, we use a unified system of notation for straightforward
applications of geometric methods for constructing exact solutions
developed in \cite{vex3,veym,vtamsuper,svvvey,tgovsv}. Such notation and
nonholonomic variables are different from that usually used in string theory
(see, for instance, \cite{lecht1}). The heterotic supergravity theory is
defined by a couple $(\mathcal{M},\check{g}),$ an NS 3-form $\check{H},$ a
dilaton field $\check{\phi}$ and a gauge connection $^{A}\check{\nabla},$
with gauge group $SO(32)$ or $E_{8}\times E_{8}.$

\subsection{N-adapted frames and coordinates}

For spacetime geometric models on a 10-d pseudo-Riemannian spacetime $\
\mathcal{M}$ with a time-like coordinate $u^3=t$ and other coordinates being
space-like, we consider conventional splitting of dimensions, $\dim \mathcal{%
M}=4+2s=10;s=0,1,2,3.$ The AFDM allows us to construct exact solutions with
arbitrary signatures of metrics $\check{g},$ but our goal is to consider
extra dimensional string gravity generalizations of the Einstein theory. In
most general forms, this is possible if we use the formalism of nonlinear
connection splitting for higher dimensional (super) spaces and strings which
was elaborated originally in (super) Lagrange-Finsler theory \cite%
{vnpfins,vapfins}. We shall not consider Finsler type (super) gravity models
in this work, but follow a similar approach with nonholonomic distributions
on (super) manifolds \cite{tgovsv,vtamsuper,vex3}.

Let us establish conventions on (abstract) indices and coordinates $%
u^{\alpha _{s}}=(x^{i_{s}},y^{a_{s}})$ by labelling the oriented number of
two dimensional, 2-d, "shells" added to a 4-d spacetime in GR. We consider
local systems of 10-d coordinates:
\begin{eqnarray}
s &=&0:u^{\alpha _{0}}=(x^{i_{0}},y^{a_{0}});\ s=1:u^{\alpha
_{1}}=(x^{i_{1}},y^{a_{1}})=(x^{i_{0}},y^{a_{0}},y^{a_{1}});
\label{coordconv} \\
s &=&2:u^{\alpha _{2}}=(x^{i_{1}}=u^{\alpha
_{1}},y^{a_{2}})=(x^{i_{0}},y^{a_{0}},y^{a_{1}},y^{a_{2}});\ s=3:u^{\alpha
_{3}}=(x^{i_{2}}=u^{\alpha
_{2}},y^{a_{3}})=(x^{i_{0}},y^{a_{0}},y^{a_{1}},y^{a_{2}},y^{a_{3}}),  \notag
\end{eqnarray}%
with values for indices: $i_{0},j_{0},...=1,2;a_{0},b_{0},...=3,4,$ when $%
u^{3}=y^{3}=t$; $a_{1},b_{1}...=5,6;a_{2},b_{2}...=7,8;$ $%
a_{3},b_{3}...=9,10;$ and, for instance, $i_{1},j_{1},...=1,2,3,4;i_{2},$ $%
j_{2},...$ $=1,2,3,4,5,6;\ i_{3},j_{3},...=1,2,3,4,5,6,7,8,$ or we shall
write only $i_{s}.$ In brief, we shall write $\ ^{0}u=(\ ^{0}x,\ ^{0}y);\
^{1}u=(\ ^{0}u,\ ^{1}y)=(\ ^{0}x,\ ^{0}y,\ ^{1}y),\ ^{2}u=(\ ^{1}u,\
^{2}y)=(\ ^{0}x,\ ^{0}y,\ ^{1}y,\ ^{2}y)$ and $\ ^{2}u=(\ ^{1}u,\ ^{2}y)=(\
^{0}x,\ ^{0}y,\ ^{1}y,\ ^{2}y).$ In order to connect these notations to
standard ones of supergravity theories (see \cite{lecht1,harl}),\footnote{%
In modern gravity, the so-called ADM (Arnowit--Deser--Misner) formalism with
3+1 splitting, or any $n+1$ splitting is used, see details in \cite{misner}.
It is not possible to develop a technique for general decoupling of the
gravitational field equations and generating off-diagonal solutions in such
cases because the conventional one dimensional "fibers" result in certain
degenerate systems of equations. To construct exact solutions in 4 to 10
dimensional theories, it is more convenient to work with non--integrable
2+2+... splitting, see details in \cite{tgovsv,vex2,vex3}. Using shell
coordinates, we are able to prove in a more "compact form" certain recurrent
formulas for integrating systems of nonlinear PDEs and understand important
nonlinear symmetries of higher dimension spacetimes in heterotic
supergravity. Such constructions are hidden in certain general statements
and sophisticated formulas if we work only with "standard" indices and
coordinates of type $x^{\mu },$ with $\mu =0,1,...,10.$} We shall consider
small Greek indices without subscripts, and respective coordinates $x^{\mu
}, $ when indices $\alpha ,\mu ,...=0,1,...,9$. The identification with
shell coordinates is of type $x^{0}$ $=u^{3}=t$ (for time-like coordinate)
and (for space-like coordinates): $%
x^{1}=u^{1},x^{2}=u^{2},x^{3}=u^{4},x^{4}=u^{5},x^{5}=u^{6},x^{6}=u^{7},x^{7}=u^{8},x^{8}=u^{9},x^{9}=u^{10}
$.

Local frames/bases, $e_{\alpha _{s}},$ on $\mathcal{M}$ are written in the
form $e_{\alpha _{s}}=e_{\ \alpha _{s}}^{\underline{\alpha }_{s}}(\
^{s}u)\partial /\partial u^{\underline{\alpha }_{s}},$ where partial
derivatives $\partial _{\beta _{s}}:=\partial /\partial u^{\beta _{s}}$
define local coordinate bases and indices are underlined if it is necessary
to emphasize that such values are defined with respect to a coordinate
frame. In general, a frame $e_{\alpha _{s}}$ is nonholonomic (equivalently,
anholonomic, or non-integrable), if it satisfies the anholonomy relations $\
e_{\alpha _{s}}e_{\beta _{s}}-e_{\beta _{s}}e_{\alpha _{s}}=W_{\alpha
_{s}\beta _{s}}^{\gamma _{s}}e_{\gamma _{s}}.$ In these formulas, the
anholonomy coefficients $W_{\alpha _{s}\beta _{s}}^{\gamma _{s}}=W_{\beta
_{s}\alpha _{s}}^{\gamma _{s}}(u)$ vanish for holonomic/integrable
configurations. The dual frames, $e^{\alpha _{s}}=e_{\ \underline{\alpha }%
_{s}}^{\ \alpha _{s}}(\ ^{s}u)du^{\underline{\alpha }_{s}},$ can be defined
from the condition $e^{\alpha _{s}}\rfloor e_{\beta _{s}}=\delta _{\beta
_{s}}^{\alpha _{s}}.$ In such conditions, the 'hook' operator $\rfloor $
corresponds to the inner derivative and $\delta _{\beta _{s}}^{\alpha _{s}}$
is the Kronecker symbol.

By using nonholonomic (non--integrable) distributions, we can define $%
2+2+... $ spacetime splitting in adapted frame and coordinate forms. For our
purpose, we shall work with certain distributions defining a nonlinear
connection, structure via a Whitney sum
\begin{equation}
\ ^{s}\mathbf{N}:T\mathcal{M}=\ ^{0}h\mathcal{M}\oplus \ ^{0}v\mathcal{M}%
\oplus \ ^{1}v\mathcal{M}\oplus \ ^{2}v\mathcal{M}\oplus \ ^{3}v\mathcal{M}.
\label{whitney}
\end{equation}%
Such a sum states a conventional horizontal (h) and vertical (v) "shell by
shell" splitting. We shall write boldface letters for spaces and geometric
objects enabled/adapted to a nonlinear connection structure. In local form,
the nonlinear connection coefficients, $N_{i_{s}}^{a_{s}},$ are defined from
a decomposition
\begin{equation}
\ ^{s}\mathbf{N}=N_{i_{s}}^{a_{s}}(\ ^{s}u)dx^{i_{s}}\otimes \partial
/\partial y^{a_{s}}.  \label{nconcoef}
\end{equation}

A manifold $\mathcal{M}$ enabled with a nonholonomic distribution (\ref%
{whitney}) is called nonholonomic (one can also use the term N--anholonomic
manifold). This definition comes form the fact that in linear form, the
coefficients (\ref{nconcoef}) determine a system of N--adapted local bases,
with N-elongated partial derivatives, $\mathbf{e}_{\nu _{s}}=(\mathbf{e}%
_{i_{s}},e_{a_{s}}),$ and cobases with N--adapted differentials, $\mathbf{e}%
^{\mu _{s}}=(e^{i_{s}},\mathbf{e}^{a_{s}}),$ For $s=0$ (on a 4-d spacetime
part)
\begin{eqnarray}
&&\mathbf{e}_{i_{0}}=\frac{\partial }{\partial x^{i_{0}}}-\ N_{i_{0}}^{a_{0}}%
\frac{\partial }{\partial y^{a_{0}}},\ e_{a_{0}}=\frac{\partial }{\partial
y^{a_{0}}},\ e^{i_{0}}=dx^{i_{0}},\mathbf{e}^{a_{0}}=dy^{a_{0}}+\
N_{i_{0}}^{a_{0}}dx^{i_{0}}\mbox{ on }\mathbf{V\simeq }h\mathcal{M}\oplus v%
\mathcal{M};\mbox{ or/ and }  \label{nadaptb} \\
&&\mathbf{e}_{i_{s}}=\frac{\partial }{\partial x^{i_{s}}}-\ N_{i_{s}}^{a_{s}}%
\frac{\partial }{\partial y^{a_{s}}},\ e_{a_{s}}=\frac{\partial }{\partial
y^{a_{s}}},e^{i_{s}}=dx^{i_{s}},\mathbf{e}^{a_{s}}=dy^{a_{s}}+\
N_{i_{s}}^{a_{s}}dx^{i_{s}}\mbox{ for }s=1,2,3.  \notag
\end{eqnarray}%
The corresponding anholonomy relations with inter--shell non--integrable
relations,
\begin{equation}
\lbrack \mathbf{e}_{\alpha _{s}},\mathbf{e}_{\beta _{s}}]=\mathbf{e}_{\alpha
_{s}}\mathbf{e}_{\beta _{s}}-\mathbf{e}_{\beta _{s}}\mathbf{e}_{\alpha
_{s}}=W_{\alpha _{s}\beta _{s}}^{\gamma _{s}}\mathbf{e}_{\gamma _{s}},
\label{anhrel}
\end{equation}%
are computed $W_{i_{s}a_{s}}^{b_{s}}=\partial _{a_{s}}N_{i_{s}}^{b_{s}}$ and
$W_{j_{s}i_{s}}^{a_{s}}=\Omega _{i_{s}j_{s}}^{a_{s}}.$ In these formulas,
the curvature of N--connection is defined as the Neijenhuis tensor, $\Omega
_{i_{s}j_{s}}^{a_{s}}:=\mathbf{e}_{j_{s}}\left( N_{i_{s}}^{a_{s}}\right) -%
\mathbf{e}_{i_{s}}\left( N_{j_{s}}^{a_{s}}\right) .$

\subsection{d--torsions and d--curvatures of d--connections}

There is a subclass of linear connections on $\mathcal{M},$ called
distinguished connections, d--connections, which preserve the N--connection
structure under parallelism (\ref{whitney}).\footnote{%
For spaces enabled with N--connection structure, terms are used like
distinguished tensor, d--tensor; distinguished spinor, d--spinor;
distinguished geometric object, d--object, if the coefficients of such
geometric/physical values are determined in a N--adapted shell form, with
respect to frames of type (\ref{nadaptb}) and their symmetric, or skew
symmetric tensor products.} With left shell labels, we write
\begin{equation*}
\ ^{s}\mathbf{D=\mathbf{\{D}}_{\alpha _{s}}\mathbf{\mathbf{\}}=}(\
^{s-1}hD;\ ^{s}vD):  \ ^{0}\mathbf{D}=(\ ^{0}hD;\ ^{0}vD),\ ^{1}%
\mathbf{D=}(\ ^{1}hD;\ ^{1}vD),\ ^{2}\mathbf{D=}(\ ^{2}hD;\ ^{2}vD),\ ^{3}%
\mathbf{D=}(\ ^{3}hD;\ ^{3}vD).
\end{equation*}%
In N--adapted form, the coefficients of a d--connection $\ ^{s}\mathbf{D=\{%
\mathbf{\Gamma }}_{\ \beta _{s}\gamma _{s}}^{\alpha _{s}}=(L_{\beta
_{s-1}\gamma _{s-1}}^{\alpha _{s-1}},L_{b_{s}\gamma _{s-1}}^{a_{s}};C_{\beta
_{s-1}c_{s}}^{\alpha _{s-1}},C_{b_{s}c_{s}}^{a_{s}})\},$ for
example:
\begin{eqnarray}
\mathbf{\Gamma }_{\ \beta _{0}\gamma _{0}}^{\alpha _{0}}
&=&(L_{j_{0}k_{0}}^{i_{0}},L_{b_{0}k_{0}}^{a_{0}};C_{j_{0}c_{0}}^{i_{0}},C_{b_{0}c_{0}}^{a_{0}}),%
\mathbf{\Gamma }_{\ \beta _{1}\gamma _{1}}^{\alpha _{1}}=(L_{\beta
_{0}\gamma _{0}}^{\alpha _{0}},L_{b_{1}\gamma }^{a_{1}};C_{\beta
_{0}c_{1}}^{\alpha _{0}},C_{b_{1}c_{1}}^{a_{1}}),\   \label{coefd} \\
\mathbf{\Gamma }_{\ \beta _{2}\gamma _{2}}^{\alpha _{2}} &=&(L_{\beta
_{1}\gamma _{1}}^{\alpha _{1}},L_{b_{2}\gamma _{1}}^{a_{2}};C_{\beta
_{1}c_{2}}^{\alpha _{1}},C_{b_{2}c_{2}}^{a_{2}}),\mathbf{\Gamma }_{\ \beta
_{3}\gamma _{3}}^{\alpha _{3}}=(L_{\beta _{2}\gamma _{2}}^{\alpha
_{2}},L_{b_{3}\gamma _{2}}^{a_{3}};C_{\beta _{2}c_{3}}^{\alpha
_{2}},C_{b_{3}c_{3}}^{a_{3}}),  \notag
\end{eqnarray}%
can be computed in N--adapted form with respect to frames (\ref{nadaptb}).
We have to consider the equations $\mathbf{D}_{\alpha _{s}}\mathbf{e}_{\beta
_{s}}=\mathbf{\Gamma }_{\ \beta _{s}\gamma _{s}}^{\alpha _{s}}\mathbf{e}%
_{\gamma _{s}}$ and covariant derivatives parameterized in the forms
\begin{eqnarray*}
\ \mathbf{D}_{\alpha _{0}} &=&(D_{i_{0}};D_{a_{0}}),\mathbf{D}_{\alpha
_{1}}=(\mathbf{D}_{\alpha _{0}};D_{a_{1}}),\ \mathbf{D}_{\alpha _{2}}=(%
\mathbf{D}_{\alpha _{1}};D_{a_{2}}),\mathbf{D}_{\alpha _{3}}=(\mathbf{D}%
_{\alpha _{2}};D_{a_{3}}), \\
\mbox{ for }hD &=&(L_{jk}^{i},L_{bk}^{a}),vD=(C_{jc}^{i},C_{bc}^{a}),\
^{1}hD=(L_{\beta \gamma }^{\alpha },L_{b_{1}\gamma }^{a_{1}}),\
^{1}vD=(C_{\beta c_{1}}^{\alpha },C_{b_{1}c_{1}}^{a_{1}}), \\
\ \ ^{2}hD &=&(L_{\beta _{1}\gamma _{1}}^{\alpha _{1}},L_{b_{2}\gamma
_{1}}^{a_{2}}),\ ^{2}vD=(C_{\beta _{1}c_{2}}^{\alpha
_{1}},C_{b_{2}c_{2}}^{a_{2}}),\ ^{3}hD=(L_{\beta _{2}\gamma _{2}}^{\alpha
_{2}},L_{b_{3}\gamma _{2}}^{a_{3}}),\ ^{3}vD=(C_{\beta _{2}c_{3}}^{\alpha
_{2}},C_{b_{3}c_{3}}^{a_{3}}),
\end{eqnarray*}%
or, in general form, $\ ^{s}hD=(L_{\beta _{s-1}\gamma _{s-1}}^{\alpha
_{s-1}},L_{b_{s}\gamma _{s-1}}^{a_{s}}),\ ^{s}vD=(C_{\beta
_{s-1}c_{s}}^{\alpha _{s-1}},C_{b_{s}c_{s}}^{a_{s}}).$ Such coefficients can
be computed with respect to mixed subsets of coordinates and/or N--adapted
frames on different shells. It is always possible to consider such frame
transforms when all shell frames are N-adapted, $\ ^{s}D_{\alpha _{s-1}}=%
\mathbf{D}_{\alpha _{s-1}}.$

Using (\ref{coefd}), we can develop an N--adapted covariant calculus on $%
\mathcal{M}$ and a corresponding differential form calculus with a
differential connection 1--form $\mathbf{\Gamma }_{\ \beta _{s}}^{\alpha
_{s}}=\mathbf{\Gamma }_{\ \beta _{s}\gamma _{s}}^{\alpha _{s}}\mathbf{e}%
^{\gamma _{s}}$ with respect to skew symmetric tensor products of N--adapted
frames (\ref{nadaptb}). For instance, the torsion $\mathcal{T}^{\alpha
_{s}}=\{\mathbf{T}_{\ \beta _{s}\gamma _{s}}^{\alpha _{s}}\}$ and curvature $%
\mathcal{R}_{~\beta _{s}}^{\alpha _{s}}=\{\mathbf{\mathbf{R}}_{\ \ \beta
_{s}\gamma _{s}\delta _{s}}^{\alpha _{s}}\}$ d--tensors of $\ ^{s}\mathbf{D}$
can be computed in explicit form following respective formulas,
\begin{eqnarray}
&&\mathcal{T}^{\alpha _{s}}:=\ ^{s}\mathbf{De}^{\alpha _{s}}=d\mathbf{e}%
^{\alpha _{s}}+\mathbf{\Gamma }_{\ \beta _{s}}^{\alpha _{s}}\wedge \mathbf{e}%
^{\beta _{s}}\   \label{dt} \\
&&\mathcal{R}_{~\beta _{s}}^{\alpha _{s}}:=\ ^{s}\mathbf{D\Gamma }_{\ \beta
_{s}}^{\alpha _{s}}=d\mathbf{\Gamma }_{\ \beta _{s}}^{\alpha _{s}}-\mathbf{%
\Gamma }_{\ \beta _{s}}^{\gamma _{s}}\wedge \mathbf{\Gamma }_{\ \gamma
_{s}}^{\alpha _{s}}=\mathbf{R}_{\ \beta _{s}\gamma _{s}\delta _{s}}^{\alpha
_{s}}\mathbf{e}^{\gamma _{s}}\wedge \mathbf{e}^{\delta _{s}},  \label{dc}
\end{eqnarray}%
see Refs. \cite{tgovsv,vtamsuper,vex3,vnpfins,vapfins} for explicit calculation
of coefficients $\mathbf{T}_{\ \beta _{s}\gamma _{s}}^{\alpha _{s}}$ and $%
\mathbf{R}_{\ \beta _{s}\gamma _{s}\delta _{s}}^{\alpha _{s}}$ in higher
dimensions.

\subsection{d--metrics and generic off--diagonal metrics}

In coordinate form, a metric $\check{g}$ on $\mathcal{M}$ is written
\begin{equation}
\ ^{s}g\mathbf{=}g_{\alpha _{s}\beta _{s}}e^{\alpha _{s}}\otimes e^{\beta
_{s}}=g_{\underline{\alpha }_{s}\underline{\beta }_{s}}du^{\underline{\alpha
}_{s}}\otimes du^{\underline{\beta }_{s}},\   \label{metr}
\end{equation}%
for $s=0,1,2,3$ corresponding to a conventional $2+2+...$ splitting. Under
general frame transforms, the coefficients of the metric transforms as $%
g_{\alpha _{s}\beta _{s}}=e_{\ \alpha _{s}}^{\underline{\alpha }_{s}}e_{\
\beta _{s}}^{\underline{\beta }_{s}}g_{\underline{\alpha }_{s}\underline{%
\beta }_{s}}.$ Similar rules can be considered for all tensor objects which
do not preserve a splitting of dimensions. The same metric can be
parameterized as a distinguished metric (d--metric, in boldface form), $\ ^{s}%
\mathbf{g}=\{\mathbf{g}_{\alpha _s\beta _s}\},$
\begin{eqnarray}
\ \ ^{s}\mathbf{g} &=&\ g_{i_{s}j_{s}}(\ ^{s}u)\ e^{i_{s}}\otimes
e^{j_{s}}+\ g_{a_{s}b_{s}}(\ ^{s}u)\mathbf{e}^{a_{s}}\otimes \mathbf{e}%
^{b_{s}}  \label{dm} \\
&=&g_{ij}(x)\ e^{i}\otimes e^{j}+g_{ab}(u)\ \mathbf{e}^{a}\otimes \mathbf{e}%
^{b}+g_{a_{1}b_{1}}(\ ^{1}u)\ \mathbf{e}^{a_{1}}\otimes \mathbf{e}%
^{b_{1}}+....+\ g_{a_{s}b_{s}}(\ ^{s}u)\mathbf{e}^{a_{s}}\otimes \mathbf{e}%
^{b_{s}}.  \notag
\end{eqnarray}%
Redefining (\ref{dm}) in coordinate frames, we find the relation between
N--connection coefficients and off-diagonal metric terms in (\ref{metr}),
\begin{eqnarray*}
\ \ \underline{g}_{\alpha \beta }\left( \ u\right) &=&\left[
\begin{array}{cc}
\ g_{ij}+\ h_{ab}N_{i}^{a}N_{j}^{b} & h_{ae}N_{j}^{e} \\
\ h_{be}N_{i}^{e} & \ h_{ab}%
\end{array}%
\right] ,\ \underline{g}_{\alpha _{1}\beta _{1}}(\ ^{1}u)=\left[
\begin{array}{cc}
\ \underline{g}_{\alpha \beta } & h_{a_{1}e_{1}}N_{\beta _{1}}^{e_{1}} \\
\ h_{b_{1}e_{1}}N_{\alpha _{1}}^{e_{1}} & \ h_{a_{1}b_{1}}%
\end{array}%
\right] ,\  \\
\ \ \underline{g}_{\alpha _{2}\beta _{2}}(\ ^{2}u) &=&\left[
\begin{array}{cc}
\ \underline{g}_{\alpha _{1}\beta _{1}} & h_{a_{2}e_{2}}N_{\beta
_{1}}^{e_{2}} \\
\ h_{b_{2}e_{2}}N_{\alpha _{1}}^{e_{2}} & \ h_{a_{2}b_{2}}%
\end{array}%
\right] ,\ \ \underline{g}_{\alpha _{3}\beta _{3}}(\ ^{3}u)=\left[
\begin{array}{cc}
\ \underline{g}_{\alpha _{2}\beta _{2}} & h_{a_{3}e_{3}}N_{\beta
_{2}}^{e_{3}} \\
\ h_{b_{3}e_{3}}N_{\alpha _{2}}^{e_{3}} & \ h_{a_{3}b_{3}}%
\end{array}%
\right] .
\end{eqnarray*}%
For extra dimensions with $y^{a_{s}},s\geq 1$, such coefficients $\ \
\underline{g}_{\alpha _{s}\beta _{s}}(\ ^{s}u)=\left[
\begin{array}{cc}
\ g_{i_{s}j_{s}}+\ h_{a_{s}b_{s}}N_{i_{s}}^{a_{s}}N_{j_{s}}^{b_{s}} &
h_{a_{s}e_{s}}N_{j_{s}}^{e_{s}} \\
\ h_{b_{s}e_{s}}N_{i_{s}}^{e_{s}} & \ h_{a_{s}b_{s}}%
\end{array}%
\right] $ are similar to those introduced in the Kaluza--Klein theory (using
cylindrical compactifications on extra dimension coordinates, $N_{\alpha
}^{e_{s}}(\ ^{s}u)\sim A_{a_{s}\alpha }^{e_{s}}(u)y^{\alpha },$ where $%
A_{a_{s}\alpha }^{e_{s}}(u)$ are (non) Abelian gauge fields). In general,
various parametrizations can be used for warped/trapped coordinates in extra
dimension (super) gravity, string and brane gravity and modifications of GR,
see examples in \cite%
{vp,vt,vsingl2,sv2001,vpars,tgovsv,vtamsuper,vex3,veym,svvvey,vcosmsol2,vcosmsol3,vcosmsol4,vcosmsol5}.

\subsection{The canonical d--connection}

There are two very important linear connection structures determined by the
same metric structure following geometric conditions: {\small
\begin{equation}
\ ^{s}\mathbf{g}\rightarrow \left\{
\begin{array}{ccccc}
\ ^{s}\nabla : &  & \ ^{s}\nabla \ (\ ^{s}\mathbf{g})\ =0;\ \ _{\nabla }^{s}%
\mathcal{T}=0, &  & \mbox{ the Levi--Civita connection;} \\
\ ^{s}\widehat{\mathbf{D}}: &  & \ ^{s}\widehat{\mathbf{D}}\ (\ ^{s}\mathbf{%
g)}=0;\ h\widehat{\mathcal{T}}=0,\ ^{1}v\widehat{\mathcal{T}}=0,\ \ ^{2}v%
\widehat{\mathcal{T}}=0,\ ^{3}v\widehat{\mathcal{T}}=0. &  &
\mbox{ the
canonical d--connection.}%
\end{array}%
\right.  \label{lcconcdcon}
\end{equation}%
} Let us explain how the above connections are defined:

The LC--connection $\ ^{s}\nabla =\{\ _{\shortmid }\Gamma _{\beta _s\gamma
_s}^{\alpha _s}\}$ can be introduced without an N--connection structure but
can always be canonically distorted to a necessary type of d--connection
completely defined by $\ ^{s}\mathbf{g}$ following certain geometric
principles. In shell N--adapted frames, we can compute $\ _{\shortmid }%
\mathcal{T}^{\alpha _{s}}=0$ using formulas (\ref{dt}) for $\ ^{s}\mathbf{%
D\rightarrow }\ ^{s}\nabla .$

To elaborate on a covariant differential calculus adapted to decomposition (%
\ref{whitney}) with a d--connection completely determined by the d--metric
and d--connection structure, we have to consider the canonical d--connection
$\ ^{s}\widehat{\mathbf{D}}$ from (\ref{lcconcdcon}). For this linear
connection, the horizontal and vertical torsions are zero, i.e. $h\widehat{%
\mathcal{T}}=\{\widehat{\mathbf{T}}_{\ jk}^{i}\}=0,$ $v\widehat{\mathcal{T}}%
=\{\widehat{\mathbf{T}}_{\ bc}^{a}\}=0,\ ^{1}v\widehat{\mathcal{T}}=\{%
\widehat{\mathbf{T}}_{\ b_{1}c_{1}}^{a_{1}}\}=0,...,\ ^{s}v\widehat{\mathcal{%
T}}=\{\widehat{\mathbf{T}}_{\ b_{s}c_{s}}^{a_{s}}\}=0.$ We can check using
straightforward computations that such conditions are satisfied by $\ ^{s}%
\widehat{\mathbf{D}}=\{\widehat{\mathbf{\Gamma }}_{\ \alpha _{s}\beta
_{s}}^{\gamma _{s}}\}$ with coefficients (\ref{coefd}) computed with respect
to N--adapted frames (\ref{nadaptb}) following formulas
\begin{eqnarray}
\widehat{L}_{j_{s}k_{s}}^{i_{s}} &=&\frac{1}{2}g^{i_{s}r_{s}}\left( \mathbf{e%
}_{k_{s}}g_{j_{s}r_{s}}+\mathbf{e}_{j_{s}}g_{k_{s}r_{s}}-\mathbf{e}%
_{r_{s}}g_{j_{s}k_{s}}\right) ,  \notag \\
\widehat{L}_{b_{s}k_{s}}^{a_{s}} &=&e_{b_{s}}(N_{k_{s}}^{a_{s}})+\frac{1}{2}%
g^{a_{s}c_{s}}\left( \mathbf{e}_{k_{s}}g_{b_{s}c_{s}}-g_{d_{s}c_{s}}\
e_{b_{s}}N_{k_{s}}^{d_{s}}-g_{d_{s}b_{s}}\ e_{c_{s}}N_{k_{s}}^{d_{s}}\right)
,  \notag \\
\widehat{C}_{j_{s}c_{s}}^{i_{s}} &=&\frac{1}{2}%
g^{i_{s}k_{s}}e_{c_{s}}g_{j_{s}k_{s}},\ \widehat{C}_{b_{s}c_{s}}^{a_{s}}=%
\frac{1}{2}g^{a_{s}d_{s}}\left(
e_{c_{s}}g_{b_{s}d_{s}}+e_{c_{s}}g_{c_{s}d_{s}}-e_{d_{s}}g_{b_{s}c_{s}}%
\right) .  \label{candcon}
\end{eqnarray}

The canonical d--connection $\ ^{s}\widehat{\mathbf{D}}$ is characterized by
nonholonomically induced torsion\ d--tensor (\ref{dt}) which is completely
defined by $\ ^{s}\mathbf{g}$ (\ref{dm}) for any chosen $\ ^{s}\mathbf{N=\{}%
N_{i_{s}}^{a_{s}}\}.$\footnote{%
It should be noted that such a torsion is different from the torsions
considered in Einstein-Cartan gauge type and string gravity theories with
absolutely antisymmetric torsion. In those theories, the torsion fields are
independent from the metric/vielbein one and may have proper sources. In our
approach, the torsion $\ ^{s}\widehat{\mathcal{T}}$ is completely determined
by the d--metric and N--connection coefficients. There are unnecessary
additional sources because such a d--torsion is determined by the
nonholonomic structures. For certain additional assumptions, we can relate
it's coefficients, for instance, to a subclass of nontrivial coefficients of
an absolute antisymmetric torsion in string gravity.} The N--adapted
coefficients can be computed if the coefficients (\ref{candcon}) are
introduced "shell by shell" into formulas
\begin{equation}
\widehat{T}_{\ j_{s}k_{s}}^{i_{s}}=\widehat{L}_{j_{s}k_{s}}^{i_{s}}-\widehat{%
L}_{k_{s}j_{s}}^{i_{s}},\widehat{T}_{\ j_{s}a_{s}}^{i_{s}}=\widehat{C}%
_{j_{s}b_{s}}^{i_{s}},\widehat{T}_{\ j_{s}i_{s}}^{a_{s}}=-\Omega _{\
j_{s}i_{s}}^{a_{s}},\ \widehat{T}_{a_{s}j_{s}}^{c_{s}}=\widehat{L}%
_{a_{s}j_{s}}^{c_{s}}-e_{a_{s}}(N_{j_{s}}^{c_{s}}),\widehat{T}_{\
b_{s}c_{s}}^{a_{s}}=\ \widehat{C}_{b_{s}c_{s}}^{a_{s}}-\ \widehat{C}%
_{c_{s}b_{s}}^{a_{s}}.  \label{dtorss}
\end{equation}%
The N-adapted formulas (\ref{candcon}) and (\ref{dtorss}) show that any
coefficient with such objects computed in 4-d can be similarly extended
"shell by shell" by any value $s=1,2$ and $3$, redefining the h- and
v-indices correspondingly.

Any (pseudo) Riemannian geometry can be equivalently formulated on
nonholonomic variables $(\ ^{s}\mathbf{g}$ (\ref{dm}), $\ ^{s}\mathbf{N}, \
^{s}\widehat{\mathbf{D}})$ or using the standard ones $(\ ^{s}\mathbf{g}$ (%
\ref{metr})$\mathbf{,}\ ^{s}\nabla ).$ This follows from the fact that both
linear connections $\ ^{s}\nabla $ and $\ ^{s}\widehat{\mathbf{D}}$ are
defined by the same metric structure via a canonical distortion relation
\begin{equation}
\ ^{s}\widehat{\mathbf{D}}=\ ^{s}\nabla +\ ^{s}\widehat{\mathbf{Z}}.
\label{distorsrel}
\end{equation}%
The distorting tensor $\ ^{s}\widehat{\mathbf{Z}}=\{\widehat{\mathbf{\ Z}}%
_{\ \beta _{s}\gamma _{s}}^{\alpha _{s}}\}$ can also be constructed from the
same metric $\ ^{s}\mathbf{g}$ (\ref{dm}). The values $\widehat{\mathbf{\ Z}}%
_{\ \beta _{s}\gamma _{s}}^{\alpha _{s}}$ are algebraic combinations of $%
\widehat{T}_{\ \beta _{s}\gamma _{s}}^{\alpha _{s}}$ and vanish for zero
torsion. The linear connections $\ ^{s}\nabla $ and $\ ^{s}\widehat{\mathbf{D%
}}$ are not tensor objects. It is possible to consider frame/ coordinate
transforms for certain parametrization $\ ^{s}\mathbf{N=\{}%
N_{i_{s}}^{a_{s}}\}$ when the conditions $\ _{\shortmid }\Gamma _{\ \alpha
_{s}\beta _{s}}^{\gamma _{s}}=\widehat{\mathbf{\Gamma }}_{\ \alpha _{s}\beta
_{s}}^{\gamma _{s}}$ are satisfied with respect to some N--adapted frames.
In general, $\ ^{s}\nabla \neq \ ^{s}\widehat{\mathbf{D}}$ and the
corresponding curvature tensors $\ _{\shortmid }R_{\ \beta _{s}\gamma
_{s}\delta _{s}}^{\alpha _{s}}\neq \widehat{\mathbf{R}}_{\ \beta _{s}\gamma
_{s}\delta _{s}}^{\alpha _{s}}.$

We note that we can always extract LC--configurations with zero torsion if
we impose additionally for (\ref{dtorss}) the conditions
\begin{equation}
\widehat{\mathbf{T}}_{\ \alpha _{s}\beta _{s}}^{\gamma _{s}}=0,
\label{zerotors}
\end{equation}%
In general, $W_{\alpha _{s}\beta _{s}}^{\gamma _{s}}$ (\ref{anhrel}) may be
not zero. We write this in the form $\ ^{s}\widehat{\mathbf{D}}_{\mid
\widehat{\mathcal{T}}=0}\rightarrow \ ^{s}\nabla $. Such nonholonomic
constraints may be stated in non--explicit forms and are not obligated to be
described by limits of certain smooth functions.

\subsection{The Riemann and Ricci d--tensors of the canonical d--connection}

The N--adapted coefficients of curvature d--tensor $\mathcal{R}_{~\beta
_{s}}^{\alpha _{s}}=\{\mathbf{\mathbf{R}}_{\ \ \beta _{s}\gamma _{s}\delta
_{s}}^{\alpha _{s}}\}$ (\ref{dc}) of $\ $the canonical d--connection $\ ^{s}%
\widehat{\mathbf{D}}$ (\ref{candcon}) and $\ ^{s}\mathbf{g}$ (\ref{dm}) are
given by formulas
\begin{eqnarray}
\widehat{R}_{\ h_{s}j_{s}k_{s}}^{i_{s}} &=&\mathbf{e}_{k_{s}}\widehat{L}_{\
h_{s}j_{s}}^{i_{s}}-\mathbf{e}_{j_{s}}\widehat{L}_{\ h_{s}k_{s}}^{i_{s}}+%
\widehat{L}_{\ h_{s}j_{s}}^{m_{s}}\widehat{L}_{\ m_{s}k_{s}}^{i_{s}}-%
\widehat{L}_{\ h_{s}k_{s}}^{m_{s}}\widehat{L}_{\ m_{s}j_{s}}^{i_{s}}-%
\widehat{C}_{\ h_{s}a_{s}}^{i_{s}}\Omega _{\ k_{s}j_{s}}^{a_{s}},  \notag \\
\widehat{R}_{\ b_{s}j_{s}k_{s}}^{a_{s}} &=&\mathbf{e}_{k_{s}}\widehat{L}_{\
b_{s}j_{s}}^{a_{s}}-\mathbf{e}_{j_{s}}\widehat{L}_{\ b_{s}k_{s}}^{a_{s}}+%
\widehat{L}_{\ b_{s}j_{s}}^{c_{s}}\widehat{L}_{\ c_{s}k_{s}}^{a_{s}}-%
\widehat{L}_{\ b_{s}k_{s}}^{c_{s}}\widehat{L}_{\ c_{s}j_{s}}^{a_{s}}-%
\widehat{C}_{\ b_{s}c_{s}}^{a_{s}}\Omega _{\ k_{s}j_{s}}^{c_{s}},  \notag \\
\widehat{P}_{\ j_{s}k_{s}a_{s}}^{i_{s}} &=&e_{a_{s}}\widehat{L}_{\
j_{s}k_{s}}^{i_{s}}-\widehat{D}_{k_{s}}\widehat{C}_{\ j_{s}a_{s}}^{i_{s}}+%
\widehat{C}_{\ j_{s}b_{s}}^{i_{s}}\widehat{T}_{\ k_{s}a_{s}}^{b_{s}},\
\label{dcurvc} \\
\widehat{P}_{\ b_{s}k_{s}a_{s}}^{c_{s}} &=&e_{a_{s}}\widehat{L}_{\
b_{s}k_{s}}^{c_{s}}-D_{k_{s}}\widehat{C}_{\ b_{s}a_{s}}^{c_{s}}+\widehat{C}%
_{\ b_{s}d_{s}}^{c_{s}}\widehat{T}_{\ k_{s}a_{s}}^{c_{s}},  \notag \\
\widehat{S}_{\ j_{s}b_{s}c_{s}}^{i_{s}} &=&e_{c_{s}}\widehat{C}_{\
j_{s}b_{s}}^{i_{s}}-e_{b_{s}}\widehat{C}_{\ j_{s}c_{s}}^{i_{s}}+\widehat{C}%
_{\ j_{s}b_{s}}^{h_{s}}\widehat{C}_{\ h_{s}c_{s}}^{i_{s}}-\widehat{C}_{\
j_{s}c_{s}}^{h_{s}}\widehat{C}_{\ h_{s}b_{s}}^{i_{s}},  \notag \\
\widehat{S}_{\ b_{s}c_{s}d_{s}}^{a_{s}} &=&e_{d_{s}}\widehat{C}_{\
b_{s}c_{s}}^{a_{s}}-e_{c_{s}}\widehat{C}_{\ b_{s}d_{s}}^{a_{s}}+\widehat{C}%
_{\ b_{s}c_{s}}^{e_{s}}\widehat{C}_{\ e_{s}d_{s}}^{a_{s}}-\widehat{C}_{\
b_{s}d_{s}}^{e_{s}}\widehat{C}_{\ e_{s}c_{s}}^{a_{s}},  \notag
\end{eqnarray}%
computed respectively for all shells $s=0,1,2,3.$

The Ricci d--tensor $\widehat{R}ic=\{\widehat{\mathbf{R}}_{\alpha _{s}\beta
_{s}}:=\widehat{\mathbf{R}}_{\ \alpha _{s}\beta _{s}\tau _{s}}^{\tau _{s}}\}$
of $\ ^{s}\widehat{\mathbf{D}}$ is introduced via a respective contraction
of coefficients of the curvature tensor (\ref{dcurvc}), when
\begin{equation}
\widehat{\mathbf{R}}_{\alpha _{s}\beta _{s}}=\{\widehat{R}_{h_{s}j_{s}}:=%
\widehat{R}_{\ h_{s}j_{s}i_{s}}^{i_{s}},\ \ \widehat{R}_{j_{s}a_{s}}:=-%
\widehat{P}_{\ j_{s}i_{s}a_{s}}^{i_{s}},\ \widehat{R}_{b_{s}k_{s}}:=\widehat{%
P}_{\ b_{s}k_{s}a_{s}}^{a_{s}},\widehat{R}_{\ b_{s}c_{s}}=\widehat{S}_{\
b_{s}c_{s}a_{s}}^{a_{s}}\}.  \label{dricci}
\end{equation}%
By contracting the N--adapted coefficients of the Ricci d--tensor with the
inverse d--metric (computed as the inverse matrix of $\ ^{s}\mathbf{g}$ (\ref%
{dm})), we define and compute the scalar curvature of $\ ^{s}\widehat{%
\mathbf{D}}\mathbf{,}$
\begin{equation}
\ ^{s}\widehat{R}:=\mathbf{g}^{\alpha _{s}\beta _{s}}\widehat{\mathbf{R}}%
_{\alpha _{s}\beta _{s}}=g^{i_{s}j_{s}}\widehat{R}%
_{i_{s}j_{s}}+g^{a_{s}b_{s}}\widehat{R}_{a_{s}b_{s}}=\widehat{R}+\widehat{S}%
+\ ^{1}\widehat{S}+\ ^{2}\widehat{S}+\ ^{3}\widehat{S},  \label{rdsc}
\end{equation}%
with respective h-- and v--components of scalar curvature, $\widehat{R}%
=g^{ij}\widehat{R}_{ij},$ $S=g^{ab}S_{ab},$ $\
^{1}S=g^{a_{1}b_{1}}S_{a_{1}b_{1}},\ ^{2}S=g^{a_{2}b_{2}}S_{a_{2}b_{2}},\
^{3}S=g^{a_{3}b_{3}}R_{a_{3}b_{3}}.$

\subsection{Conventions for 10-d nonholonomic manifolds, d--tensor and
d--spinor indices}

\label{assclifford}In heterotic superstring theory, one considers domain
walls for nonholonomic splitting transform, in general, in nonholonomic
domains. In the framework of conventions for indices and coordinates on
shells (\ref{coordconv}), we consider additional parameterizations {\small
\begin{eqnarray*}
\mu _{s},\alpha _{s},... &=&1,2,...10,%
\mbox{ on a 10-d nonholonomic manifold
with shell coordinates }u^{\mu
_{s}}=(x^{i},y^{a},y^{a_{1}},y^{a_{2}},y^{a_{3}})\mbox{ on }\mathcal{M}; \\
\check{\mu},\check{\alpha},... &=&1,2,...10,%
\mbox{ equivalently, with general
indices and coordinates  }u^{\check{\mu}}\mbox{ on }\mathcal{M}; \\
\ \check{i},\check{j},... &=&1,2,3,%
\mbox{  for a 3-d pseudo-Euclidean
signature }(++-)\mbox{ and coordinates }u^{\check{i}}=x^{\check{i}%
}=(x^{i},y^{3}=t)=(x^{1},x^{2},t); \\
\check{a},\check{b},... &=&5,6,7,8,9,10,\mbox{ with }u^{\check{a}}=y^{\check{%
a}}\mbox{ on N-anholonomic }\ ^{6}\mathbf{X}%
\mbox{ with Euclidian
signature }; \\
\widetilde{a},\widetilde{b},... &=&4,5,6,7,8,9,10,\mbox{ with }u^{\widetilde{%
a}}=y^{\widetilde{a}}\mbox{ on N-anholonomic }\ ^{7}\mathbf{X}%
\mbox{ with
Euclidian signature };
\end{eqnarray*}%
} We shall underline such indices in order to emphasize that we work in a
coordinate base $\partial _{\underline{\check{a}}}$: For instance, $A^{%
\check{a}}\mathbf{e}_{\check{a}}=A^{\underline{\check{a}}}\partial _{%
\underline{\check{a}}}$ if we consider decompositions of a d--vector.

Anti--symmetrization is performed with a factorial factor, for instance, $%
A_{[\check{\mu}}B_{\check{\alpha}]}:=\frac{1}{2}(A_{\check{\mu}}B_{\check{%
\alpha}}-B_{\check{\alpha}}A_{\check{\mu}})$. A p--form $\omega $ is
expressed $\omega =\frac{1}{p!}\omega _{\check{\mu}_{1}...\check{\mu}_{p}}%
\mathbf{e}^{\check{\mu}_{1}}\wedge ...\wedge \mathbf{e}^{\check{\mu}_{p}},$
see (\ref{nadaptb}). The Clifford action of such a form on a spinor $%
\epsilon $ can be defined in N--adapted form as $\ \omega \cdot \epsilon :=%
\frac{1}{p!}\omega _{\check{\mu}_{1}...\check{\mu}_{p}}\gamma ^{\check{\mu}%
_{1}...\check{\mu}_{p}}\epsilon ,$ for $\gamma ^{\check{\mu}_{1}...\check{\mu%
}_{p}}:=\gamma ^{\lbrack \check{\mu}_{1}}...\gamma ^{\check{\mu}_{p}]}$ if
the Clifford d--algebra is introduced following convention%
\begin{equation*}
\{\gamma ^{\check{\mu}},\gamma ^{\check{\nu}}\}:=\gamma ^{\check{\mu}}\gamma
^{\check{\nu}}+\gamma ^{\check{\nu}}\gamma ^{\check{\mu}}=2\mathbf{g}^{%
\check{\mu}\check{\nu}}
\end{equation*}%
for 10-d gamma matrices $\gamma ^{\check{\mu}}$ with $\mathbf{g}^{\check{\mu}%
\check{\nu}}=\{\mathbf{g}^{\mu _{s}\nu _{s}}\}$ admitting a shell
decomposition (\ref{dm}). A theory of N--adapted spinors and Dirac operators
is elaborated upon \cite{vnpfins,vp,vt,vnrflnc}, see also references
therein. We omit such considerations in this work.

It is proven in \cite{harl} that at first order in $\alpha ^{\prime },$ the
BPS equations are solved by $^{A}\check{\nabla}=\ ^{c}\nabla ,\check{H}=0,%
\check{\phi}=const,$ where $\ ^{c}\nabla $ is the LC--connection on $\check{c%
}(\ ^{6}X).$ A series of less trivial solutions with $\check{H}\neq 0$ have
been studied in \cite{harl,gemmer,klaupt,gray,lukas,lecht1} under
assumptions that the gauge field is chosen to be an instanton and within the
framework of dynamic $SU(3)$ structures. In this work, we shall extend those
results in string theory by proving that the equations of motion of
heterotic string supergravity can be decoupled and solved in very general
off--diagonal forms with dependence, in principle, of all 10-d spacetime
cooridinates (using the AFDM for higher dimensions\ \cite%
{tgovsv,vtamsuper,vex3}). In order to preserve certain relations to former
"holonomic" solutions, we shall work with N--anholonomic manifolds $\mathcal{M%
}$ and $\ ^{6}\mathbf{X}$ determined by d--metric (\ref{dm}). We shall prove
that for any solution of certain generalized/modified 10-d Einstein
equations for $\ ^{s}\widehat{\mathbf{D}}_{\mid \widehat{\mathcal{T}}%
=0}\rightarrow \ ^{s}\nabla $ and additional extensions with gauge and
scalar fields, the submanifold $\ ^{6}\mathbf{X}$ $\ $can be endowed with
almost-K\"{a}hler variables \cite{vjgp,vdq1,vmedit}. As a result, the
curvature $\widetilde{\mathbf{R}}$ of d--connection $\widetilde{\mathbf{D}}$
can be determined for arbitrary solutions of the equations of motion of
heterotic supergravity with generic off--diagonal interactions,
nonholonomically deformed connections and various parameterizations of
effective sources.

\end{document}